\documentclass[twocolumn,trackchanges]{aastex631}
\usepackage{amsmath}
\usepackage{url}
\usepackage{xcolor}
\usepackage{soul}
\turnoffedit

\shorttitle{Two Color Component Model}
\shortauthors{Hand et. Al}
\graphicspath{{./}{figures/}}

\DeclareMathOperator{\clone}{\mathbf{L}_1}
\DeclareMathOperator{\cltwo}{\mathbf{L}_2}
\DeclareMathOperator{\clbv}{{\mathcal{L}}}

\DeclareMathOperator{\cbv}{\hat{\mathbf{e}}}

\begin{document}

\title{An Agnostic Approach To Building Empirical Type Ia Supernova Light Curves: Evidence for Intrinsic Chromatic Flux Variation using Nearby Supernova Factory Data}
\author[0000-0001-7260-4274]{Jared Hand}
\affiliation{
    Pittsburgh Particle Physics, Astrophysics, and Cosmology Center (PITT PACC)\\
    Department of Physics and Astronomy, University of Pittsburgh, Pittsburgh, PA 15260, USA
}

\author[0000-0001-6315-8743]{A.~G.~Kim}
\affiliation{Physics Division, Lawrence Berkeley National Laboratory, 1 Cyclotron Road, Berkeley, CA, 94720}

\author{G.~Aldering}
\affiliation{Physics Division, Lawrence Berkeley National Laboratory, 1 Cyclotron Road, Berkeley, CA, 94720}

\author[0000-0002-0389-5706]{P.~Antilogus}
\affiliation{Laboratoire de Physique Nucl\'eaire et des Hautes Energies, CNRS/IN2P3, \\ Sorbonne Universit\'e, Universit\'e de Paris, 4 place Jussieu, 75005 Paris, France}

\author[0000-0002-9502-0965]{C.~Aragon}
\affiliation{Physics Division, Lawrence Berkeley National Laboratory, 1 Cyclotron Road, Berkeley, CA, 94720}
\affiliation{College of Engineering, University of Washington 371 Loew Hall, Seattle, WA, 98195}

\author{S.~Bailey}
\affiliation{Physics Division, Lawrence Berkeley National Laboratory, 1 Cyclotron Road, Berkeley, CA, 94720}

\author[0000-0003-0424-8719]{C.~Baltay}
\affiliation{Department of Physics, Yale University, New Haven, CT, 06250-8121}

\author{S.~Bongard}
\affiliation{Laboratoire de Physique Nucl\'eaire et des Hautes Energies, CNRS/IN2P3, \\ Sorbonne Universit\'e, Universit\'e de Paris, 4 place Jussieu, 75005 Paris, France}

\author[0000-0002-5828-6211]{K.~Boone}
\affiliation{Physics Division, Lawrence Berkeley National Laboratory, 1 Cyclotron Road, Berkeley, CA, 94720}
\affiliation{Department of Physics, University of California Berkeley, 366 LeConte Hall MC 7300, Berkeley, CA, 94720-7300}
\affiliation{DIRAC Institute, Department of Astronomy, University of Washington, 3910 15th Ave NE, Seattle, WA 98195, USA}

\author[0000-0002-3780-7516]{C.~Buton}
\affiliation{Univ Lyon, Universit\'e Claude Bernard Lyon~1, CNRS/IN2P3, IP2I Lyon, F-69622, Villeurbanne, France}
   
\author[0000-0002-5317-7518]{Y.~Copin}
\affiliation{Univ Lyon, Universit\'e Claude Bernard Lyon~1, CNRS/IN2P3, IP2I Lyon, F-69622, Villeurbanne, France}

\author[0000-0003-1861-0870]{S.~Dixon}
\affiliation{Physics Division, Lawrence Berkeley National Laboratory, 1 Cyclotron Road, Berkeley, CA, 94720}
\affiliation{Department of Physics, University of California Berkeley, 366 LeConte Hall MC 7300, Berkeley, CA, 94720-7300}

\author[0000-0002-7496-3796]{D.~Fouchez}
\affiliation{Aix Marseille Univ, CNRS/IN2P3, CPPM, Marseille, France}

\author[0000-0001-6728-1423]{E.~Gangler}  
\affiliation{Univ Lyon, Universit\'e Claude Bernard Lyon~1, CNRS/IN2P3, IP2I Lyon, F-69622, Villeurbanne, France}
\affiliation{Universit\'e Clermont Auvergne, CNRS/IN2P3, Laboratoire de Physique de Clermont, F-63000 Clermont-Ferrand, France}

\author[0000-0003-1820-4696]{R.~Gupta}
\affiliation{Physics Division, Lawrence Berkeley National Laboratory, 1 Cyclotron Road, Berkeley, CA, 94720}

\author[0000-0001-9200-8699]{B.~Hayden}
\affiliation{Physics Division, Lawrence Berkeley National Laboratory, 1 Cyclotron Road, Berkeley, CA, 94720}
\affiliation{Space Telescope Science Institute, 3700 San Martin Drive Baltimore, MD, 21218}

\author{W.~Hillebrandt}
\affiliation{Max-Planck-Institut f\"ur Astrophysik,  Karl-Schwarzschild-Str. 1, D-85748 Garching, Germany}

\author[0000-0003-2495-8670]{Mitchell Karmen}
\affiliation{Physics Division, Lawrence Berkeley National Laboratory, 1 Cyclotron Road, Berkeley, CA, 94720}

\author[0000-0001-8594-8666]{M.~Kowalski}
\affiliation{Institut f\"ur Physik,  Humboldt-Universitat zu Berlin, Newtonstr. 15, 12489 Berlin}
\affiliation {DESY, D-15735 Zeuthen, Germany}

\author[0000-0002-9207-4749]{D.~K\"usters}
\affiliation {Department of Physics, University of California Berkeley, 366 LeConte Hall MC 7300, Berkeley, CA, 94720-7300}
\affiliation {DESY, D-15735 Zeuthen, Germany}

\author[0000-0002-8357-3984]{P.-F.~L\'eget}
\affiliation{Laboratoire de Physique Nucl\'eaire et des Hautes Energies, CNRS/IN2P3, \\ Sorbonne Universit\'e, Universit\'e de Paris, 4 place Jussieu, 75005 Paris, France}

\author{F.~Mondon}  
\affiliation{Universit\'e Clermont Auvergne, CNRS/IN2P3, Laboratoire de Physique de Clermont, F-63000 Clermont-Ferrand, France}

\author[0000-0001-8342-6274]{J.~Nordin}
\affiliation{Physics Division, Lawrence Berkeley National Laboratory, 1 Cyclotron Road, Berkeley, CA, 94720}
\affiliation{Institut f\"ur Physik,  Humboldt-Universitat zu Berlin, Newtonstr. 15, 12489 Berlin}

\author[0000-0003-4016-6067]{R.~Pain}
\affiliation{Laboratoire de Physique Nucl\'eaire et des Hautes Energies, CNRS/IN2P3, \\ Sorbonne Universit\'e, Universit\'e de Paris, 4 place Jussieu, 75005 Paris, France}

\author{E.~Pecontal}
\affiliation{Centre de Recherche Astronomique de Lyon, Universit\'e Lyon 1, 9 Avenue Charles Andr\'e, 69561 Saint Genis Laval Cedex, France}

\author{R.~Pereira}
\affiliation{Univ Lyon, Universit\'e Claude Bernard Lyon~1, CNRS/IN2P3, IP2I Lyon, F-69622, Villeurbanne, France}

\author[0000-0002-4436-4661]{S.~Perlmutter}
\affiliation{Physics Division, Lawrence Berkeley National Laboratory, 1 Cyclotron Road, Berkeley, CA, 94720}
\affiliation{Department of Physics, University of California Berkeley, 366 LeConte Hall MC 7300, Berkeley, CA, 94720-7300}

\author[0000-0002-8207-3304]{K.~A.~Ponder}
\affiliation{Department of Physics, University of California Berkeley, 366 LeConte Hall MC 7300, Berkeley, CA, 94720-7300}

\author{D.~Rabinowitz}
\affiliation{Department of Physics, Yale University, New Haven, CT, 06250-8121}

\author[0000-0002-8121-2560]{M.~Rigault}
\affiliation{Univ Lyon, Universit\'e Claude Bernard Lyon~1, CNRS/IN2P3, IP2I Lyon, F-69622, Villeurbanne, France}

\author[0000-0001-5402-4647]{D.~Rubin}
\affiliation{Department of Physics and Astronomy, University of Hawai`i, 2505 Correa Rd, Honolulu, HI, 96822}
\affiliation{Physics Division, Lawrence Berkeley National Laboratory, 1 Cyclotron Road, Berkeley, CA, 94720}

\author{K.~Runge}
\affiliation{Physics Division, Lawrence Berkeley National Laboratory, 1 Cyclotron Road, Berkeley, CA, 94720}

\author[0000-0002-4094-2102]{C.~Saunders}
\affiliation{Physics Division, Lawrence Berkeley National Laboratory, 1 Cyclotron Road, Berkeley, CA, 94720}
\affiliation{Department of Physics, University of California Berkeley, 366 LeConte Hall MC 7300, Berkeley, CA, 94720-7300}
\affiliation{Princeton University, Department of Astrophysics, 4 Ivy Lane, Princeton, NJ, 08544}
\affiliation{Sorbonne Universit\'es, Institut Lagrange de Paris (ILP), 98 bis Boulevard Arago, 75014 Paris, France}

\author{N.~Suzuki}   
\affiliation{Physics Division, Lawrence Berkeley National Laboratory, 1 Cyclotron Road, Berkeley, CA, 94720}
\affiliation{Kavli Institute for the Physics and Mathematics of the Universe, The University of Tokyo Institutes for Advanced Study, \\ The University of Tokyo, 5-1-5 Kashiwanoha, Kashiwa, Chiba 277-8583, Japan}

\author{C.~Tao}
\affiliation{Tsinghua Center for Astrophysics, Tsinghua University, Beijing 100084, China}
\affiliation{Aix Marseille Univ, CNRS/IN2P3, CPPM, Marseille, France}

\author[0000-0002-4265-1958]{S.~Taubenberger}
\affiliation{Max-Planck-Institut f\"ur Astrophysik, Karl-Schwarzschild-Str. 1, D-85748 Garching, Germany}

\author{R.~C.~Thomas}
\affiliation{Physics Division, Lawrence Berkeley National Laboratory, 1 Cyclotron Road, Berkeley, CA, 94720}
\affiliation{Computational Cosmology Center, Computational Research Division, \\ Lawrence Berkeley National Laboratory, 1 Cyclotron Road, Berkeley, CA, 94720}

\author{M.~Vincenzi}
\affiliation{Physics Division, Lawrence Berkeley National Laboratory, 1 Cyclotron Road, Berkeley, CA, 94720}
\affiliation{Institute of Cosmology and Gravitation, University of Portsmouth, Portsmouth, PO1 3FX, UK}

\collaboration{50}{The Nearby Supernova Factory}

\begin{abstract}
    We present a new empirical type Ia supernova (SN~Ia) model with three chromatic flux variation templates: one phase-dependent and two phase-independent. 
    No underlying dust extinction model or patterns of intrinsic variability are assumed.    
    Implemented with {\tt Stan} and trained using spectrally-binned Nearby Supernova Factory spectrophotometry, we examine this model's two-dimensional, phase-independent flux variation space using two motivated basis representations. 
    In both, the first phase-independent template captures variation that appears dust-like, while the second captures a combination of effectively intrinsic variability and second order dust-like effects. 
    We find that $\approx13$\% of modeled phase-independent flux variance is not dust-like. 
    Previous empirical SN~Ia models either assume an effective dust extinction recipe in their architecture, or only allow for a single mode of phase-independent variation. 
    The presented results demonstrate such an approach may be insufficient, because it could `leak' noticeable intrinsic variation into phase-independent templates. 
\end{abstract}

\section{Introduction}\label{sec:intro}
The discovery of dark energy with standardized type~Ia supernovae (SNe~Ia) solidified these transient objects' importance to cosmology \citep{Perlmutter99,Riess98}. 
SNe~Ia exhibit markedly similar peak \textit{B}-band brightness dispersion of only $\sim1$~mag, and this dispersion can be reduced further with multi-filter photometric time series (or light curves) by exploiting correlations between light curve duration (also called shape, width, or stretch) and color with \textit{B}-band maximum brightness \citep{Phillips93, Riess1996, Perlmutter1997, Tripp1998}.
These standardization results reduce brightness dispersion by an order of magnitude to $\approx0.12$~mags \citep{Scolnic18}. 
Standardization using spectra can reduce the dispersion even further, to $\approx0.08$~mags \citep{Fakhouri2015,Boone2021b}. 

There are two errors that limit SNe~Ia's capacity to constrain cosmological parameters: the number of observed SNe~Ia (statistical uncertainty), and errors resulting from observation bias, modeling error, and calibration (systematics). 
Recent SN~Ia cosmology analyses have significantly reduced statistical uncertainty by utilizing over $10^{3}$ spectroscopically confirmed SNe~Ia \citep{Betoule14,Scolnic18}. 
LSST, via the Vera Rubin Observatory, will further increase our usable SN~Ia sample by at least an order of magnitude over ten years after its commissioning \citep{Ivezic2019}. 
Although internal photometric calibration will remain an important systematic to account for, LSST will alleviate tedious inter-survey photometric calibration systematics performed in many past analyses while still providing impressive statistics. 
As a result, LSST will increase the relative importance of systematics arising from SN~Ia light curve modeling and standardization for constraining cosmological parameters.  

SNe~Ia result from a catastrophic disruption of carbon-oxygen white dwarfs \citep{Hoyle1960}. 
Potential progenitor scenarios include accretion from a white dwarf's companion star or merging/collision of white dwarf binary constituents \citep{Whelan+Iben1973,Iben+Tutukov1984}. 
Unfortunately, accounting for observed SN~Ia spectral variation while simultaneously recovering established empirical relations within the context of a detonating (or deflagrating) white dwarf framework remains a daunting and incomplete task. 
For example, it is suggested that progenitor mass can extend below an otherwise expected white dwarf mass limit, or Chandrasekhar mass, of $M_{\odot}=1.4$ \citep{Scalzo2014}.  
See~\cite{Soker2019} for a recent review of and summary of SN~Ia theoretical modeling progress and unanswered questions. 
Usable parametric theoretical SN~Ia models remain elusive, leaving us reliant on empirical models trained from SN~Ia observations. 

\subsection{Photometric Variation and Empirical Models}\label{sec:intro:photvar}
As mentioned, light curve width correlates with \textit{B}-band maximum brightness so that longer duration SNe~Ia are systematically brighter \citep{Phillips93}. 
We refer to this as the width-luminosity relation (WLR). 
Similarly, bluer SN~Ia light curves are systematically brighter at \textit{B}-band maximum, which we similarly refer to as the color-luminosity relation (CLR) \citep{Riess1996}. 

One can interpret SN~Ia empirical models as transforming high-dimensional sets of observations to a lower-dimensional set of parametric latent templates that inscribe dominant modes of SN~Ia variability. 
The first generation of models attempted to describe light curves in standard rest-frame filters using templates specific to those filters 
\citep{Riess1996, Jha2007, Burns2011}, which when applied to measurements made with different rest-frame filters required a K-correction
\citep{Kim1996,Nugent2002} to convert the data to the
model photometric system.
The limitations of models based on broad-band templates were quickly recognized: K-corrections carry their own set of uncertainties and biases that are correlated with model parameters \citep{Mandel2022}.
More importantly, by restricting the templates to broadband, these models were insensitive to substantially more complex SN Ia variability revealed through spectroscopy.

SN~Ia spectral variability is either intrinsic to SN~Ia populations or is extrinsic, instead arising from processes external to the explosion, such as dust extinction or an SN~Ia's interaction with its circumstellar environments \citep{Nugent2002,Jha2007}. 
Furthermore, photometric relationships emerge from spectral variation, with the temperature dependence of Fe line blanketing at least partly driving the WLR and explaining its wavelength dependence being such an example \citep{Kasen2007}. 
Variation in progenitor mass also contributes to the WLR, as lower-mass SN~Ia progenitors are systematically dimmer and have faster-declining light curves compared to their more massive counterparts \citep{Scalzo2014}. 
Certain spectral features directly correlate with photometric SN~Ia properties, such as the F(6420~\r{A})/F(4430~\r{A}) line ratio correlating with maximum $B$-band brightness \citep{Bailey2009}, and the ratio of Si~II $\lambda\lambda$5972 and 6355~\r{A} (or Si~II $\lambda\lambda$5972 and 3858~\r{A}) correlating with light curve width \citep{Nugent1995}. 

Many SN~Ia subtypes have been categorized by their spectral variation \citep{Branch2009, Blondin2012}. 
Grouping SNe~Ia based on Si~II velocity at maximum brightness during a Tripp standardization procedure \citep{Tripp1998} has been shown to reduce SN~Ia dispersion post-standardization more than use of only color and stretch alone \citep{Wang2009,Foley2013b}. 
Furthermore, spectral information can improve effective total-to-selective extinction $R_V$ estimation, with~\cite{Chotard2011} using spectral features to recover an effective $R_V$ value consistent the Milky Way average $R_V=3.1$. 
Given this plethora of spectral variety within SNe~Ia, and this variety's potential to further improve standardization, commonly used and recent SN~Ia models make heavy use of spectroscopic observations in training.

Most recent SN~Ia models reduce the dimensionality of SN~Ia observations by constructing combinations of underlying spectral or color variation templates, with one template capturing the average, or fiducial, spectral evolution of SNe~Ia. 
This approach removes any need for K-corrections and related uncertainty propagation \citep{Guy2007, Mandel2022}. 
The ubiquitous SALT model family (and its cousin SiFTO) are the canonical example of the spectral template technique \citep{Guy2007, Conley2008, Betoule14,Pierel2022}. 
This family of linear SN~Ia models capture variation beyond a mean spectral surface with a first-order flux variation template and a phase-independent color template (with per-SN contribution parameters $x_1$ and $c$, respectively). 
SALT2's success over prior models saw its widespread adoption and continuous improvement, with the most recent version SALT3 extending its wavelength coverage to the near-infrared (NIR) \citep{Kenworthy2021}. 
More statistically rigorous linear spectral template models have also been developed, such as BayeSN with its potent hierarchical Bayesian framework \citep{Mandel2017, Mandel2022, Thorp2021}. 

A plethora of sophisticated models have recently been developed using SNfactory's spectrophotometric time series \citep{Aldering2002}. 
\cite{Saunders2018} and their SNEMO model extracts up to 15 linear principal functional components from a set of SN~Ia spectral surfaces trained using Gaussian processes to maximally explain SN~Ia variation. 
Alternatively,~\cite{Leget2020} in their SUGAR model treats SN~Ia variation as a linear combination of spectral index templates, extending the initial work of~\cite{Chotard2011} into a fully generative model. 
SNEMO and SUGAR, along with all the models mentioned so far, utilize linear dimensionality reduction techniques. 

\cite{Boone2021} apply the Isomap method with their Twins Embedding model to train a nonlinear parameterization of intrinsic SN~Ia variation at maximum brightness, while~\cite{Stein2022} introduce a nonlinear probabilistic autoencoder (PAE) that captures intrinsic variation across both wavelength and phase. 
Both find that a nonlinear approach requires only three intrinsic model components to describe SN~Ia spectral variation where more traditional linear principal component analysis model would require seven components or more \citep{Saunders2018}. 
These two models also demonstrate noticeable improvements over SALT2 in standardized SN~Ia dispersion from $\approx 0.12$~mags to $\lesssim 0.09$~mags.
Improvements through nonlinear technique application are not limited to light curve models:~\cite{Rubin2015} introduce the hierarchical Bayesian framework UNITY that allows for nonlinear standardization, leading again to improved SN~Ia dispersion post-standardization relative to the linear Tripp approach. 

\subsection{Shortcomings in Phase-Independent Modeling}\label{sec:intro:ourmodel}
All past models either assume a dust extinction model for explicit phase-independent templates (MCLS2k2, SNooPY, SNEMO, SUGAR, BayeSN, and current nonlinear models) or include a single phase-independent template which does not differentiate between intrinsic and extrinsic variation (the SALT family). 
Excluding the maximum brightness model Twins Embedding, each of these models characterize phase-independent variability with only a single model component. 
Physical considerations alone demonstrate this to be an insufficient treatment. 
As summarized in~\cite{Weingartner2001}, one would expect at least two extrinsic variation parameters per SN~Ia: one gauging dust column density or optical depth (i.e. $A_V$) and the other probing second order characteristics such as dust grain properties (i.e. $R_V$). 
Furthermore, it is plausible that empirical SN~Ia models could extract intrinsic variation into a phase-independent template set.  
In this era of precision cosmology, modeling and standardization systematics remain stubborn obstacles to maximizing current and future SN~Ia survey utility.  
Better understanding underlying extracted modes of phase-independent variation could answer outstanding questions about the SN~Ia population (i.e. the low SN~Ia $R_V$ debate or the source of the bias associated with host properties), and improve both SN~Ia modeling and standardization. 

We present a new SN~Ia empirical model to more deeply explore the phase-independent variability of SNe~Ia. 
This model features three chromatic flux variation templates: one phase-dependent and two phase-independent. 
These two phase-dependent components provide the flexibility to account for multi-parameter dust models while also absorbing an intrinsic time-averaged flux variation beyond that accounted for by the phase-dependent component. 
All templates are physics-agnostic, as no assumptions are made about expected spectral features or dust treatment. 
This new model is trained on SNfactory's rest-frame spectrophotometric time series.

Our model bears some similarities with BayeSN. 
Both models are linear models implemented with {\tt Stan}. 
How phase-independent variability is accounted for varies in approach, though. 
BayeSN implements a single-component dust extinction recipe within a hierarchical model framework, from which they recover an effective $R_V$ that is largely consistent with the Milky Way average.
In contrast, our model has two phase-independent templates, providing it two model degrees of freedom for which no physical assumptions are made. 
Unlike BayeSN, our model does not model spectral surface residuals.  

In Section~\ref{sec:data} we introduce the training set and its quality cuts.  
We describe our model and fitting technique in Section~\ref{sec:method}, with global model template and refit per-SN parameter results being presented in Section~\ref{sec:results}.  
Finally, we provide concluding remarks in Section~\ref{sec:conclusion}.

\section{Data}\label{sec:data}
Between 2004 and 2014 SNfactory observed spectrophotometric time series of nearly 300 SNe~Ia \citep{Aldering2002} with the SuperNova Integral Field Spectrograph (SNIFS, Lantz \citeyear{Lantz2004}). 
SNIFS is continuously mounted at the University of Hawaii 2.2~m telescope, using dual-channel, moderate resolution ($R\sim600-1300$) spectrographs to simultaneously observe transient events from 3200 to 5200~\r{A} and 5100 to 10000~\r{A}, respectively. 
This unique and homogeneous SN~Ia data set is calibrated with CALSPEC and Hamuy standard stars \citep{Bohlin2014,Hamuy1992,Hamuy1994}. 
The photometric calibration method is largely summarized in~\cite{Buton2013} with~\cite{Pereira2013} further describing non-photometric-night calibration. 
Host-galaxy subtraction methodology is presented in~\cite{Bongard2011}. 
Each SNe~Ia has also been fit using SALT2.4 \citep{Betoule14}. 

For our SN~Ia training sample we generate synthetic SN-frame photometry using $n_{\lambda}$ log-distributed top-hat filters from published rest-frame SNfactory spectra 
\citep{Aldering2020}. 
These spectra were first corrected for Milky Way extinction, then their wavelengths were de-redshifted (from an initial range of approximately $0.01 < z < 0.08$), and then they are placed on a relative luminosity scale (i.e., that equivalent to having been observed at $z = 0.05$). Cosmological time dilation is accounted for during de-redshifting.
Observed spectra are in units of $10^{10}\text{erg}\; \text{s}^{-1}\; \text{cm}^{-2}$.
This work does not attempt to fit absolute magnitudes or fit for cosmological parameters, so per-SN redshifts are not used. 
Due to high flux variance at wavelength boundaries, and because most objects have a higher redshift than $z=0.05$, the per-spectra reference frame wavelength range is truncated to between 3350~\r{A} and 8030~\r{A}.

The spectral resolution of this top-hat filter synthetic photometry is flexible --- for this work, we use a modest $n_{\lambda}=10$ filter count. 
These $n_{\lambda}=10$ bins are spaced at constant spectral resolution $R$, providing a wavelength bin size of $\approx 400$~\r{A} resolution for bluer bands and $\approx 500$~\r{A} for redder bands. 
SNfactory data consists of flux densities along a uniform grid of wavelengths. 
Because of this uniform spacing, we simply sum flux densities along the wavelength range defining our top-hat filters and then multiply said sum by the filter's wavelength range to calculate top-hat synthetic photometry:
\begin{equation}
    F_{\text{bin}}=[\lambda_{\text{max}} - \lambda_{\text{min}}]\sum_{i=i_{\text{min}}}^{\text{max}}f_{\lambda_i}.
\end{equation}
Similarly, each corresponding variance spectrum is summed in quadrature to calculate synthetic photometry uncertainties. 

For every observation, for the binned synthetic photometry a signal-to-noise ratio (SNR) of at least $\text{SNR} > 5$ is required. 
It is also required that each SN~Ia have at least eight separate days of observations. 
Where a single SN~Ia has multiple spectra observed within a few hours, the weighted average of these flux values is used as a single effective observation.
This decision was made to avoid introducing two timescales in the sampling of our light curves. 
Given the difficulty in constraining date of maximum in SN~Ia empirical models, we demand that there exist at least one observation two days before SALT2 maximum phase. 
Furthermore, no SNe~Ia with an observation gap greater than four days within a four-day range before and after SALT2 maximum are used. 
SN~Ia light curves do not have significant structure less than four days, so gaps of this size or smaller have no discernible impact on results. 
For consistency, the chosen maximum gap size of four days is the same as our fixed Gaussian process mean predictor length scale hyperparameter later described in Section~\ref{sec:method}. 
These cuts leave 80 SNe~Ia in the training sample. 

\begin{figure}
    \plotone{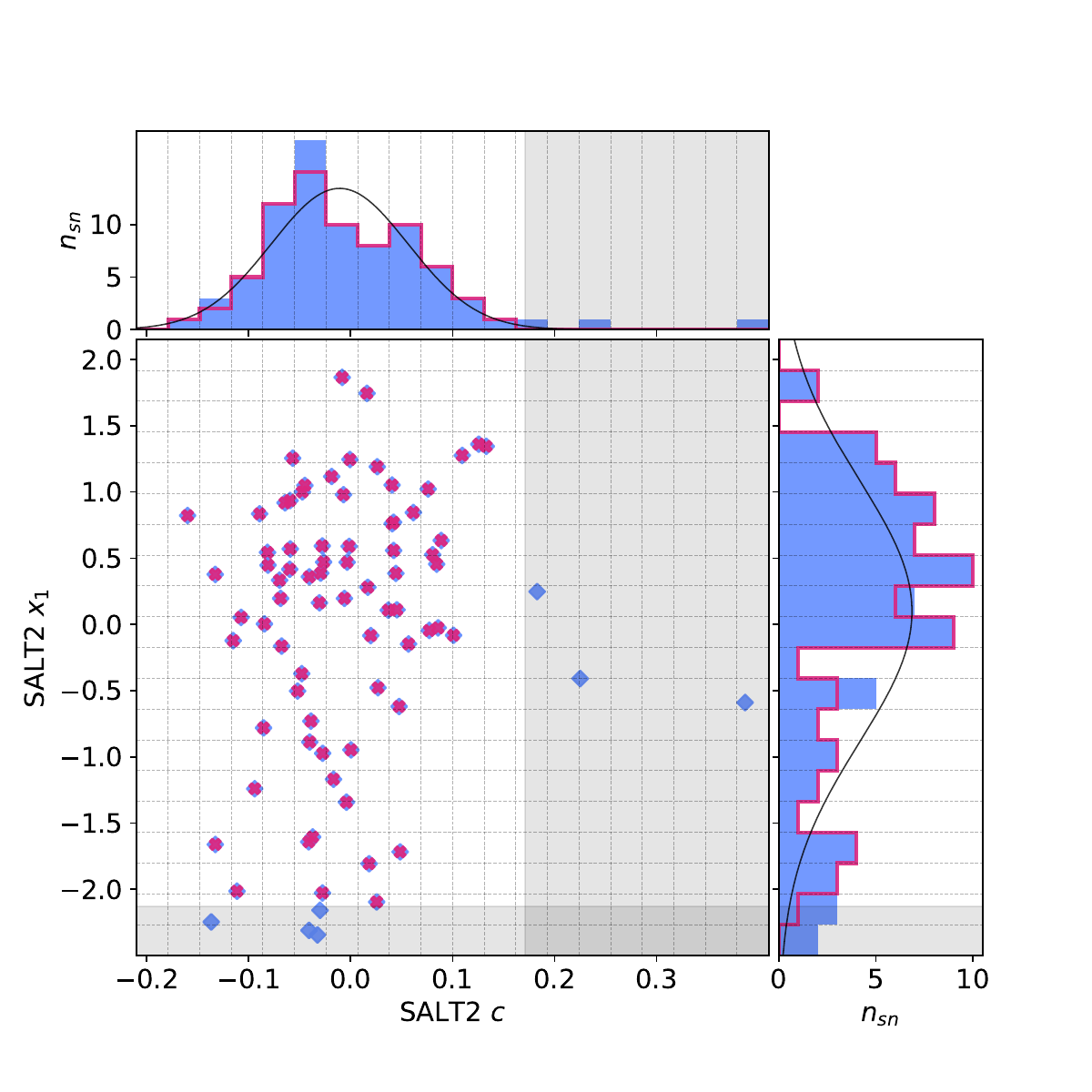}
    \caption{SALT2 $c$ and $x_1$ cuts to better capture a Gaussian `core' for training. 
    The shaded regions correspond to $2\sigma$ clippings along the longer tail of each respective $c$ and $x_1$ distribution; blue points correspond to sigma-clipped supernovae.}
    \label{fig:salt2cutdata}
\end{figure}

The distribution of SN~Ia color parameters for any model is asymmetric due to the positive-definite nature of dust extinction \citep{Scolnic2014,Mandel2017,Brout2020}. 
SN~Ia stretch parameters such as SALT2's stretch proxy $x_1$ are also best modeled with asymmetric distributions \citep{Scolnic2014}. 
We are interested in the Gaussian core of these distributions and partly `symmetrize' the data set by clipping extended tails of our SALT2 $c$ and $x_1$ samples. 
Specifically, a $2\sigma$ clipping is done on each SALT2 $c$ and $x_1$ parameter samples in the direction each parameter's longer tail (Figure~\ref{fig:salt2cutdata}).  
The $c$ clipping prevent heavily reddened SNe~Ia from dominating recovered dust-like behavior and obscuring dust properties of the average SN~Ia in training set --- this $c$ cut removes particularly reddened SNe~Ia with peak apparent $B-V>0.18$. 
The clipping is motivated by our interest in the core of the SN distribution that is used for cosmology, but comes at the expense of removing rarer objects that potentially provide more information on modeling SN colors. 
A total of 73 SNe~Ia remain after SALT2-parameter $\sigma$ clippings, consisting of 1155 individual spectra. 

Relative to~\cite{Aldering2020}, and similar in spirit to \cite{Boone2021}, we remove spectra having poor extractions caused by SNR$<3$; adjusting this threshold higher up to $10$ did not affect our results. 
This step removes seven spectra, leaving 1148 to train the model.

\section{Model}\label{sec:method}
Global template parameters and per-SN parameters are differentiated by upper-case and lower-case characters, respectively. 
This model discretizes phase and wavelength space, using $n_p=16$ phase nodes ranging from $-16\leq t_{p,i}\leq44$ in four-day intervals; as mentioned in Section~\ref{sec:data}, $n_{\lambda}=10$ with bins of constant $R$. 
Each phase-dependent template is an $n_p\times n_{\lambda}$ matrix of parameter nodes, while each phase-independent template is a length-$n_{\lambda}$ vector. 

\begin{figure*}
    \plotone{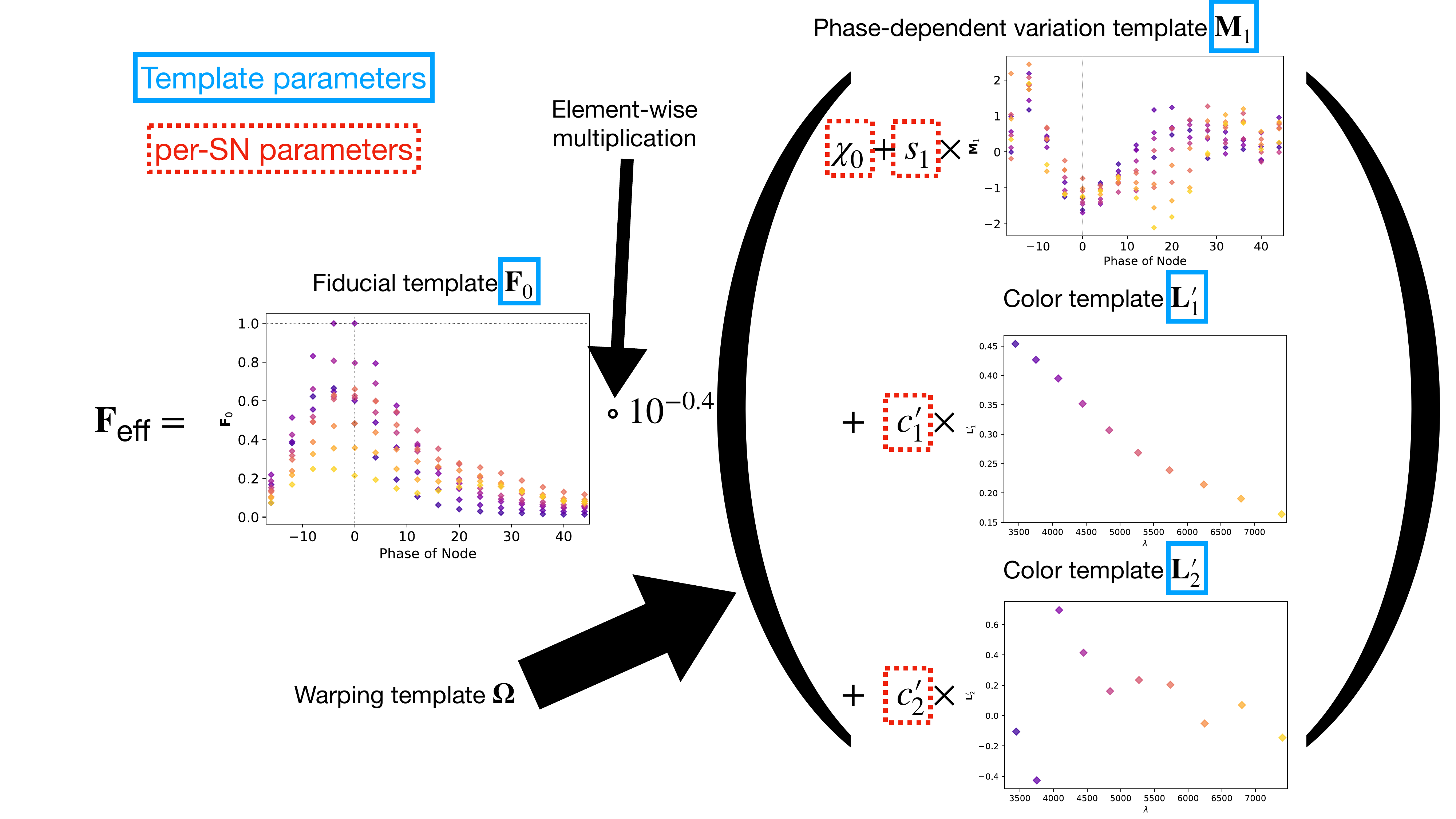}
    \caption{A schematic of our model's flux nodes. 
    Each SN~Ia has an effective flux node matrix $\mathbf{F}_{\text{eff}}$ that is an element-wise product of the sample's fiducial flux template $\mathbf{F}_0$ and a warping matrix $\mathbf{\Omega}$. 
    This warping matrix includes the phase-dependent chromatic flux variation template $\mathbf{M}_1$, two phase-independent chromatic flux variation templates $\clone^{\prime}$ and $\cltwo^{\prime}$ (which make up the two dimension phase-independent chromatic variation model), and per-SN parameters $\chi_0$ (achromatic offset), $s_1$ (phase-dependent chromatic flux variation contribution), and $c_1$ and $c_2$ (phase-independent chromatic flux template contributions). 
    Presented here are the $\clone^{\prime}$ and $\cltwo^{\prime}$ basis representation of the phase-independent templates (see Section~\ref{sec:method:postprocessing} for more information).
    Per-band nodes are presented in this plot having the same figure color.}
    \label{fig:model_cartoon}
\end{figure*}
The model prediction of the time-dependent SED evolution of an individual SN~Ia is based on a temporal interpolation over a set of wavelength-dependent light curves at fixed phases $F_{\lambda,\text{eff}}$ that are specific to that supernova.  
The interpolation is controlled by a kernel $K$, which operates on $F_{\lambda,\text{eff}}$ as shown in Eq.~\ref{eq:fullmodel1}.  
This equation is the same used to predict the mean in a Gaussian process, so we refer to this interpolation scheme as the Gaussian Process Mean Predictor (GPMP)\footnote{The use of GPMPs over other techniques (e.g. spline interpolation) is partly historical.  Our original intention was to treat model functions probabilistically, a use which Gaussian processes would have been suited.  We later opted to use the more conventional approach of using a deterministic function as is done in SALT2.}:
\begin{equation}\label{eq:fullmodel1}
    f_{\lambda}(t) = \mathbf{K}(t - t_0,\mathbf{t}_p)\mathbf{K}^{-1}(\mathbf{t}_p,\mathbf{t}_p)\mathbf{F}_{\lambda,\text{eff}}.
\end{equation} 
These GPMP kernel matrices $\mathbf{K}$ specifically are calculated with a stationary $p=2$ Matérn covariance function $C_{5/2}$ to ensure interpolated curves are twice-differentiable:
\begin{align}
    K_{ij} &= C_{5/2}(|t_{p,i} - t_{p,j}|;\rho, \sigma^2).
\end{align} 
$\mathbf{F}_{\lambda,\text{eff}}$ is the light curve at wavelength $\lambda$ on a grid of phase nodes; all $n_{\lambda}=10$ light curves form the flux node matrix $\mathbf{F}_{\text{eff}}$. 
$t_{p,i}\in \mathbf{t}_p$ is a vector indexing the model's $n_p$ phase nodes and the per-SN parameter $t_0$ aligns said SN~Ia's observations with the model's phase grid. 
Intuitively, the first kernel matrix $\mathbf{K}$ in Eq.~\ref{eq:fullmodel1} maps each observation to our phase node space after said observation phase $t_i$ is translated by $t_0$, while the second accounts for $\mathbf{F}_{\lambda,\text{eff}}$ flux node covariance at grid phases $t_{p,i}$ and $t_{p,j}$. 
GPMPs provide a natural framework to translate observed phase $t_i$ by the per-SN $t_0$ parameter to the model grid's phase zero-point.

Note that $t_0$ is not the fit date of maximum brightness---instead, $t-t_0$ aligns observation phase with the model's phase grid $\mathbf{t}_p$. 
As we train the model using rest-frame transformed spectrophotometry, each $t_0$ is fit in its SN~Ia's reference $z=0.05$ frame. 
The kernel length scale hyperparameter $\rho$ is fixed to match the phase node interval resolution of $4$~days, although the model is insensitive to any reasonable choice in $\rho$ (for example, $\rho\approx 1$~week). 
Furthermore, by fixing $\rho=4$, the matrix $\mathbf{K}$ is calculated and inverted only once during sampling. 
The uncertainty hyperparameter $\sigma^2$ is fixed to unity since it is divided out in Eq.~\ref{eq:fullmodel1}. 

The $\mathbf{F}_{\text{eff}}$ of each SN is decomposed via element-wise multiplication (also called the Hadamard operation $\circ$) from a fiducial flux template matrix $\mathbf{F}_0$ and a warping matrix $\mathbf{\Omega}$:
\begin{equation}\label{equ:fullmodel2}
    \mathbf{F}_{\text{eff}}=\mathbf{F}_0 \circ \mathbf{\Omega}. 
\end{equation}
$\mathbf{F}_0$ encodes the training sample's mean flux evolution via a set of fiducial light curves, while $\mathbf{\Omega}$ encodes deviations from these fiducial light curves for a given SN~Ia. 
Specifically, each column $\mathbf{\Omega}_{\lambda}$ of matrix $\mathbf{\Omega}$, for a given $\lambda$ node, is defined as:
\begin{equation}\label{eq:fullmodel3}
    \log(\mathbf{\Omega}_{\lambda}) = -0.4(\chi_0 + s_1\mathbf{M}_{\lambda,1} + c_1L_{\lambda,1} + c_2L_{\lambda,2}).
\end{equation}
$\mathbf{M}_{\lambda,1}$ is the $\lambda$-node column of the phase-dependent chromatic flux variation template matrix $\mathbf{M}_{1}$. 
$\mathbf{M}_{1}$ therefore encodes training sample light curve variation. 
$L_{\lambda,1}$ and $L_{\lambda,2}$ are the $\lambda$-node elements of the two phase-independent chromatic flux variation template vectors $\clone$ and $\cltwo$, respectively (these appear as scalars in Equation~\ref{eq:fullmodel3} because of their phase independence). 
Each explicit per-SN parameter is contained within the warping template: the achromatic offset parameter $\chi_0$, the phase-dependent chromatic flux variation parameter $s_1$, and the two phase-independent chromatic flux variation parameters $c_1$ and $c_2$. 
Both differences in intrinsic brightness and peculiar velocity effects are accounted for by the $\chi_0$ parameters. 

\begin{figure*}
    \plotone{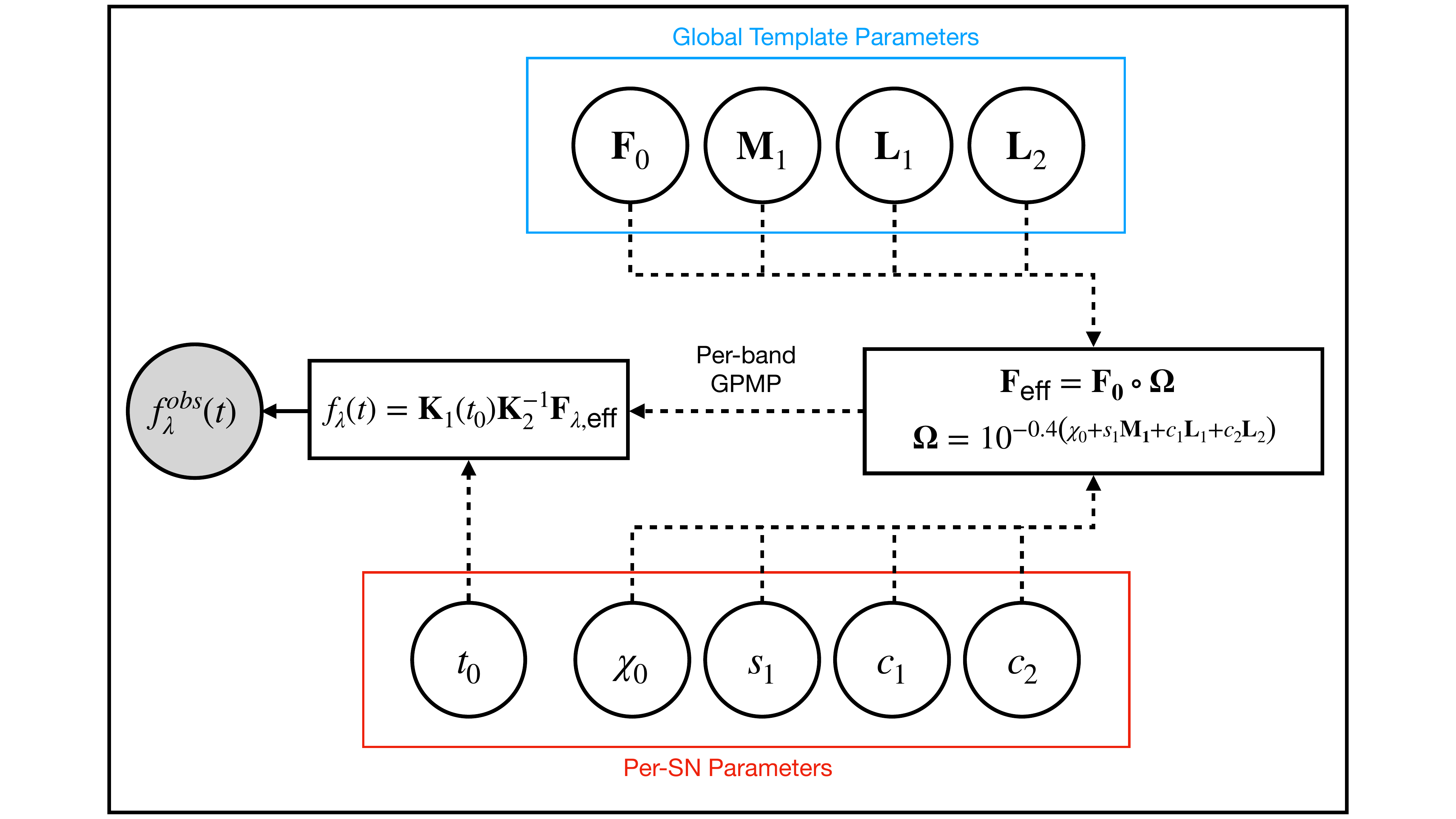}
    \caption{A directed acyclic graph representation of our model. 
    Per-SN model parameters are located in the bottom red box; global template parameters are in the top blue box. 
    The dashed arrows are deterministic relations (transformations and definitions). 
    The only explicit conditional probability in the model's architecture relates observations $f_{\lambda}^{obs}(t)$ to modeled flux $f_{\lambda}(t)$. 
    We perform Gaussian process mean predictor (GPMP) interpolation per-band in mapping effective template nodes $\mathbf{F}_{\text{eff}}$ nodes to a predicted flux $f_{\lambda}(t)$.}
    \label{fig:dag1}
\end{figure*}
Figure~\ref{fig:model_cartoon} illustrates this model's architecture, specifically displaying the transformed phase-independent template basis $\clone^{\prime}$ and $\cltwo^{\prime}$ vectors later discussed in Section~\ref{sec:method:postprocessing}. 
Figure~\ref{fig:dag1} is a directed acyclic graph of our model. 
Its conditional probability structure only connects observations to the model flux --- deterministic connections (transformations and definitions) are presented with dashed arrows. 
Global template parameters are located in the top blue box and per-SN parameters in the bottom red box. 

\subsection{Template Constraints and Per-SN parameter Models}\label{sec:method:constraints}
The third $\lambda$ band, $t_{p,i}=0$ phase node of $\mathbf{F}_0$ is fixed to one --- it is referred to here as fixed band~3 and corresponds to the wavelength node at $4084$~\r{A}, the top-hat filter band closest to a standard \textit{B}-band. 
This constraint prevents any scaling degeneracy between $\mathbf{F}_0$ and the model's $\chi_0$ parameters while also setting the specific phase node that the $t_0$ parameter aligns observations to.
Physical consideration further requires all $\mathbf{F}_0$ flux node parameters be bound to greater than or equal zero, so we enforce nonnegative values for all flux values. 

All chromatic flux variation templates $\clone$, $\cltwo$, $\mathbf{M}_1$ have scaling degeneracies with their respective per-SN parameters $c_1$, $c_2$, and $s_1$. 
For example, the transformations $s_1 \rightarrow s_1 / \alpha$ and $\mathbf{M}_1 \rightarrow \alpha \mathbf{M}_1$ leave the model unchanged; identical degeneracies exist for $c_1$-$\clone$ and $c_2$-$\cltwo$. 
Each scaling degeneracy is removed by requiring these three templates be normalized.   
This is a straightforward procedure for $\clone$ and $\cltwo$, where each are instantiated in {\tt Stan} as unit vectors, but a more involved process is used to normalize template matrix $\mathbf{M}_1$. 
We first define a unit vector of length $n_p \times n_{\lambda}$ that is then transformed into $\mathbf{M}_1$ by `chopping' said unit vector into $n_{\lambda}$ column vectors (each of length $n_p$) that forms the column space of a now normalized $\mathbf{M}_1$\footnote{The unit vector constraints for $\clone$, $\cltwo$, and $\mathbf{M}_1$ are what prevent these templates from being described as color variation templates.
If instead we constrained these templates at a reference wavelength node to be fixed to zero, then these templates would have units color, per the usual astronomical definition.}. 
No further constraints or bounds are placed on template parameters. 

Zero-mean constraints are placed on per-SN parameter sets $c_1$, $c_2$, and $s_1$. 
A reference SN~Ia could be selected to serve as the $c_1$, $c_2$, and $s_1$ zero points, but we opt instead to require these per-SN parameter sets to always have a mean of zero.
These constraints are enforced structurally by instantiating these parameter sets as centered vectors (Appendix~\ref{appendix:centeredvectors}). 

\subsection{Fitting the Model}\label{sec:method:fitting}
As described in, e.g., \citet{Saunders2018,Rubin2022}, the SNfactory data are extracted from the $15 \times 15$ spaxels of the SuperNova Integral Field Spectrograph \citep{Aldering2002, Lantz2004} that are projected as spectra onto $2k \times 4k$ CCD detectors. In this extraction process the Poisson noise and the readout noise for each pixel are included with appropriate weights. The host
galaxy is subtracted from this datacube using a reference datacube taken more than 1\,yr later, as described in \citet{Bongard2011}. Gray dimming by clouds is corrected as described in \citet{Pereira2013}  et al 2013. The remaining SN-only light is fit with a PSF model, as described in \citet{Buton2013, Rubin2022}, again propagating the Poisson and detector noise uncertainty. Due to small unmodeled PSF shape variations, there is additional uncertainty, which is empirically determined to be approximately 3\% \citep{Leget2020}. The final calculated uncertainties are dominated either by this PSF shape noise or the Poisson plus detector noise.

Initial trial runs confirmed that nominal uncertainties were underestimated, not having included galaxy-subtraction error, and that the overall distribution had broader tails than a Normal distribution.  We thus added a further uncertainty equal to 2\% of each SN\,Ia’s maximum observed flux and treat measurement uncertainties as having a Cauchy distribution, which led to stable convergence of the fit.  The credibility of this ansatz was checked though inspection of the pull distribution of photometric residuals around the final best fit.
This approach has previously been used to represent model uncertainty in \citet{Rose2020}.

The first source of error preferentially increases high-flux observational error, while the second largely affects low-flux observations, particularly those 20 or more days after peak brightness. 
All added uncertainty is diagonal: no covariance is injected into our data before training. 

The nominal SNfactory measurement and uncertainty are used as the Cauchy distribution location and scale. 
For each flux observation $f_{\lambda}^{obs}(t)$ with its corresponding measurement uncertainty $\sigma_{f_{\lambda}(t)}$, our likelihood function takes the form: 
\begin{equation}
    f_{\lambda}^{obs}(t) \sim \text{Cauchy}\big[f_{\lambda}(t), \sigma_{f_{\lambda}(t)}\big]. 
\end{equation}
No correlations are added between measurement errors.  

The model is implemented and trained using the statistics programming language {\tt Stan} \citep{Stan}. 
Built into {\tt Stan} is a No U-Turns (NUTS) Hamiltonian Monte Carlo sampler well suited for sampling our model's high-dimensional posterior. 
No explicit priors are placed on templates or per-SN parameter sets, instead leaving them with default implicit flat priors along any aforementioned bounds (Section~\ref{sec:method:constraints}). 

{\tt Stan} is informed with initial conditions estimated by first running simpler versions of the model. 
We do this only to improve sampling efficiency --- it is not necessary for our model's convergence. 
This process is done iteratively, starting with the simplest model that only obtains $t_0$, $\mathbf{F}_0$, and $\chi_0$ parameters (a mean light curve model). 
The results of this simplest model then become the initial conditions for a more complex model that includes $\clone$ and $c_1$ parameters. 
Other components (specifically, $\cltwo$ and $c_2$, and then $\mathbf{M}_1$ and $s_1$) are then added and trained using prior model iteration's fit as initial conditions until all the described model's components are incorporated. 
Note that we use SALT2's $t_{\max}$ as initial conditions for $t_0$ parameters when training the simplest of these models. 

With {\tt Stan}'s default NUTS we pull 4000 samples for each of type 16 instantiated samplers: 2000 warm-up followed by 2000 samples iterations per-chain. 
{\tt Stan} is run on the University of Pittsburgh's Computational Resource Center\footnote{\url{https://crc.pitt.edu/}}.
The convergence metrics `split-$\hat{R}$' and `effective sample size' were calculated after post-processing using techniques provided in~\cite{Vehtari2019}. 
After training and post-processing, each SN~Ia is refit with template parameters fixed ($\mathbf{F}_0$, $\mathbf{M}_1$, $\clone$, and $\cltwo$) to determine final values for per-SN~Ia $\chi_0$, $s_1$, $c_1$, and $c_2$, permitting a direct comparison of these per-SN parameters against other empirical SN~Ia models. 

There is no selected standard $\Delta M_B=0$ SN~Ia identified in the training sample, leaving a nontrivial linear degeneracy between achromatic offset parameters $\chi_0$ and phase-independent parameters $c_1$ and $c_2$. 
For physical reasons, $c_1$ and $c_2$ should not correlate with intrinsic magnitude, which ideally should only be captured by $\chi_0$. 

Linear transformations from~\cite{Leget2020} are used to remove correlations between both the $c_1$ and $\chi_0$ parameters, and the $c_2$ and $\chi_0$ parameters, as summarized in Appendix~\ref{appendix:degens}. 
Implementing this directly in the {\tt Stan} model leaves results unchanged but does reduce sampling efficiency, so this step is performed after sampling. 

\subsection{Interpreting the Phase-independent Templates}\label{sec:method:postprocessing}
Each sampler from {\tt Stan} explores a plane spanned by the template vectors $\clone$ and $\cltwo$\footnote{In general, a plane is an affine space, not a vector space. 
All planes described in this paper do pass through the embedding vector space's origin, which ensures they are proper 2D vector subspaces. 
Because of this, all planes discussed either intersect or are parallel. 
For brevity, in this paper we refer to any 2D subspaces as planes.}.
Even after decorrelating $\chi_0$ from $c_1$ and $c_2$, the output basis \{$\clone,\cltwo$\} is not unique, a consequence of this model's physics-agnostic architecture. 
This is because for any nonsingular linear transformation the output basis vectors (i.e. $a\clone + b\cltwo$ and $c\clone + d\cltwo$ for $a,b,c,d\in\mathcal{R}$ and $ab -cd\neq 0$) necessarily span the aforementioned plane. 
To quantify this plane's convergence (as opposed to only its basis vectors), for each posterior sample we calculate a bivector $\clbv=\clone \wedge \cltwo$ that, by definition,  spans the plane of interest. 
Importantly, the bivector representation $\clbv$ is no longer ambiguous. 
\begin{figure}
    \plotone{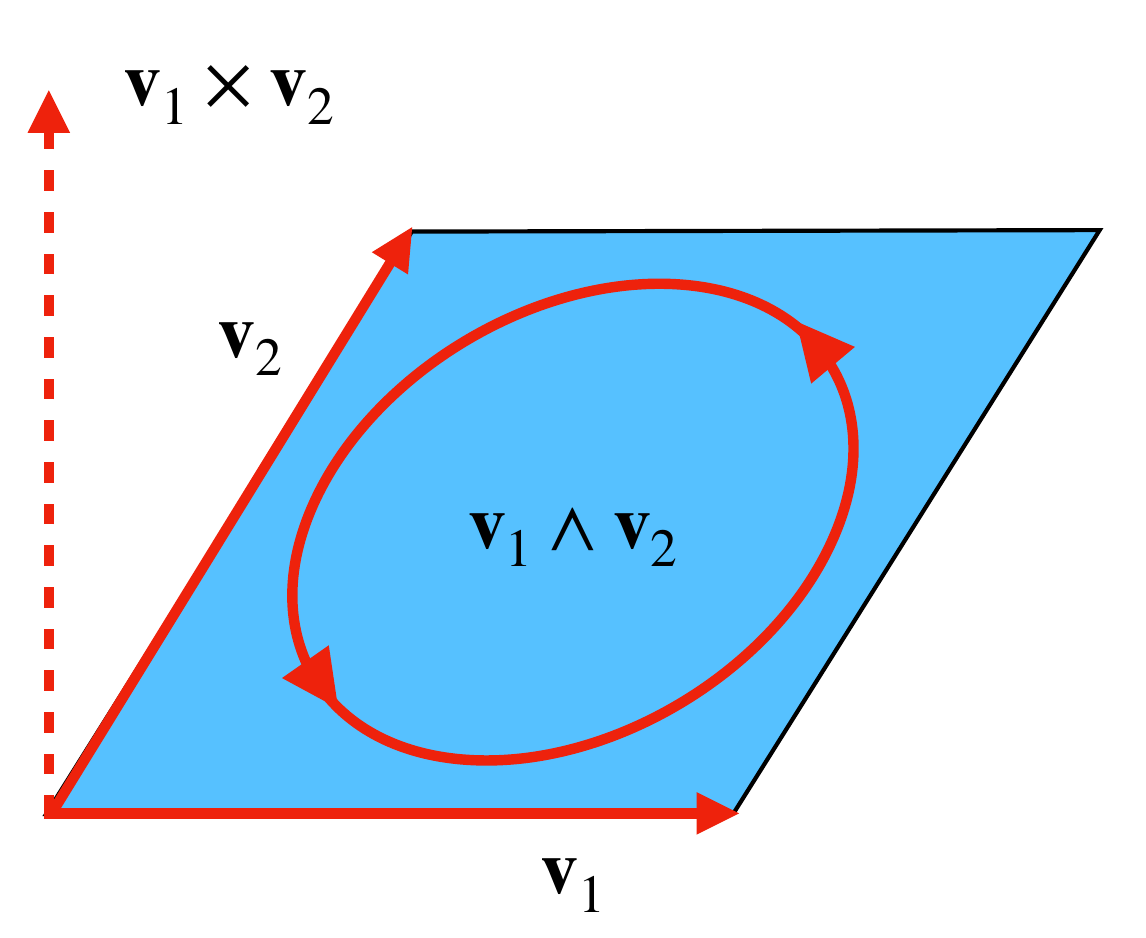}
    \caption{This is an illustration of a bivector (the blue parallelogram) $\mathbf{v}_1\wedge \mathbf{v}_2$ constructed by the vectors $\mathbf{v}_1$ and $\mathbf{v}_2$ (the red vectors) in three dimensions. 
    A three-dimensional space allows for the corresponding cross product to be included for reference (the red dashed vector). 
    Bivectors, like vectors, are oriented objects, with the bivector $\mathbf{v}_1\wedge \mathbf{v}_2$ having a counter-clockwise orientation determined by component vector ordering (here represented with an oriented red loop). 
    Reversing the product order reverses a bivector's `circulation' or orientation: $\mathbf{v}_2\wedge \mathbf{v}_1 = -\mathbf{v}_1\wedge \mathbf{v}_2$.
    Note that the cross product does not generalize to all finite-dimensional vector spaces, while the wedge product $\wedge$ does. }
    \label{fig:bv_illustration}
\end{figure}

A bivector is a geometric object representing an oriented plane element constructed from the wedge product (see Figure~\ref{fig:bv_illustration} for an illustration). 
Intuitively, a bivector corresponds to a plane like a vector corresponds to a line, and the wedge product is the dual to a cross product in three dimensions. 
Unlike the cross product, the wedge product generalizes to any finite-dimensional vector space greater than two, meaning bivectors are well-defined in this model's 10-dimensional wavelength node space.   
Now any SN~Ia'a phase-independent chromatic flux variation curve $\mathbf{c}=\mathbf{c}_1\clone + \mathbf{c}_2\cltwo$ can be interpreted as residing in the plane spanned by $\clbv$, regardless of the selected ${\clone,\cltwo}$ basis. 

Each component of the bivector $\clbv$ is calculated as follows:
\begin{equation}\label{eq:bv_comp}
    \mathcal{L}_{ij}=\frac{L_{1,i}L_{2,j}-L_{1,j}L_{2,i}}{\sqrt{\sum^{n_{\lambda}}_{k=1} \sum^{n_{\lambda}}_{m>k}\big(L_{1,k}L_{2,m}-L_{1,m}L_{2,k}\big)}},
\end{equation}
where $\hat{L}_{1,i}$ is the $i$th wavelength component of template $\clone$ and $\hat{L}_{2,i}$ is the $i$th wavelength component of template $\cltwo$. 
Importantly, this representation is independent of ${\clone,\cltwo}$ choice. 
Note that $\clbv$ is normalized so as to represent a unit plane element. 
With these transformed parameters $\mathcal{L}_{ij}$, the model can unambiguously be tested for convergence and the best-fit templates be determined. 

\subsection{Bases for the Phase-Independent Chromatic Variation Model}\label{sec:method:bases}
We now seek a pair of vectors, $\mathbf{L}_1$
and $\mathbf{L}_2$ that span $\mathcal{L}$ and readily provide insight into the physical origin of the model's two-dimensional phase-independent chromatic flux model.  
Two such bases are considered.

\subsubsection{Maximum Variance Ratio Basis}\label{sec:method:bases:mvr}
The first basis, called the maximum variance ratio (MVR\footnote{We developed this technique and are not aware of any previous equivalent use.}) basis, is derived directly from the corresponding $c_1$--$c_2$ distribution.

New and uncorrelated parameter sets $c_1^{\text{mvr}}$ and $c_2^{\text{mvr}}$ with their relative variance maximized are found via a linear transformation. 
The result is a basis for $\clbv$ where $\clone^{\text{mvr}}$ accounts for the most chromatic flux diversity by a single template in $\clbv$, while $\cltwo^{\text{mvr}}$ captures any remaining variation.
This basis amounts to the assumption that two independent physical effects affect chromatic flux variation (e.g.\ amount of dust, intrinsic SN~Ia diversity) 
while designating as much of the variance as possible to one source (e.g.\ dust).
This assumption provides some useful insights even
if it does not totally satisfy our
physical expectations, as it
does not consider additional possible  effects (e.g.\ kind of dust), nor that supernovae and their progenitor environments likely correlate with said effects. 
This basis's solution is found by numeric minimization; it is not orthogonal.

The MVR basis is determined as follows.  
A basis centered at the origin can be described by the angle between unit vectors and their orientation.
Starting with an orthogonalized output basis from {\tt Stan} $\mathbf{L}=[\clone^{\text{orth}}, \cltwo^{\text{orth}}]$ for $\clbv$ and a per-SN coefficient matrix $\mathbf{c}$ (here arranged as an $n_{\text{sn}}\times 2$ matrix), adding an extra angle $\theta$ between the basis is achieved with the transformation
\begin{align}
\begin{split}
    \tilde{\mathbf{L}} & = M \mathbf{L} \\
    & =
    \begin{pmatrix}
    1 & 0\\
    \sin{\theta} & \cos{\theta}
    \end{pmatrix} \mathbf{L}.
\end{split}
\end{align}
In this $\tilde{\mathbf{L}} $ basis, the per-SN coefficients $\mathbf{c}^{\prime}=\mathbf{c}M^{-1}$ have an orientation given by $V^T$ from the singular value decomposition (SVD) of $\mathbf{c}^{\prime}=U\Sigma V^T$. 
Taking $V^TM$ and normalizing its rows to be unit vectors
gives the properly oriented basis given $\theta$. 
Note that these primed components here are unrelated to those introduced below. 

For this basis, the ellipticity of its corresponding $c$ distribution is again found using SVD, given by $\log{\Sigma_{11}}-\log{\Sigma_{22}}$.
An optimizer is used to determine the $\theta$ that maximizes the ellipticity. 

\subsubsection{CCM89-Derived Basis}\label{sec:method:bases:ccm89}
We also desire a basis that readily separates rapidly changing chromatic flux variation (i.e. that akin to SN absorption/emission features) from continuum-like chromatic flux variation (i.e. dust-like behavior). 
This continuum-like variation is assumed to be dust-like, given dust extinction's ubiquitous contribution to SN~Ia color/chromatic flux variation. 
The two phase-independent plots found in Figure~\ref{fig:model_cartoon} provide an example of what we want from this basis --- specifically, one template being more smooth with respect to wavelength than the other. 

For the second basis, the vector $\clone^{\prime}$  simultaneously resides within the planes spanned by both $\clbv$ and the Cardelli-Clayton-Mathis (CCM89) dust model (\citeyear{Cardelli1989}), defining the first basis component. 
This intersection ensures it captures smooth, CCM89-like chromatic variability.
The other basis vector $\cltwo^{\prime}$ is chosen to be perpendicular to $\clone^{\prime}$, while still residing in the $\clbv$ plane; this basis is orthogonal by construction.  
Note that $\cltwo^{\prime}$ will not necessarily be perpendicular to the plane spanned by CCM89, but is still guaranteed to provide the least continuum-like variability w.r.t. wavelength (and therefore, the most spectral-feature-like behavior) allowed by $\clbv$. 
This basis provides a useful representation not because it recovers a mathematically valid dust extinction curve, but instead because it clearly separates rapidly changing chromatic flux variation from SN continuum-like variability. 
It is important to remember that intrinsic variability could still affect the direction of the first basis vector $\clone^{\prime}$, which means this basis does not guarantee a physical decomposition into exclusive dust and intrinsic components. 
As such, any dust-like properties inferred from $\clone$ in isolation are physically ambiguous. 

The CCM89-derived basis is calculated as follows.
Since CCM89 has two basis curves $a(\lambda)$ and $b(\lambda)$ (one for each parameter $A_V$ and $A_V/R_V$, respectively), one can construct a CCM89 unit bivector $\clbv_{\text{ccm}}$ from discretized curves $a(\lambda) \rightarrow \mathbf{a}$ and $b(\lambda) \rightarrow \mathbf{b}$ using an appropriately modified version of Equation~\ref{eq:bv_comp}: $\hat{L}_{1,i} \rightarrow a_i$ and $\hat{L}_{2,i} \rightarrow b_i$.
The two planes spanned by $\clbv$ and $\clbv_{\text{ccm}}$ then intersect within the $n_{\lambda}$-dimensional wavelength vector space along a line. 
It is the vector which spans this intersecting line that defines the new $\clone^{\prime}$ template:
\begin{equation}
    \clone^{\prime}=\text{Intersection}[\clbv, \clbv_{\text{ccm}}].
\end{equation}
We are free to choose a new $\cltwo$ as long as it resides within the subspace represented by $\clbv$. 
To minimize this $\cltwo$ template's dust-like properties, $\cltwo^{\prime}$ is defined as a $\theta=\pi/2$~radian rotation of $\clone^{\prime}$ within the plane spanned by $\clbv$ via a rotation operator $\mathbf{R}_{\clbv}(\theta/2)$:
\begin{equation}
    \cltwo^{\prime}=\mathbf{R}_{\clbv}(\pi/4)\clone^{\prime}\mathbf{R}_{\clbv}^{-1}(\pi/4).
\end{equation}
This rotation maximizes the component of $\cltwo^{\prime}$ that is perpendicular to the plane spanned by $\clbv_{\text{ccm}}$. 
The new $\{\clone^{\prime},\cltwo^{\prime}\}$ also transforms the $c_1$ and $c_2$ parameters sets, here labeled $c_1 \rightarrow c_1^{\prime}$ and $c_2 \rightarrow c_2^{\prime}$. 

\begin{figure}
    \plotone{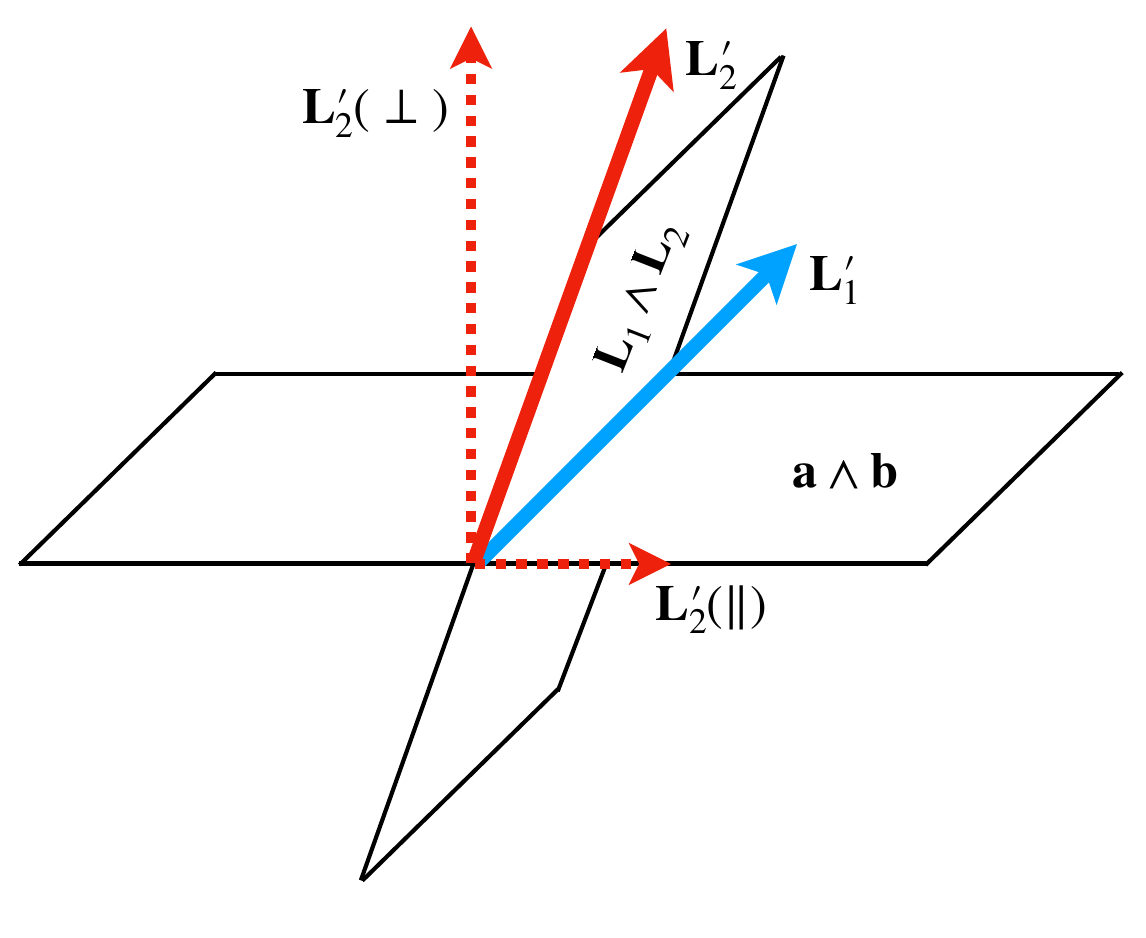}
    \caption{A three-dimensional representation the CCM89-basis's geometric intuition. 
    The blue solid vector is the transformed $\clone^{\prime}$ template spanning the intersection of two planes, one spanned by the model bivector $\clbv=\clone \wedge \cltwo$ and the other spanned by the CCM89 bivector $\clbv_{\text{ccm}}=\mathbf{a} \wedge \mathbf{b}$. 
    The red solid vector is a $\pi/2$ rotation in the plane spanned by $\clbv$ that defines the transformed $\cltwo^{\prime}$ template. 
    The decomposition $\cltwo^{\prime} = \cltwo^{\prime}(\parallel) + \cltwo^{\prime}(\perp)$ with respect to the CCM89 plane spanned by $\clbv_{\text{ccm}}$ is given with the red dashed arrows. }
    \label{fig:geoalg_intuition}
\end{figure}

Figure~\ref{fig:geoalg_intuition} provides a three-dimensional view of the geometric intuition involved in finding the CCM89-derived basis. 
All calculations discussed here are implemented using geometric algebra, which provides a novel approach to study oriented subspaces. 
Geometric algebra implementations for calculating intersections, projections, and rotation operations are summarized in Appendix~\ref{appendix:geoalg}. 
Although we exploit geometric algebra's elegance and interpretability to perform all said operations, each operation could be done using more classical linear algebra techniques if desired. 

\subsection{Handling Similar Solutions}\label{sec:method:handling}
In training, the samplers converge to one of three sets of very similar solutions,
as is apparent in the
per-chain values of $\mathbf{L}_1$ shown in
Fig~\ref{fig:twogroups}.
We find that simultaneously obtaining global template and per-SN parameters leaves the resulting samplers sensitive to the three least representative SNe~Ia in the training sample (based on residuals).  
These groups are associated with slightly different solutions for these three SNe~Ia, as seen by eye and quantitatively through their $\chi_0$ solutions.
Consider the following analogy: consider the inertia tensor of a space station as being the model's global templates and the coordinates of its crew members  corresponding to per-SN parameters. 
The moment of inertia changes only slightly if crew members move to another location on the station (assuming the space station is much more massive than the crew's total weight). 
Similarly, it is this `change in position' of per-SN solutions that is causing a very subtle change in template parameters, preventing complete convergence. 
Obviously this is not a fair comparison, but the resulting effects are very similar, hence this comparison.

The worst performing SN~Ia PTF11mkx consistently sees a $1\sigma$ difference between $\chi_0$ values between different chain groups --- considering that this $\chi_0$ parameter fit uncertainty is only $\approx1$\%, this tail wagging is very subtle.  
Indeed, when refitting per-SN parameters with a one of the group solutions as a fixed global template (Section~\ref{sec:results:refit}), all chains converge to the same solutions for the three SNe~Ia. 
Similarly, if two different group solutions are separately held constant and refit individually, the two resulting per-SN parameter sets are indistinguishable. 
Because each group's solution are consistently very similar, we opt to use the weighted average of all solutions for the best-fit template. 
This decision to use an average of all groups, as opposed to using a specific solution, has no impact on the remaining analysis. 

Note that refitting with fixed $t_0$ parameters does not prevent this subtle grouping, nor does cutting these three SNe~Ia from the training set.
If these three aforementioned troublesome SNe~Ia are removed, the next few SNe~Ia that are the least representative of this newly trimmed sample start `wagging'. 
This instability may be a feature of our model. 
In particular, the flat prior we use for the $c_1$ and $c_2$ distributions may bias the results for the least representative SNe~Ia of a given training sample.
Considering that global template parameters again were effectively unchanged with their removal, we retain those three SNe~Ia for the remainder of the analysis. 

\section{Best-fit Model Results}\label{sec:results}
\begin{figure*}
    \plotone{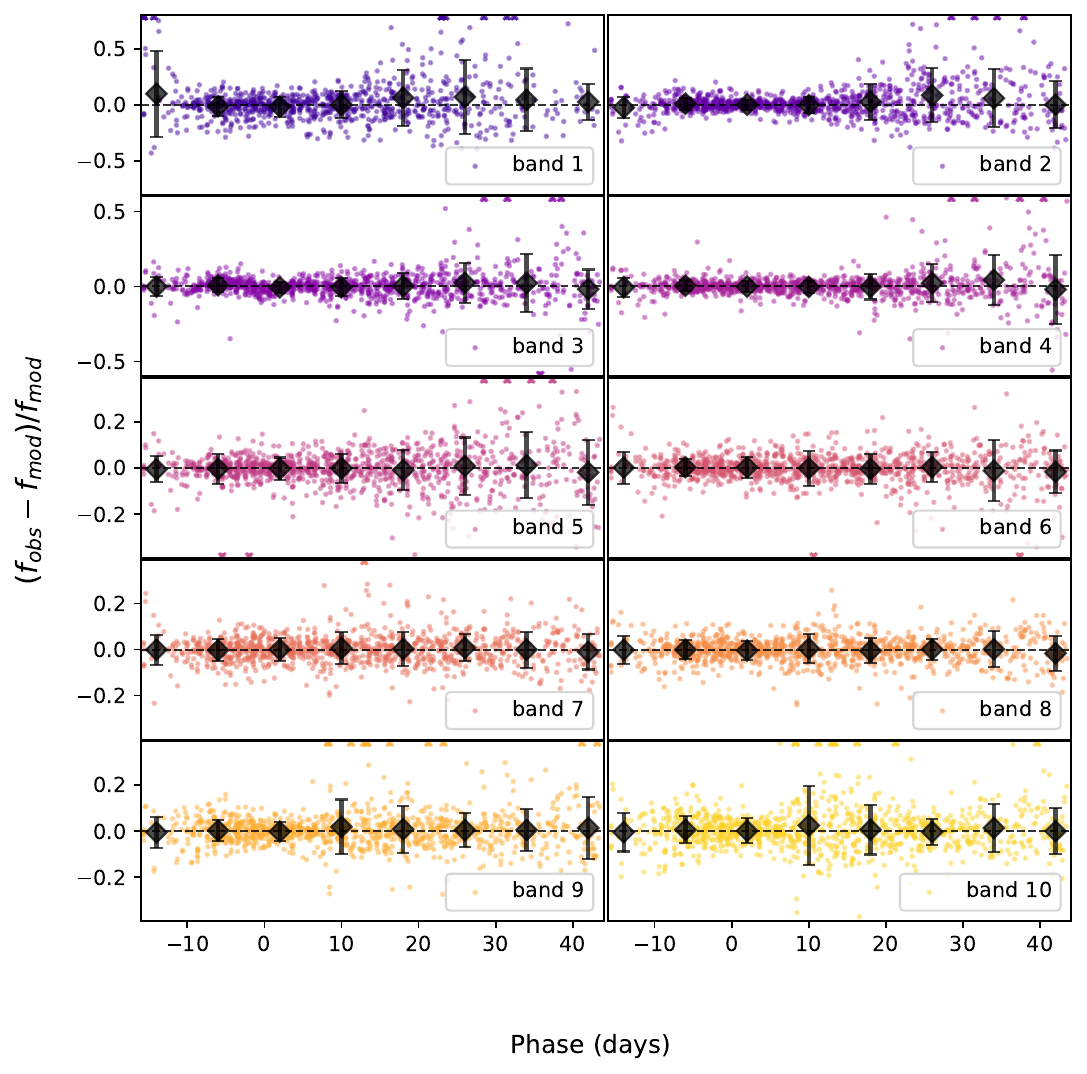}
    \caption{Best-fit model residuals with respect to observations presented for each of our ten bands.  
    Eight day binned averages for each band are presented as black diamonds, with error bars being binned standard deviations. 
    The carets at the top and
    bottom edges represent points
    that lie outside the range of the plots.}
    \label{fig:residuals}
\end{figure*}

Our model, as implemented with {\tt Stan}, consists of global template parameters and per-SN parameters. 
We define the best fit solution to be the mode of a 365-dimensional template parameter space\footnote{159 from $\mathbf{F}_0$, 159 from $\mathbf{M}_1$, 9 from $\clone$ and 9 from $\cltwo$, specifically.}, with each per-SN parameter marginalized before estimating from HMC sampling the posterior's maximum. 
This mode is estimated using a mean shift clustering algorithm implemented in the scikit-learn package using the default flat kernel \citep{scikit-learn}. 
To estimate a consistent best-fit solution, bivector components $\mathbf{L}=\clone \wedge \cltwo$ (Equation~\ref{eq:bv_comp}) are used instead of $\clone$ and $\cltwo$ components directly. 
This process is analogous to maximum \textit{a posteriori} estimation of the HMC-sampled posterior, but allows first for the aforementioned post-processing. 
Marginal posterior dispersion for each parameter are presented as 68\%-th percentile error bars. 
Residuals of the best-fit model, alongside binned averages, are presented in Figure~\ref{fig:residuals}.

Two of the 16 HMC samplers were rejected after fitting because (1) parameter values were systematically inconsistent by 4 to 5 days with SALT2 date of maximum, (2) the corresponding light-curve shapes were nonphysical, and (3) these two samplers have notably inconsistent $\ln{p}$ values relative to the 14 retained samplers. 
The 14 remaining samplers converge to three very similar group solutions of which the weighted average is taken; see Section~\ref{sec:method:handling} for more details.  

Table~\ref{table:table1} provides median values and 68\textsuperscript{th} percentile upper and lower values for per-SN light curve model parameters.

\subsection{Phase-independent Chromatic Flux Variation Templates}\label{sec:results:colorlaw} 
Ostensibly, if $\clbv$ were to only capture CCM89 dust-like behavior, then the planes spanned by $\clbv$ and $\clbv_{\text{ccm}}$ would be effectively parallel and any intersection poorly constrained. 
This turns out to not be the case, with the best-fit solution recovering a planar separation angle of about $80^{\circ}$. 

\subsubsection{Maximum Variance Ratio Basis}\label{sec:results:colorlaw:mvr}
\begin{figure}
    \plotone{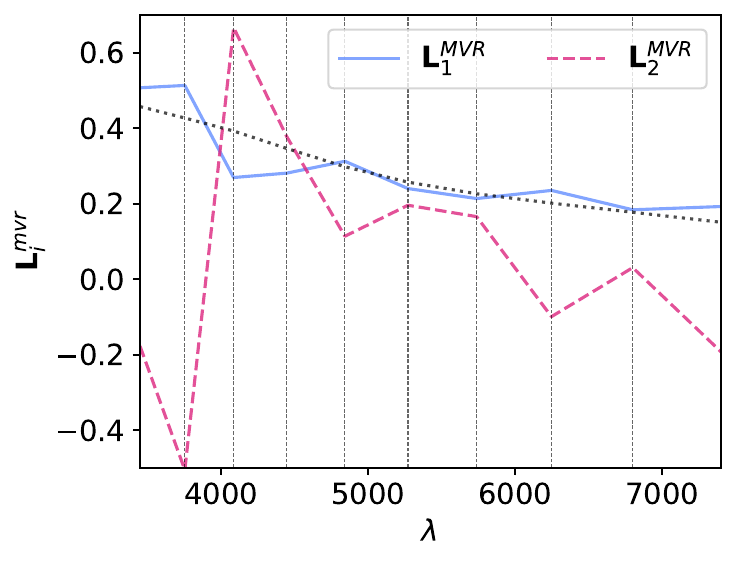}
    \caption{The blue solid line corresponds to the first MVR component $\clone^{\text{mvr}}$, which appears nominally more consistent with dust-like variation than its counterpart $\cltwo^{\text{mvr}}$, given as the magenta dashed.
    $\clone^{\text{mvr}}$ has a best-fit $R_V^{\text{mvr}}=2.18$ given as the gray dotted line, with most of its divergence from a CCM89 curve occurring blueward of $5000$~\r{A}. 
    $\cltwo^{\text{mvr}}$ captures variation not readily describable as dust-like. }
    \label{fig:l1l2_maxvarratio}
\end{figure}

Figure~\ref{fig:l1l2_maxvarratio} presents the MVR basis as described in Section~\ref{sec:method:bases}. 
Qualitatively, $\clone^{\text{mvr}}$ appears nominally more dust-like than its counterpart $\cltwo^{\text{mvr}}$.
Figure~\ref{fig:l1l2_maxvarratio} includes the best fit CCM89 curve $R_V^{\text{mvr}}=2.18$ for reference, showcasing that its most extreme divergence from CCM89 is blueward of $5000$~\r{A}. 
$\cltwo^{\text{mvr}}$, on the other hand, captures variation that is not readily describable as dust-like: normal dust extinction should be absorptive across the optical wavelength range whereas the sign flip in $\cltwo^{\text{mvr}}$ produces simultaneous brightening/dimming on either side of $4000$~\r{A}.
Although the degree of variation increases as wavelength decreases, there is a distinct flip in behavior around $4000$~\r{A}. 
Such behavior is inconsistent with dust extinction.

The conditions for the target parameter set distributions is to assign maximum variance to one component while keeping the second component uncorrelated.
These conditions yield a basis consistent with the expectation that dust-like variation is the primary contributor to SN~Ia phase-independent chromatic flux variation while intrinsic SN~Ia diversity uncorrelated with dust accounts for additional variability.

\subsubsection{CCM89-derived Basis}\label{sec:results:colorlaw:ccm89}
\begin{figure}
    \plotone{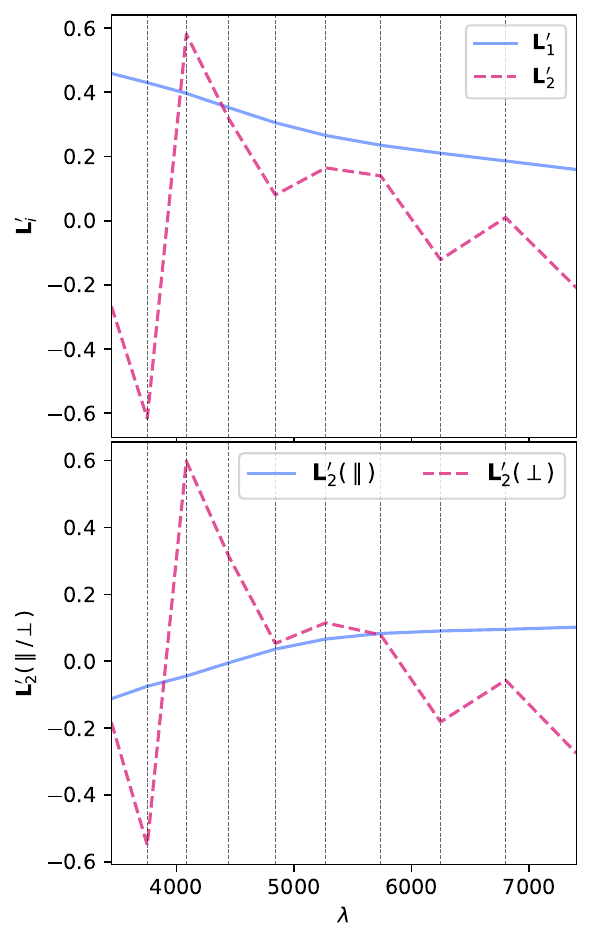}
    \caption{The top plot presents phase-independent chromatic flux variation templates $\clone^{\prime}$ and $\cltwo^{\prime}$. 
    $\clone^{\prime}$ has a recovered total-to-selective extinction of $R_V^{\text{int}}=2.4$. 
    The bottom plot presents a decomposition of $\cltwo^{\prime}$ into its parallel and perpendicular components with respect to the CCM89 plane. 
    $\cltwo^{\prime}$ clearly captures some dust-like variability, despite being dominated by intrinsic modes. 
    Although the low-resolution wavelength binning prevents quantification of spectral features, the most impressive $\cltwo^{\prime}$ variability appears in the Ca~II~H\&K regime. }
    \label{fig:l1l2_l2decomp}
\end{figure}

The template $\clone^{\prime}$ is presented in the top plot of Figure~\ref{fig:l1l2_l2decomp}. 
As summarized in Section~\ref{sec:method:bases}, the intersection of the plane spanned by $\clbv$ with the plane spanned by $\clbv_{\text{ccm}}$ defines the first phase-independent chromatic flux template $\clone^{\prime}$. 
This $\clone^{\prime}$ template, as expected, captures continuum-like variation akin to dust extinction. 

From $\clone^{\prime}$ we find an intersection total-to-selective extinction ratio of $R_V^{\text{int}}=2.4$. 
As mentioned in~\ref{sec:method:bases}, this $R_V$ is not immediately physically interpretable without an $A_V$ model.  

Also presented in the top plot of Figure~\ref{fig:l1l2_l2decomp} is template $\cltwo^{\prime}$. 
Unlike $\clone^{\prime}$, $\cltwo^{\prime}$ captures the more rapidly changing flux variability allowable by $\clbv$.  
Specifically, $\cltwo$ is capturing wavelength variation at scales smaller than that expected by continuum dust variability, at least within the optical regime. 
Indeed, its features nominally align with spectral feature. 
Also note the similarity between $\cltwo^{\prime}$ and $\cltwo^{\text{mvr}}$ despite their vastly different constructions. 

$\cltwo^{\prime}$ is not perpendicular to the CCM89 plane because the plane spanned by $\clbv$ is itself not perpendicular.  
As such, $\cltwo^{\prime}$ also captures a dust-like component alongside its dominating spectral-like component.   
The bottom panel presents this template's decomposition into parallel and perpendicular components defined with respect to CCM89's plane for reference. 

\begin{figure}
    \plotone{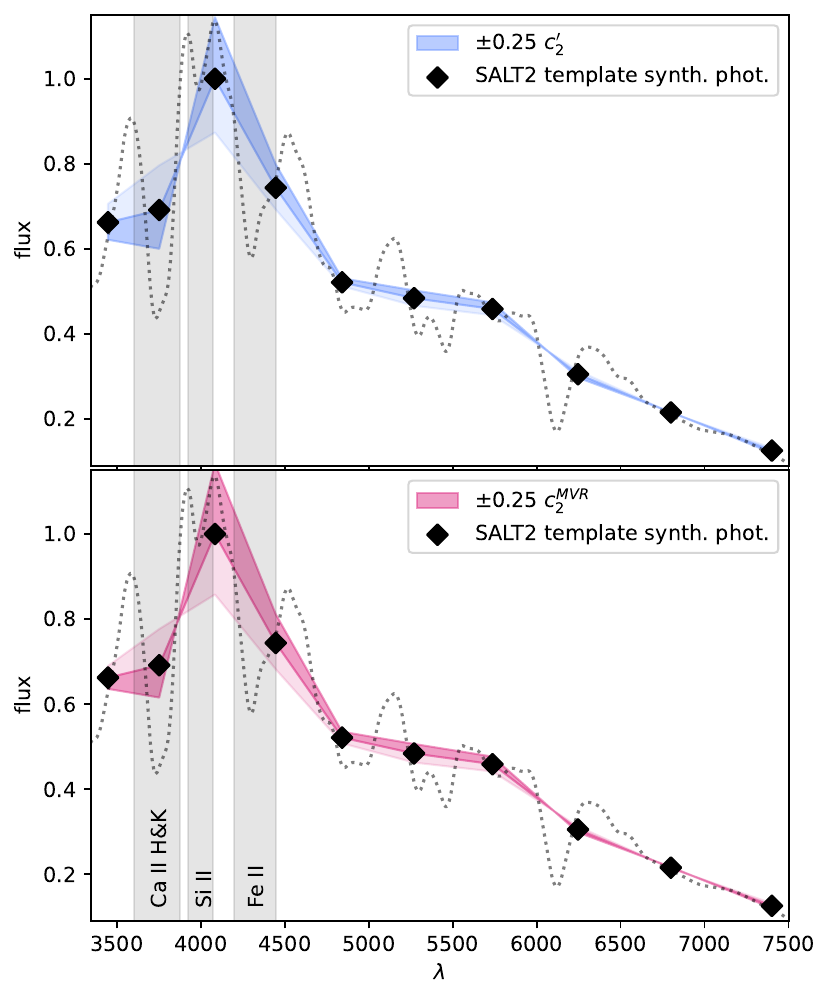}
    \caption{This figure demonstrates a $\pm 0.2$~mag $c_2$ variation (blue for CCM89-basis $c_2^{\prime}$, maroon for $c_2^{\text{MVR}}$) of $\cltwo^{\prime}$ overlaid on SALT2's mean template $t=0$ phase spectrum (dashed black line). 
    The spectrum is binned via synthetic photometry with top hat filters, presented as black diamonds. 
    Flux units are normalized by synthetic photometry wavelength $4048$~\r{A} value to unity and example spectral features blueward of $4500$~\r{A} are presented for reference.} 
    \label{fig:l2prime_phasemaxvar}
\end{figure}
\subsubsection{Intrinsic Variation}\label{sec:results:colorlaw:intrinsicvariation}
Both of the best-fit template representations $\cltwo^{\prime}$ and $\cltwo^{\text{mvr}}$ capture pronounced phase-independent chromatic flux variation blueward of $4500$~\r{A} that is inconsistent with dust (see Figure~\ref{fig:l2prime_phasemaxvar}). 
Indeed, no extrinsic phenomena readily describe this behavior. 
Variation blueward of 4500~\r{A} includes the prominent Ca~II~H\&K feature and its Si~III counterpart, Si~II $\lambda 4130$, C~II $\lambda 4267$, Fe~II $\lambda 4404$, and Mg~II $\lambda 4481$. 
With our choice of splitting the spectral range into $n_{\lambda}=10$ synthetic filters, the model cannot completely distinguish between said features, although at least one node for both $\cltwo^{\prime}$ and $\cltwo^{\text{mvr}}$ and their corresponding variation seemingly align with 
the wavelength bin where Ca II H\&K is the dominant contributor to a strong spectral feature; Si\,II and Fe\,II features also contribute to this variation. 
Hints of chromatic flux variability ostensibly aligns with SNe~Ia's signature Si~II $\lambda 6347$ feature and O~I $\lambda 7774$ are visible in Figure~\ref{fig:l1l2_l2decomp}, but in practice do not affect any resulting model flux predictions (again, see Figure~\ref{fig:l2prime_phasemaxvar}). 

We use the same dataset used by~\cite{Saunders2018} for their SNEMO analysis. 
Comparing both $\cltwo^{\prime}$ and $\cltwo^{\text{mvr}}$ to SNEMO2 and SNEMO7 eigenvectors (with two and seven components, respectively) yields indecisive insight, though. 
Figure~6 from Saunders et~al.~(2018) shows SNEMO eigenvectors describing similar $\cltwo^{\prime}$ or $\cltwo^{\text{mvr}}$ behavior at maximum, but all of these eigenvectors are clearly phase dependent --- no eigenvector's evolution seems approximately phase independent. 
Figures~9 through 12 from Saunders et~al.~(2018) present some unexplained variation for SNEMO2 blueward of $4500$~\r{A} that is nominally phase-independent up to six days post-maximum; this all but disappears with SNEMO7. 
That SNEMO2, itself similar to SALT2, sees unexplained phase-independent variability around maximum that aligns with $\cltwo^{\prime}$ and $\cltwo^{\text{mvr}}$ features seems indicative of our model's performance, but such a claim taken alone is likely an excessive interpretation. 

SNfactory data is also used by~\cite{Boone2021} with their Twins Embedding nonlinear model. 
More interesting insight is gained in comparing their findings with our template representations $\cltwo^{\prime}$ and $\cltwo^{\text{mvr}}$, since Twins Embedding is currently a phase-independent, maximum-phase model. 
Blueward of $4500$~\r{A}, spectral variation recovered by Twins Embedding loosely aligns with the both $\cltwo^{\prime}$ and $\cltwo^{\text{mvr}}$ templates (see there Figures~4, 6, and 10 from Boone et~al. (2021a) for reference). 
Recovering this consistent variation in our phase-independent template, albeit at lower wavelength bin resolution, lends credibility that $\clbv$ as a whole is capturing intrinsic variation. 

Past analyses by~\cite{Branch1993} and~\cite{Riess1998} find Ca~II~H\&K features are relatively stable in the week before and weeks after peak \textit{B}-band brightness, with this effective phase-independence being sufficiently stable to exploit for the latter's `snapshot' methodology from which they constrain luminosity distances. 
Both $\clbv$ basis representations do capture intrinsic, phase-independent variation around Ca~II~H\&K, but this alone is not profound evidence of fundamental Ca~II~H\&K time independence in the SN~Ia population. 
SN~Ia Ca~II~H\&K features, as with all spectral features, demonstrably evolve with time. 
In its current form, this model cannot distinguish between phase-averaged spectral variation or truly phase-independent intrinsic variation. 
Indeed, a goal of this project is to demonstrate that intrinsic chromatic flux variation can `leak' into phase-independent components, something that is occurring in these results. 

\subsection{Fiducial Template}\label{sec:results:meantemplate}
Figure~\ref{fig:stretchtemplate} presents GPMP interpolations of each band's best-fit $\mathbf{F}_{\lambda,0}$'s nodes as solid curves. 
Fixed band~3's template is the third solid curve presented in the top plot of Figure~\ref{fig:stretchtemplate}. 
The ubiquitous NIR bump is recovered for redder bands (bottom plot of Figure~\ref{fig:stretchtemplate}), with this second maximum occurring $\approx25$~days after our fixed-band peak brightness as expected \citep{Jha2019}. 
Apart from the reddest template curve centered at $7401$~\r{A}, peak-brightness phase per-band occurs earlier for bluer bands and later for redder band, again consistent with established trends \citep{Jha2019}. 
As would be expected by its \textit{I}-band overlap, the reddest template curve exhibits somewhat more complex behavior than other curves, such as an inflection point between its two local maxima (bottom plot, Figure~\ref{fig:stretchtemplate}). 

Note that each fiducial template light curve's peak brightness phase does not align with our $t_{p,i}=0$ flux node, meaning $t_0$ should not be interpreted as the fixed band~3's peak brightness phase. 
This is ultimately inconsequential, requiring only that each light curve's peak brightness phase be calculated deterministically after fitting, and has no effect on this analysis or its conclusions. 

\subsection{Phase-dependent Chromatic flux Variation  Template}\label{sec:results:stretchtemplate}
\begin{figure}
    \plotone{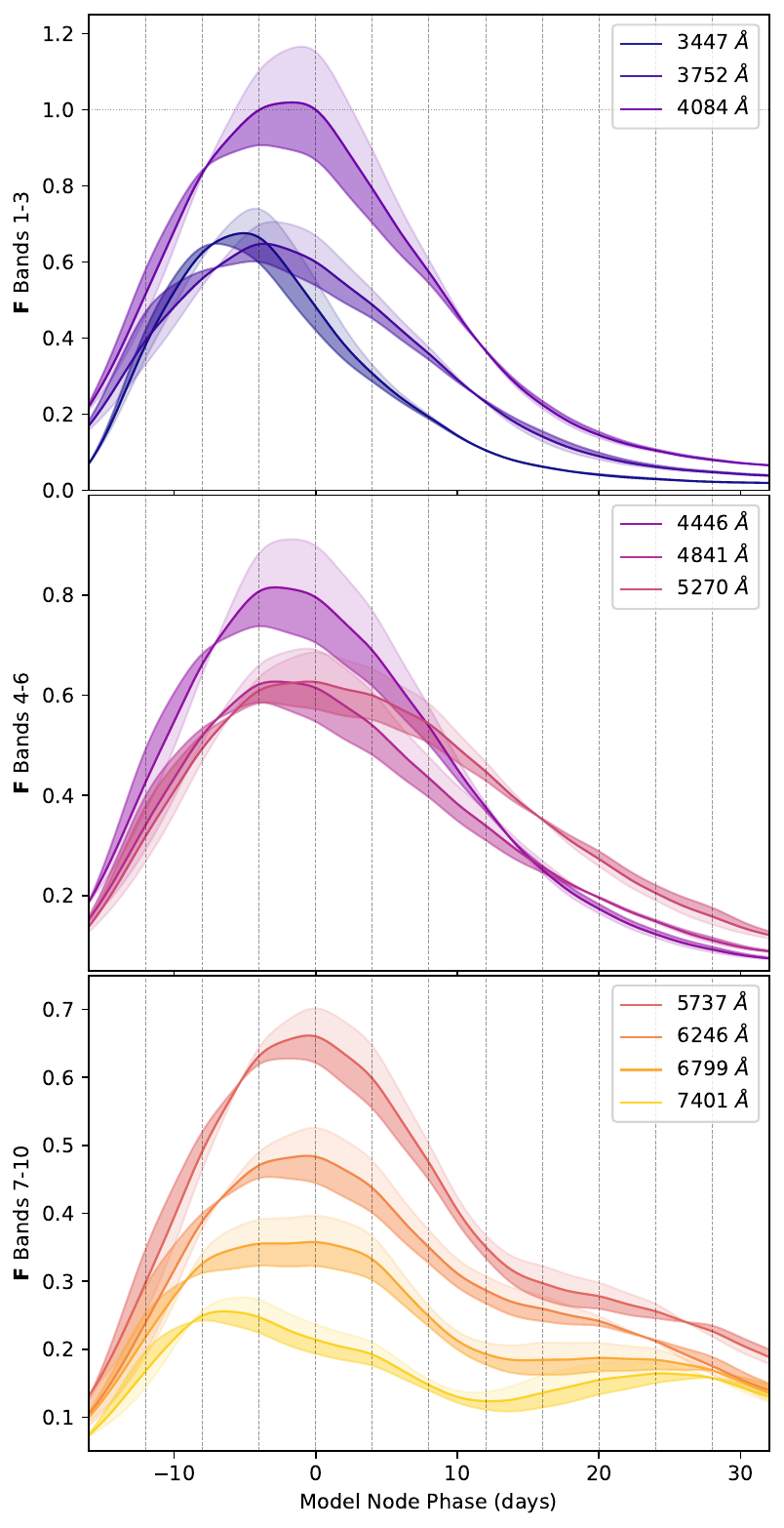}

    \caption{A visualization of $\pm 0.09$~mag variation in $s_1$ on the model's fiducial flux template $\mathbf{F}_0$, as warped by the phase-dependent chromatic flux template $\mathbf{M}_1$. 
    Positive $s_1$ contribution is given by light shaded regions, while negative $s_1$ contribution is given by the dark shaded regions.
    Solid lines are the GPMP interpolated light curve for that band's fiducial template nodes,
    which is deterministic (not stochastic) in our model. 
    The top two plots illustrate recovered stretch-like behavior by the template $\mathbf{M}_1$, with broadening to narrowing of effective light curves as $s_1$ increases in value. 
    The bottom plot captures stretch-like behavior further convolved with NIR bump variational modes (bump location and size).
    Note that these figures are not portraying model uncertainty, only model response to $s_1$ parameter variation.}
    \label{fig:stretchtemplate}
\end{figure}

As shown in Figure~\ref{fig:stretchtemplate}, the best-fit phase-dependent variation template $\mathbf{M}_1$ exhibits stretch-like behavior across all bands. 
The shaded regions in these plots show phase-dependent light curve variation from$-0.09<s_1<0.09$~mag, which approximately captures the dispersion of the fit $s_1$ parameter set (the set's standard deviation is $0.09$~mag). 
The lighter shaded regions correspond to positive $s_1$ values, while the darker correspond to negative values. 
As $s_1$ increases (decreases), $\mathbf{F}_{\text{eff}}$ node values increases (decreases) with respect to $\mathbf{F}_0$'s node values, resulting in each GPMP interpolation curve global maximum decreasing (increasing). 
This change in $\mathbf{F}_{\text{eff}}$ node scaling is offset for by a change in $\chi_0$, correlating $\chi_0$ and $s_1$ parameter sets (see Section~\ref{sec:results:refit}). 

\begin{figure}
    \plotone{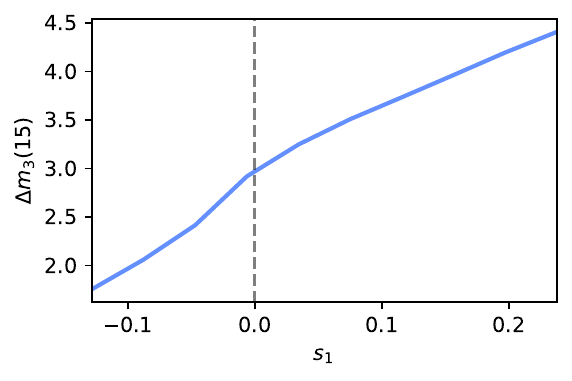}
    \caption{The model's $\Delta m_3(15)$ (the $\Delta m_B(15)$ analog for the fixed band~3) as a function of $s_1$ calculated for the fixed band~3 along the training sample's obtained $s_1$ value range.}
    \label{fig:delm15}
\end{figure}

The sign of $\mathbf{M}_1$ template's contribution is a function of phase: for each curve there are two phases where $\mathbf{M}_1$ template's contribution reverses in sign.  
For positive (negative) $s_1$, the result is a narrowing (broadening) of the effective flux curve. 
The phase and degree of this broadening varies between bands in a manner consistent with~\cite{Kasen2007}, being more extreme for bluer wavelengths. 
Furthermore, Figure~\ref{fig:delm15} plots our model's $\Delta m_3(15)$ as a function of $s_1$, demonstrating our $\mathbf{M}_1$ template indeed recovers stretch-like behavior for this model's \textit{B}-band analog fixed band~3. 

\begin{figure}
    \plotone{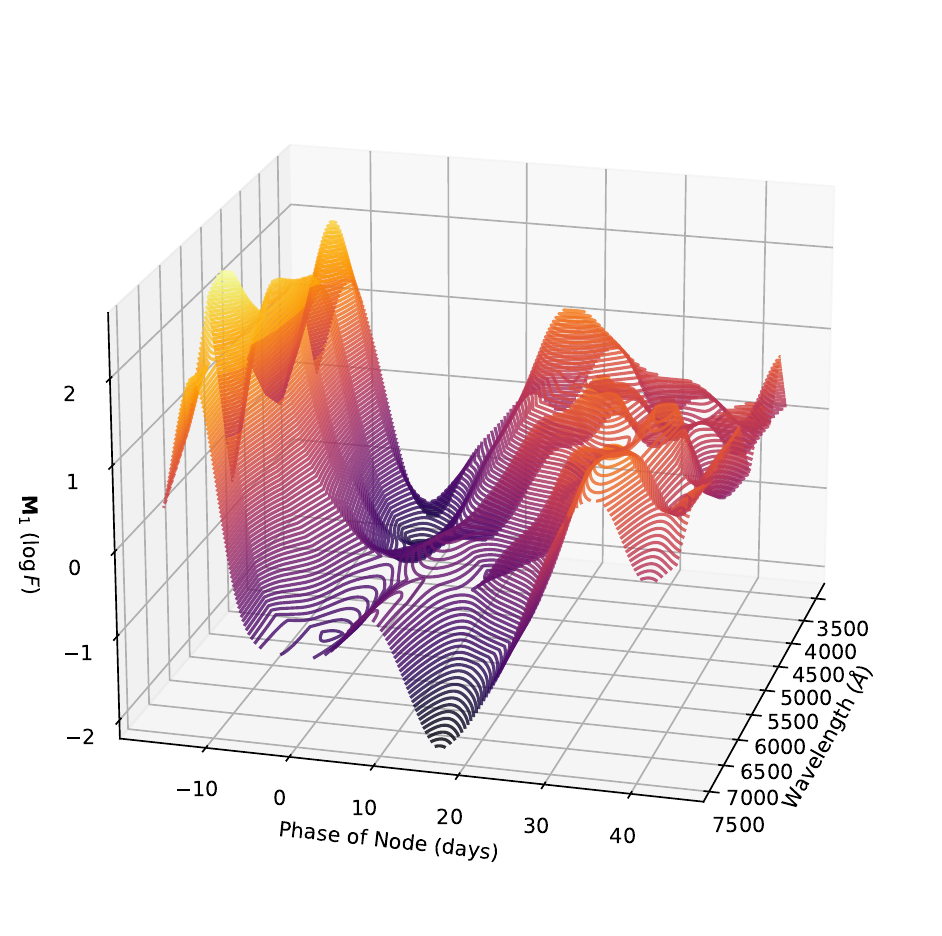}
    \caption{A contoured three-dimensional view of our phase-dependent variation template --- this is our model's equivalent to SALT2's $M_1$ stretch template}. 
    The valley-like structure corresponds to stretch-like behavior extracted by our $\mathbf{M}_1$ template.
    \label{fig:stretchtemplate3D_backend}
\end{figure}

For redder bands, stretch-like behavior is convolved with NIR bump variation (bottom plot of Figure~\ref{fig:stretchtemplate}). 
A 3D mesh plot Figure~\ref{fig:stretchtemplate3D_backend} best illustrates these two modes of NIR variation: bump depth and bump location.
As expected, these variation modes also correlate with stretch \citep{Burns2011, Dhawan2015}, with stretch appearing as the valley-like feature in Figure~\ref{fig:stretchtemplate3D_backend}. 

For fixed band~3, the phase of maximum brightness relative to our zero-phase node is a function of $s_1$. 
This movement in maximum brightness location, made clear in Figure~\ref{fig:stretchtemplate}, ranges from $+1$~day for our most negative $s_1=-0.15$~mag SN~Ia to $-3$~days for our most positive $s_1=0.22$~mag. 
Again, this has no effect on our analysis, requiring only an \textit{a posteriori} calculation at phase of maximum brightness if desired. 

\subsection{Per-Supernova Results}\label{sec:results:refit}
Each SN~Ia is refit with template parameters fixed to the previously discussed best-fit solution. 
Scatter plots comparing parameter sets include Spearman rank correlation coefficient (SCC) calculations alongside corresponding $p$-values. 
For color, only the CCM89-derived basis parameter sets $c_1^{\prime}$ and $c_2^{\prime}$ are presented.
As the MVR and CCM89-derived bases yield parallel $\mathcal{L}$ planes, we choose to present results from the latter since the CCM89-derived basis is more readily interpreted --- its first component is, by definition, a mathematically valid CCM89 curve.
Note that we do not compare per-SN parameter sets from different basis decompositions.

\begin{figure}
    \plotone{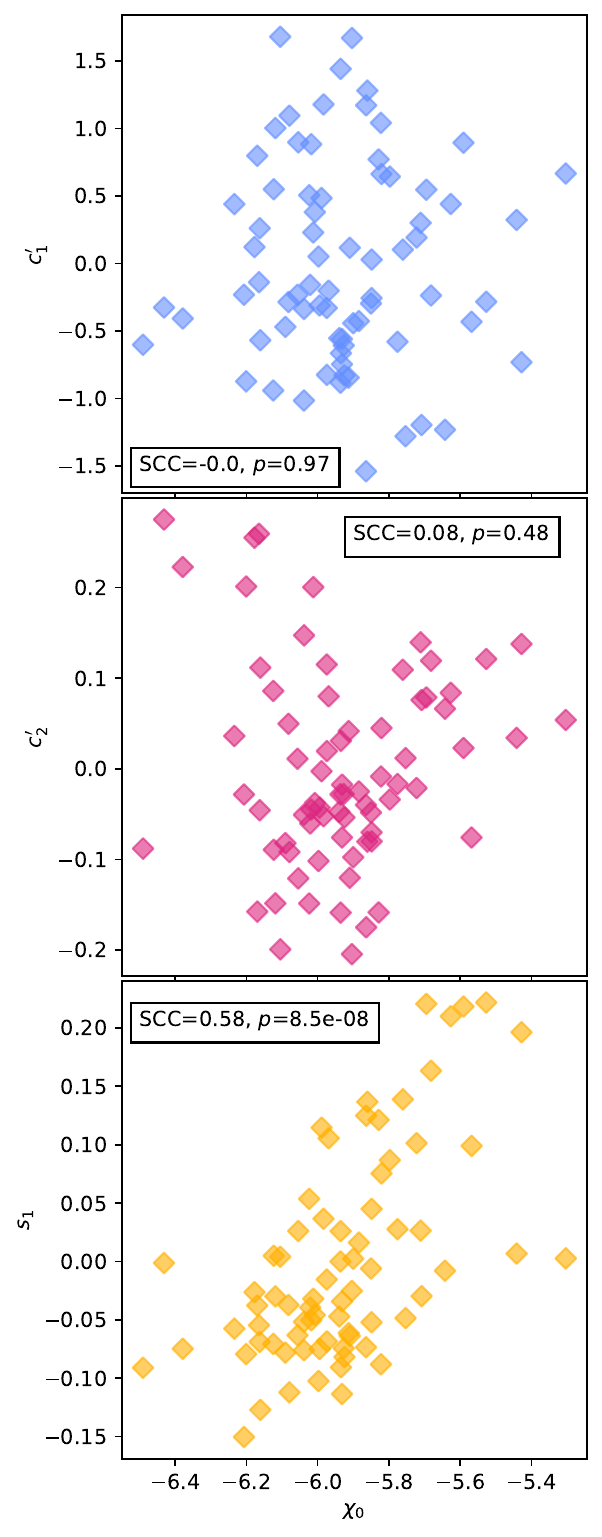}
    \caption{Comparison of fit $c_1^{\prime}$ (top, blue), $c_2^{\prime}$ (middle, magenta), and $s_1$ (bottom, yellow) samples against our $\chi_0$ samples. 
    The correlation between $\chi_0$ and $s_1$ arises from $s_1$'s changing of $\mathbf{F}_{\text{eff}}$'s scale, which is then compensated for by a change in $\chi_0$.
    There are no correlations between either $c_1^{\prime}$ and $\chi_0$ or $c_2^{\prime}$ and $\chi_0$ by construction.} 
    \label{fig:chi0_c1c2s1}
\end{figure}

\begin{figure*}
    \plotone{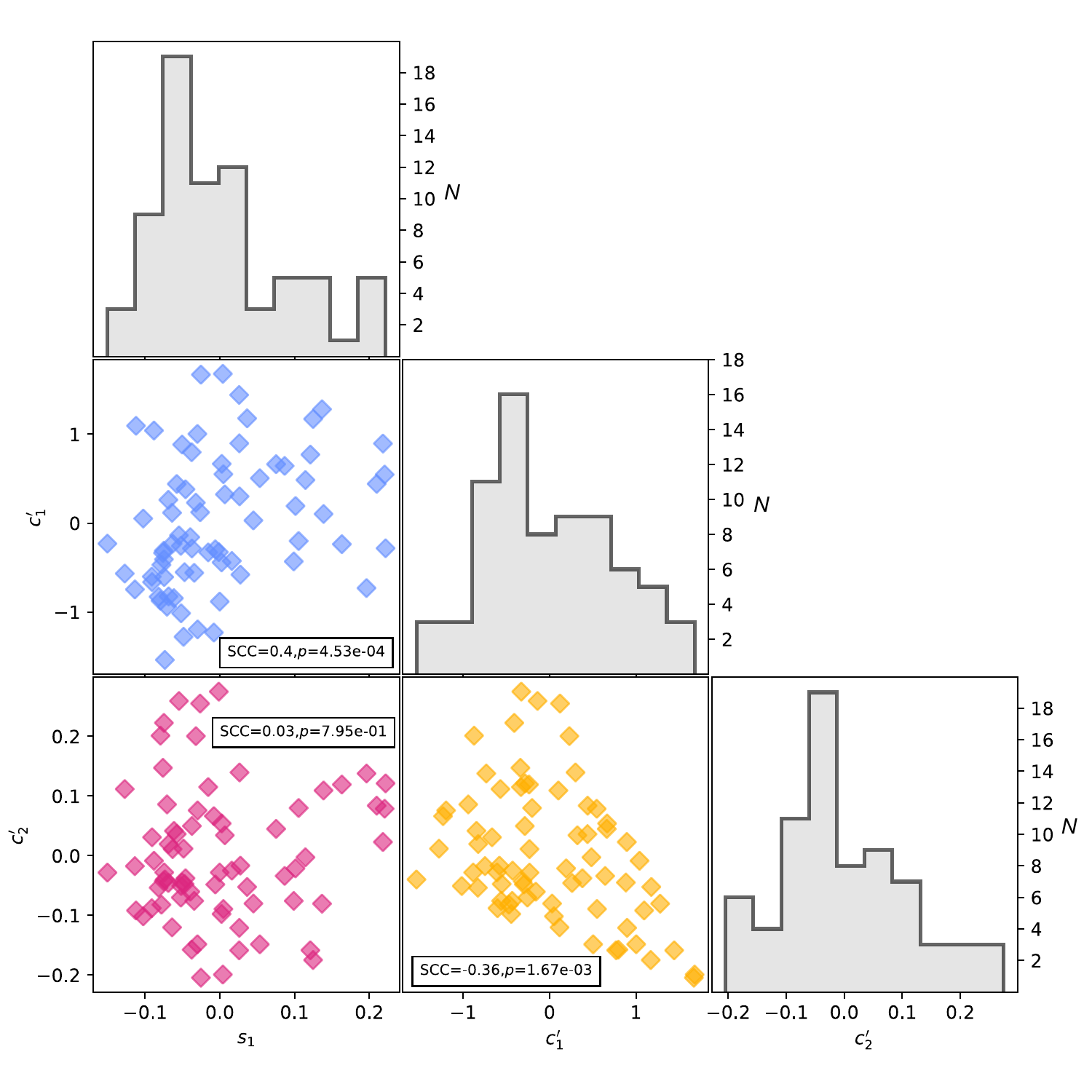}
    \caption{Corner plot for per-SN parameters $s_1$, $c_1^{\prime}$, $c_2^{\prime}$.  
    Magenta points compare $s_1$ and $c_1^{\prime}$, yellow points compare $c_2^{\prime}$ and $c_1^{\prime}$, and blue points compare $c_2^{\prime}$ and $c_1^{\prime}$ parameter sets, respectively.
    We measure only marginal rank correlations between both $c_1^{\prime}$ vs. $s_1$ and $c_1^{\prime}$ vs $c_2^{\prime}$. $N$ is the number of SNe~Ia.}
    \label{fig:c1c2s1_corner}
\end{figure*}

In Figure~\ref{fig:chi0_c1c2s1}, per-SN parameter sets for $c_1^{\prime}$, $c_2^{\prime}$, and $s_1$ are compared with corresponding $\chi_0$ values. 
The measured SCC value of $0.59$ between the training sample's $s_1$ and $\chi_0$ parameter sets results from a varying $s_1$ changing the resulting $\mathbf{F}_{\text{eff}}$ scaling, requiring a compensating change in $\chi_0$ to offset (see Section~\ref{sec:results:stretchtemplate}). 
By construction, $c_1^{\prime}$ and $c_2^{\prime}$ sets are decorrelated with $\chi_0$ (see Section~\ref{sec:method:postprocessing}). 
Higher rank correlations are recovered for $c_1^{\prime}$ vs $s_1$ and $c_1^{\prime}$ vs $c_2^{\prime}$ compared to $c_2^{\prime}$ vs $s_1$, as seen in the scatter plots of Figure~\ref{fig:c1c2s1_corner}. 

We also quantify the fractional variance of the $c_1^{\prime}$ and $c_2^{\prime}$ bivariate distribution not explained by CCM89-like behavior. 
Each SN~Ia phase-independent chromatic flux variation vector $\mathbf{c}=c_1^{\prime}\clone^{\prime} + c_2^{\prime}\cltwo^{\prime}$ is first normalized. 
The perpendicular component with respect to the CCM89 plane spanned by $\clbv^{\text{ccm}}$ of each normalized $\mathbf{c}$ is then calculated via a projection operation (see Appendix~\ref{appendix:geoalg:proj_rej}). 
This resulting distribution has a median value of $0.13$ with 68\textsuperscript{th} percentiles $[0.05, 0.4]$ and provides a measure of our sample's fractional variance attributable to captured phase-independent chromatic variability which is not dust-like. 
Unsurprisingly, dust-like variation, which explains the remaining $\approx87$\% variance, dominates captured phase-independent variability. 
Even if this dust-like variation was the exclusive result of actual dust extinction (no `leaking' of intrinsic variability into dust-like behavior), the remaining $\approx13$\% variance, which instead arises from intrinsic variability in the sample, is not negligible. 
$\clbv$, with $\cltwo^{\prime}$ in particular, is capturing a discernible addition of SN~Ia variation over past two-component  models (i.e. SALT2). 

\subsubsection{Comparison to SALT2}\label{sec:results:refit:comparison}
\begin{figure}
    \plotone{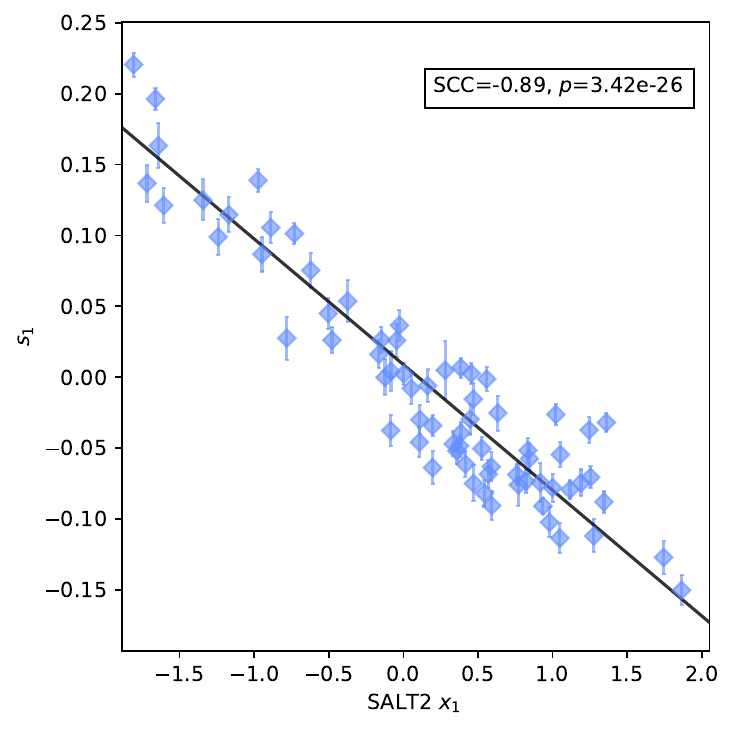}
    \caption{Per-SN comparison of our stretch parameter $s_1$ versus SALT2's stretch proxy $x_1$. 
    An ordinary least squares linear best fit is provided with a solid black line. 
    Error bars correspond to 68\textsuperscript{th} percentiles.}
    \label{fig:s1x1}
\end{figure}

This new SN~Ia model and SALT2 are trained using optical wavelength observations, with neither making assumptions about dust extinction, making it an obvious comparator. 
One technical difference is our model's accounting for phase-dependent variability with a multiplicative variation template as opposed to SALT2's flux variation linear component, which is obviously additive in flux space. 
Nonetheless, $s_1$ and $x_1$ should correlate. 
This is the case as seen in Figure~\ref{fig:s1x1}, with a rank correlation of $-0.89$ between $x_1$ and $s_1$.  
As presented in Section~\ref{sec:results:stretchtemplate}, this model's $\mathbf{M}_1$ templates obtains stretch-like behavior, just as SALT2's first-order variation template $M_1(t,\lambda)$ does.  

\begin{figure}
    \plotone{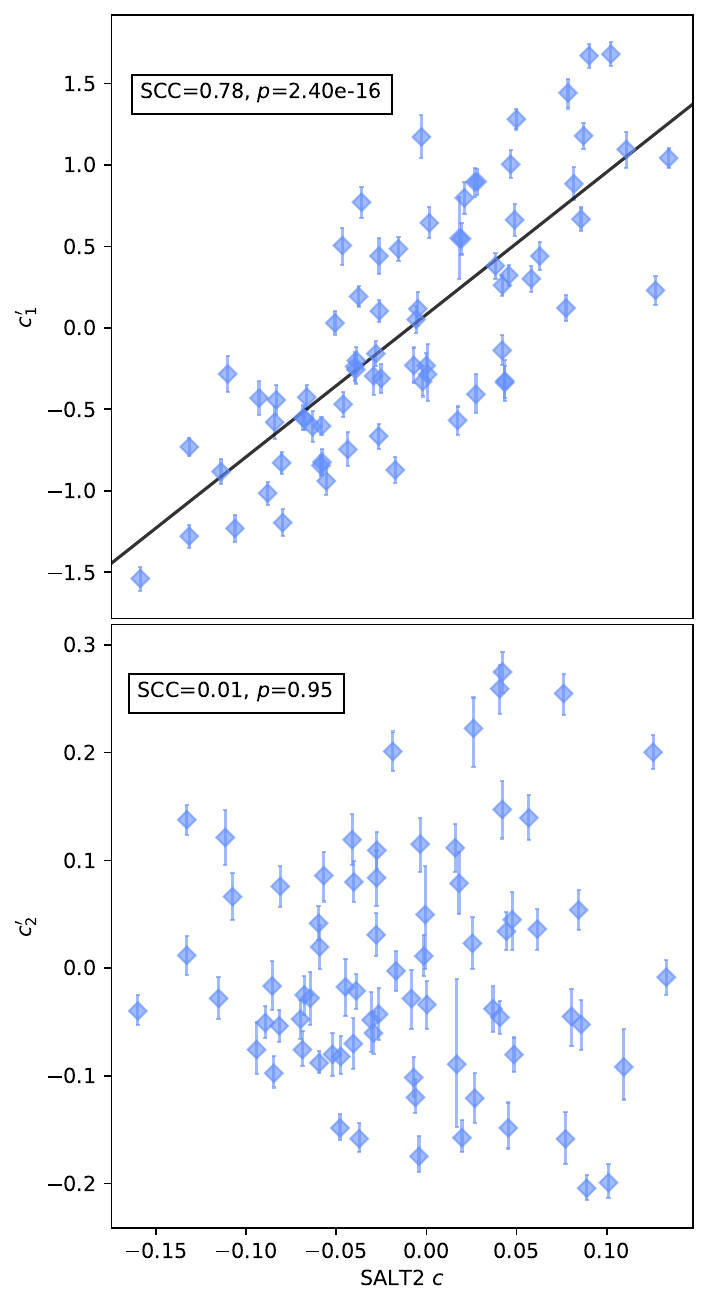}
    \caption{A per-SN comparison of our per-SN chromatic flux variation parameters $c_1^{\prime}$ (top) and $c_2^{\prime}$ (bottom) against SALT2's $c$ parameter. 
    We measure a clear rank anticorrelation between $c_1^{\prime}$ and SALT2 $c$, but measure no correlation between $c_2^{\prime}$ and SALT2 $c$. 
    We interpret this as template $\cltwo^{\prime}$ capturing chromatic flux variation not modeled by SALT2. 
    An ordinary least squares linear best fit between $c_1^{\prime}$ and SALT2 $c$ is provided by a solid black line. 
    Error bars correspond to 68\textsuperscript{th} percentiles.}
    \label{fig:c1c2c}
\end{figure}

The phase-independent chromatic flux variation template basis $\{\clone^{\prime},\clone^{\prime}\}$ is selected without consideration of SALT2's $CL(\lambda)$ phase-independent chromatic variation model. 
As such, any correlations between $c_1^{\prime}$ or $c_2^{\prime}$ and SALT2 $c$ are nontrivial --- as seen in the top plot of Figure~\ref{fig:c1c2c}, only SALT2 $c$ and $c_1^{\prime}$ are correlated with a rank correlation of $0.78$. 
Considering $\clone^{\prime}$ is the maximal CCM89 dust-like vector allowed, that SALT2 $c$ and $c_1^{\prime}$ are strongly correlated is because SALT2's $CL(\lambda)$ template predominately captures dust-like variation. 
Indeed, the latest SALT3 recovers a $CL(\lambda)$ curve that is consistent with SALT2 and similarly aligns with CCM89 between $4000$~\r{A} and $7000$~\r{A} \citep{Kenworthy2021}. 
As seen in Figure~13 of~\cite{Kenworthy2021}, both of SALT2 and SALT3 $CL(\lambda)$ templates begin to diverge from CCM89 near where $\cltwo^{\prime}$ starts to exhibit most of its variability. 

The bottom plot of Figure~\ref{fig:c1c2c} demonstrates that $c_2^{\prime}$ and SALT2 $c$ parameter sets are uncorrelated. 
As one would hope, the presented model's two-component phase-independent chromatic variation model captures SN~Ia variation beyond that of SALT2. 
For reference, we provide best fit linear relationships between SALT2 $x_1$ and $s_1$, and between SALT2 $c$ and $c_1^{\prime}$:
\begin{align}
    c_1^{\prime}(c) &= 8.74(\pm0.88)c+0.094(\pm0.056)\\
    s_1(x_1) &=-0.089(\pm0.004)x_1 + 0.0089(\pm0.0035).
\end{align} 

\section{Conclusion}\label{sec:conclusion}
In this article, we introduce a new empirical SN~Ia linear model that expands beyond similar analyses by introducing a second phase-independent chromatic flux variation template into its architecture. 
Our model converges to a solution with three variation components: one phase-dependent chromatic flux variation template $\mathbf{M}_1$ and two phase-independent chromatic flux variation templates $\clone$ and $\cltwo$, which are represented together by the bivector $\clbv$. 
The phase-dependent model component recovers stretch and NIR bump variation; it along with the phase-independent templates are consistent with SALT2's two model components. 

The model's extended phase-independent architecture captures a nuanced combination of variability. 
Analysis of the two-dimensional, phase-independent chromatic flux variation plane spanned by $\clbv$ is done two bases: one that maximizes variation ratios (MVR) between $c_1$ and $c_2$ parameter sets, and another that is defined with respect to the CCM89 dust plane. 
The MVR approach recovers nominally dust-like variation for its dominant component $\clone^{\text{mvr}}$, with a fit $R_V^{\text{MVR}}=2.18$, while $\cltwo^{\text{mvr}}$ trace potentially low-resolution intrinsic features. 
The CCM89 basis yields similar, albeit somewhat more physically-interpretable, results. 
Specifically, this basis naturally decomposes $\clbv$ into one template capturing continuum-like wavelength variation ($\clone^{\prime}$) and another capturing rapidly changing chromatic flux variation w.r.t. wavelength ($\cltwo^{\prime}$). 
Component $\clone^{\prime}$ is defined as the intersection of our model's phase-independent chromatic flux variation plane with CCM89 dust-like behavior. 
$\cltwo^{\prime}$ is orthogonal to $\clone^{\prime}$ and can contain phase-averaged intrinsic spectral variation not related to the $\mathbf{M}_1$ template; by construction, it can also contain residual CCM89-like variation. 

Despite very different construction methods, $\cltwo^{\prime}$ and $\cltwo^{\text{mvr}}$ are remarkably similar. 
As derived from the CCM89 basis, the fractional variance of intrinsic spectral variation for the two-dimensional phase-independent chromatic flux variation plane (with 68\textsuperscript{th} percentiles) is $0.13_{-0.05}^{+0.4}$ --- dust-like variation dominates $\clone^{\prime}$ and $\cltwo^{\prime}$ in combination, but residual spectroscopic-feature-like variation is not negligible. 
Despite this model's coarse $n_{\lambda}=10$ wavelength bin count, we find that the features in $\cltwo^{\prime}$ and $\cltwo^{\text{mvr}}$ both align with known SN~Ia spectral features. 
Because of intrinsic variation `leaking' into these two phase-independent components, we make no strong conclusion about the recovered dust extinction curve from either $\clone^{\prime}$ or $\clone^{\text{mvr}}$. 

In our approach of supernovae standardization, we avoid setting conditions on the properties of unextinguished SNe or of dust and hence are unable to directly make conclusions about foreground dust, apart from recovering a model component whose behavior is like that of dust (specifically, CCM89).  
Either forward modeling via additional post-processing, or directly integrating a dust model into an improved version of our SN~Ia model are options to consider --- such augmentations will be the focus of work in the future.
Such future work will also increase synthetic photometry spectral resolution to allow for better identification of spectral features.  

\begin{acknowledgements}
The authors thank Prof.\ Michael Wood-Vasey for his continuous feedback about model implementation and interpretation of results. This material is based upon work supported by the U.S. Department of Energy, Office of Science, Office of Workforce Development for Teachers and Scientists, Office of Science Graduate Student Research (SCGSR) 2018 S2 program. The SCGSR program is administered by the Oak Ridge Institute for Science and Education for the DOE under contract number DE-SC0014664.  
This work is also supported in part by the US Department of Energy Office of Science under DE-SC0007914. 
This research is supported in part by the University of Pittsburgh Center for Research Computing through the resources provided. 

This work was supported in part by the Director, Office of Science, Office of High Energy Physics of the US Department of Energy under contract No. DE-AC02-05CH11231. Support in France was provided by CNRS/IN2P3, CNRS/INSU, and PNC and French state funds managed by the National Research Agency within the Investissements d'Avenir program under grant reference numbers ANR-10-LABX-0066, ANR-11-IDEX-0004-02 and ANR-11-IDEX-0007. Additional support comes from the European Research Council (ERC) under the European Union's Horizon 2020 research and innovation program (grant agreement No. 759194–USNAC). Support in Germany was provided by DFG through TRR33 "The Dark Universe" and by DLR through grants FKZ 50OR1503 and FKZ 50OR1602. In China support was provided by Tsinghua University 985 grant and NSFC grant No. 11173017. Some results were obtained using resources and support from the National Energy Research Scientific Computing Center, supported by the Director, Office of Science, Office of Advanced Scientific Computing Research of the US Department of Energy under contract No. DE-AC02-05CH11231. We also thank the Gordon \& Betty Moore Foundation for their support.

M.R. has received funding from the European Research Council (ERC) under the European Union's Horizon 2020 research and innovation program (grant agreement No. 759194—USNAC).

The first author extends their gratitude to Prof.\ Wood-Vasey and Dr.\ Kim for their constant, unconditional, and unwavering support during a period of health struggles. 
This project would not have come to fruition without their mentorship, constant advocating, and kindness. 
\end{acknowledgements}

\appendix



\section{Centered Vectors for SN~Ia Parameters}\label{appendix:centeredvectors}
A centered vector is a vector $\mathbf{v}$ with the mean of its components being zero. 
There are numerous ways to structural enforce centered vectors in Stan\footnote{\url{https://mc-stan.org/docs/2_18/stan-users-guide/parameterizing-centered-vectors.html}}. 
For performance reasons we opt for a shifted simplex method. 

A simplex $\mathbf{\Delta}$ of dimension $n$ is a vector with the constraint that its components sum to one: $\sum^i_n \Delta_i=1$.
Dividing the sum of the simplex by its component count $n$ gives that simplex's mean: $\langle\mathbf{\Delta}\rangle = 1/n$.
Therefore, any simplex $\mathbf{\Delta}$ instantiated in {\tt Stan} can be transformed via translation by a factor $-1/n$ into a centered simplex $\mathbf{\Delta^{\langle 0 \rangle}}$ with a mean of zero:
\begin{equation}
    \Delta^{\langle 0 \rangle}_i = \Delta_i - 1/n.
\end{equation}
Finally, any centered vector $\mathbf{r}$ with component $r_i$ can be defined as the product of a centered simplex $\mathbf{\Delta^{\langle 0 \rangle}}$ and a scaling parameter $r$:
\begin{equation}
    r_i = r\Delta^{\langle 0 \rangle}_i = r\big(\Delta_i - 1/n\big).
\end{equation}

\section{Achromatic Offset and Chromatic Parameter Degeneracy}
\label{appendix:degens}
As described in~\cite{Leget2020}, for all SNe~Ia there is a degeneracy between an achromatic offset parameter (i.e. $\chi_0$) and a phase-independent chromatic variation vector (i.e. $c_1\clone + c_2\cltwo$). 
With an explicit length-$n_{\lambda}$ vector of ones $\mathbf{1}$ to represent achromatic offset behavior in wavelength space, one can define a vector for each SN~Ia:
\begin{equation}
    \mathbf{v} \equiv \chi_0\mathbf{1} + c_1\clone + c_2\cltwo.
\end{equation} 
For two arbitrary constants $\{\alpha_1, \alpha_2\}$, the vector $\mathbf{v}$ is invariant under the following transformations:
\begin{align*}
    \mathbf{v} &= \chi_0^{\prime}\mathbf{1} + c_1\clone^{\prime} + c_1\cltwo^{\prime}\\
    \clone^{\prime} &= \clone - \alpha_1 \mathbf{1} \\
    \cltwo^{\prime} &= \cltwo - \alpha_2 \mathbf{1} \\
    \chi_0^{\prime} &= \chi_0 + \alpha_1c_1 + \alpha_2c_2.
\end{align*}

At each iteration during sampling or in post-processing, this degeneracy can be removed by choosing a fixed $\{\alpha_1, \alpha_2\}$ and then recalculating the components of $\mathbf{v}$. 
Both $c_1$ and $c_2$ parameter sets should be uncorrelated with peak brightness dispersion, so $\alpha_1$ or $\alpha_2$ are calculated so that both $c_1$ and $\chi_0$, and $c_2$ and $\chi_0$, are uncorrelated. 
Representing parameter sets as length-$n_{\text{sn}}$ vectors $\mathbf{c}_1$, $\mathbf{c}_2$ and $\boldsymbol{\chi}_0^{\prime}$, and starting from the transformation definition for $\chi_0^{\prime}$ above, we require:
\begin{equation}
    \begin{bmatrix}
        \sigma_{\mathbf{c}_1 \boldsymbol{\chi}_0^{\prime}}\\
        \sigma_{\mathbf{c}_2 \boldsymbol{\chi}_0^{\prime}}
    \end{bmatrix}
    =
    \begin{bmatrix}
        \sigma_{\mathbf{c}_1 \boldsymbol{\chi}_0}\\
        \sigma_{\mathbf{c}_2 \boldsymbol{\chi}_0}
    \end{bmatrix}
    +
    \begin{bmatrix}
        \alpha_1\sigma_{\mathbf{c}_1\mathbf{c}_1} + \alpha_2\sigma_{\mathbf{c}_1\mathbf{c}_2}\\
        \alpha_1\sigma_{\mathbf{c}_2\mathbf{c}_1} + \alpha_2\sigma_{\mathbf{c}_2\mathbf{c}_2}
    \end{bmatrix}
    =
    \begin{bmatrix}
        \sigma_{\mathbf{c}_1 \boldsymbol{\chi}_0}\\
        \sigma_{\mathbf{c}_2 \boldsymbol{\chi}_0}
    \end{bmatrix}
    +
    \begin{bmatrix}
        \alpha_1\\
        \alpha_2
    \end{bmatrix}
    \begin{bmatrix}
        \sigma_{\mathbf{c}_1\mathbf{c}_1} & \sigma_{\mathbf{c}_1\mathbf{c}_2}\\
        \sigma_{\mathbf{c}_2\mathbf{c}_1} & \sigma_{\mathbf{c}_2\mathbf{c}_2}
    \end{bmatrix}
    =0,
\end{equation}
or after solving for the vector $[\alpha_1, \alpha_2]^T$:
\begin{equation}
    \begin{bmatrix}
        \alpha_1\\
        \alpha_2
    \end{bmatrix}
    =-
    \begin{bmatrix}
        \sigma_{\mathbf{c}_1 \boldsymbol{\chi}_0}\\
        \sigma_{\mathbf{c}_2 \boldsymbol{\chi}_0}
    \end{bmatrix}
    \begin{bmatrix}
        \sigma_{\mathbf{c}_1\mathbf{c}_1} & \sigma_{\mathbf{c}_1\mathbf{c}_2}\\
        \sigma_{\mathbf{c}_2\mathbf{c}_1} & \sigma_{\mathbf{c}_2\mathbf{c}_2}
    \end{bmatrix}^{-1},
\end{equation}
where the matrix on the right-hand side of this equation as the precision matrix for our $c_1$ and $c_2$ samples.

Both $\clone^{\prime}$ and $\cltwo^{\prime}$ are maintained as unit vectors. 
This is enforced after each transformation by normalizing $\clone^{\prime}$ and $\cltwo^{\prime}$ by factors $|\clone - \alpha_1 \mathbf{1}|$ and $|\cltwo - \alpha_2 \mathbf{1}|$. 
Each parameter set for $c_1$ and $c_2$ is multiplied by their respective factors to preserve the products $c_1\clone^{\prime}$ and $c_2\cltwo^{\prime}$, respectively. 

\section{Brief Introduction to Geometric Algebra}\label{appendix:geoalg} 
Geometric algebra provides a robust, elegant, and easy to interpret framework to perform geometric operations. 
It extends linear algebra by introducing the geometric product, a combination of the inner product and outer product.  
Although we directly calculate intersections, rotations, and projections/rejections using geometric algebra, these operations can be performed using linear algebra instead. 

We will assume through that our vector space finite-dimensional and that each vector's elements is real-valued (specifically, we assume our vector space's field are the real numbers $\mathbb{R}$).  
We will also utilize a canonical basis representation of our vector, such a Cartesian basis for $\mathbb{R}^n$, $\{\cbv_1,\cbv_2,\ldots,\cbv_n\}$. 
All geometric algebra calculations are implemented with the clifford package (\citeauthor{clifford}). 
For a more detailed introduction, see~\cite{Hitzer2013}. 

Geometric algebra extends elementary vector algebra using the geometric product. 
The geometric product of two vectors $\{\mathbf{x},\mathbf{y}\}\in\mathbb{R}^n$ combines a symmetric inner (or dot) product with an antisymmetric outer (or wedge) product. 
\begin{align}
    \mathbf{x}\mathbf{y} &= \mathbf{x}\cdot\mathbf{y} + \mathbf{x}\wedge\mathbf{y}  = \sum^{n}_{i=1}x_iy_i + \sum^{n}_{i=1}\sum^{n}_{j>i}(x_iy_j - y_ix_j)\cbv_i\wedge\cbv_j\\
    &=|\mathbf{x}||\mathbf{y}|\cos\theta_{xy} + |\mathbf{x}||\mathbf{y}|\sin\theta_{xy}\hat{\mathbf{x}}\wedge\hat{\mathbf{y}}.
\end{align} 
In the last step we define the angle $\theta_{xy}$ as the angle between the two vectors $\mathbf{x}$ and $\mathbf{y}$. 
The inner product maps our vectors to our underlying field $\mathbb{R}$, yielding a scalar and is a symmetric operation: $\mathbf{x}\cdot\mathbf{y}=\mathbf{y}\cdot\mathbf{x}$. 
The wedge product of two vectors $\mathbf{x}$ and $\mathbf{y}$ is an antisymmetric operation: $\cbv_i\wedge\cbv_j=-\cbv_j\wedge\cbv_i=0\:(\text{if}\:i=j)$. 
The object output by the wedge product of two vectors is called a bivector and is an oriented plane element. 

It is common in geometric algebra to refer to a scalar as the 0-blade and a bivector as the 2-blade.
Similarly, vectors such as $\mathbf{x}$ and $\mathbf{y}$ are referred to as 1-blades. 
In general, a k-blade is a k-dimensional object with can be generated by k-wedge (or k-geometric) products of k independent vectors. 
These k-vectors are themselves oriented k-dimensional subspace elements. 
The produced object $\mathbf{x}\mathbf{y}$ is a linear combination of two grades of k-blades (a 0-vector and a 2-vector, specifically) that we call a multivector. 
In general, a multivector is a linear combination k-blades. 
Also, each multivector $\mathbf{M}$ has an inverse such that via the goemtric product $\mathbf{M}\mathbf{M}^{-1}=1$. 
Furthermore, any multivector defined as a product of vectors $\mathbf{M}=\mathbf{a}_1\mathbf{a}_2...\mathbf{a}_m$ has a corresponding multivector called its reverse: $\mathbf{M}^{\dagger}=\mathbf{a}_m\mathbf{a}_{m-1}...\mathbf{a}_1$.

We can derive a canonical basis of our geometric algebra using the geometric product on our starting vector space's canonical basis as a generating set.
Because canonical basis is orthogonal, the geometric product of canonical basis components reduces to a wedge product for $i\neq j$: $\cbv_i\cbv_j = 0 + \cbv_i\wedge\cbv_j$, or to an inner product for $i=j$: $\cbv_i\cbv_i=\cbv_i \cdot \cbv_i + 0$. 
As an example, for a vector space of dimension $n=3$, the generated geometric algebra canonical basis has 8 basis elements: $\{1;\cbv_1,\cbv_2,\cbv_3;\cbv_1\cbv_2,\cbv_1\cbv_3,\cbv_2\cbv_3;\cbv_1\cbv_2\cbv_3 \}$. 
This set of basis elements naturally fall into grades of equal subspace dimension: one scalar element, three one-dimensional basis vectors (the starting vector space's canonical vector basis), three two-dimensional basis bivectors, and one three-dimensional basis trivector. 
The geometric algebra over $\mathbb{R}^3$ basis elements therefore prescribes corresponding sets of orthogonal bases for all possible subspaces that exist within $\mathbb{R}^3$. 
The generation of basis elements generalizes naturally for any finite dimension vector space $\mathbb{R}^n$, with the resulting geometric algebra $\mathbb{G}^n$ having $2^n$ generated basis components. 
The respective subspaces, or grades, of the geometric algebra (i.e. the bivector or grade-2 components) have dimensionality $n \choose k$, where $k$ is the grade of interest. 
Note that the last element in this geometric algebra basis is called the pseudoscalar $i$, since $ii=-1$ for any dimension $n$ in a way analogous with an imaginary number. 
Finally, we can decompose any multivector $\mathbf{M}$ using a grade projection operation:
\begin{equation}
    \mathbf{M} = \sum_{k=1}^{n}\langle \mathbf{M} \rangle_{k},
\end{equation} 
where the $k$-grade projection operator $\langle \mathbf{M} \rangle_{k}$ returns the $k$-blade component of $\mathbf{M}$. 

\subsection{Reflections and Rotations}\label{appendix:geoalg:rotations}
Consider two vectors $\mathbf{r},\mathbf{x}\in\mathbb{R}^n$.  
Say we reflect $\mathbf{r}$ along $\mathbf{x}$.  
This operation can be interpreted as negating the component of $\mathbf{r}$ perpendicular to $\mathbf{x}$. 
The geometric product can be used to decompose $\mathbf{r}$ into its parallel and perpendicular components relative to the unit vector $\hat{\mathbf{x}}$:
\begin{align}
    \mathbf{r}\hat{\mathbf{x}} &= \mathbf{r}\cdot\hat{\mathbf{x}} + \mathbf{r}\wedge\hat{\mathbf{x}} \\
    \mathbf{r} &= (\mathbf{r}\cdot\hat{\mathbf{x}})\hat{\mathbf{x}}^{-1} + (\mathbf{r}\wedge\hat{\mathbf{x}})\hat{\mathbf{x}}^{-1}.
\end{align}
The component $(\mathbf{r}\cdot\hat{\mathbf{x}})\hat{\mathbf{x}}^{-1}$ is proportional to the projection of $\mathbf{r}$ onto $\mathbf{x}$, while the component $(\mathbf{r}\wedge\hat{\mathbf{x}})\hat{\mathbf{x}}^{-1}$ is the perpendicular (or rejected) remainder. 
From this decomposition we can define a reflection $\mathbf{r}^{\prime}$ of $\mathbf{r}$ along $\mathbf{x}$:
\begin{align*}
    \mathbf{r}^{\prime} &= -(\mathbf{r}\cdot\hat{\mathbf{x}})\hat{\mathbf{x}}^{-1} + (\mathbf{r}\wedge\hat{\mathbf{x}})\hat{\mathbf{x}}^{-1}\\
    &= -(\hat{\mathbf{x}}\cdot\mathbf{r})\hat{\mathbf{x}}^{-1} - (\hat{\mathbf{x}}\wedge\mathbf{r})\hat{\mathbf{x}}^{-1}\\
    &= -(\hat{\mathbf{x}}\cdot\mathbf{r} + \hat{\mathbf{x}}\wedge\mathbf{r})\hat{\mathbf{x}}^{-1}\\
    &=-\hat{\mathbf{x}}\mathbf{r}\hat{\mathbf{x}}^{-1}\\
    &=-\hat{\mathbf{x}}\mathbf{r}\hat{\mathbf{x}}.
\end{align*}
Here we exploit a unit vector's geometric algebra inverse being itself ($\hat{\mathbf{x}}\hat{\mathbf{x}}=\hat{\mathbf{x}}\cdot\hat{\mathbf{x}}=1$). 

Any rotation can be decomposed into two reflections, with the rotational plane being spanned by two vectors we reflect against. 
Picking another vector $\mathbf{y}\in\mathbb{R}^n$, we write the rotation $\mathbf{r}^{\prime}$ of $\mathbf{r}$ in the plane corresponding to grade-2 (or bivector component) of the multivector $\mathbf{R}=\hat{\mathbf{x}}\hat{\mathbf{y}}$ as:
\begin{align*}
    \mathbf{r}^{\prime} &=\hat{\mathbf{y}}\hat{\mathbf{x}}\mathbf{r}\hat{\mathbf{x}}\hat{\mathbf{y}}\\
    &=\mathbf{R}^{\dagger}\mathbf{r}\mathbf{R}.
\end{align*}
This bilinear operation on $\mathbf{r}$ by the unit 2-vector $\mathbf{R}$ is indeed a rotation.  
To see this, let us insert a more illuminating form of $\mathbf{R}$:
\begin{equation}\label{eq:almost_rotation}
    \mathbf{r}^{\prime} =\Big(\cos\frac{\theta_{xy}}{2} - \sin\frac{\theta_{xy}}{2}\hat{\mathbf{x}}\wedge\hat{\mathbf{y}}\Big)\mathbf{r}\Big(\cos\frac{\theta_{xy}}{2} + \sin\frac{\theta_{xy}}{2}\hat{\mathbf{x}}\wedge\hat{\mathbf{y}}\Big).
\end{equation}
To interpret this result, consider first the geometric algebra of vector space $\mathbb{R}^2$ with its four generated basis elements $\{1;\cbv_1,\cbv_2;\cbv_1\cbv_2\}$. 
In two dimensions any bivector $\hat{\mathbf{x}}\wedge\hat{\mathbf{y}}$ is proportional our unit pseudoscalar $\cbv_1\cbv_2$, itself which behaves identically to the imaginary number (similarly denoted as $i$). 
Substituting $i$ for $\hat{\mathbf{x}}\wedge\hat{\mathbf{y}}=\cbv_1\cbv_2$ into Equation~\ref{eq:almost_rotation}, we recover the canonical form of a rotation by an angle $\theta$ in two dimensions:
\begin{equation}
    \mathbf{r}^{\prime} = e^{-i\theta/2}\mathbf{r}e^{i\theta/2}.
\end{equation}
For higher dimensions we can similarly treat the bivector $\hat{\mathbf{x}}\wedge\hat{\mathbf{y}}$ as being the imaginary number of the two-dimensional subspace spanned by said bivector. 
Therefore, any rotation of a vector $\mathbf{r}$ by an angle $\theta$ within a plane spanned by the unit bivector$\hat{\mathcal{A}}=\hat{\mathbf{x}}\wedge\hat{\mathbf{y}}$ can be written:
\begin{equation}
    \mathbf{r}^{\prime} = e^{-\hat{\mathcal{A}}\theta/2}\mathbf{r}e^{\hat{\mathcal{A}}\theta/2} = \mathbf{R}^{\dagger}\mathbf{r}\mathbf{R}.
\end{equation}
This unit bivector is interpreted as a generator of rotation within the plane spanned by $\hat{\mathcal{A}}$; the multivector $\mathbf{R}$ (which has scalar and bivector components) is called a rotor.

\subsection{Projection and Rejection of Vectors onto a Bivector}\label{appendix:geoalg:proj_rej}
Consider a vector $\mathbf{v}$ and a bivector $\mathcal{A}$. 
Similar to Subsection~\ref{appendix:geoalg:rotations}, we can decompose a $\mathbf{v}$ into components parallel and perpendicular to the plane spanned by $\mathcal{A}$:
\begin{align}
    \mathbf{v}\mathcal{A} &= \mathbf{v}\cdot\mathcal{A} + \mathbf{v}\wedge\mathcal{A}\\
    \mathbf{v} &= (\mathbf{v}\cdot\mathcal{A})\mathcal{A}^{-1} + (\mathbf{v}\wedge\mathcal{A})\mathcal{A}^{-1}.
\end{align}
If the wedge product of any vector with a $\mathcal{A}$ is zero, then said vector must exist within the plane spanned by that $\mathcal{A}$. 
Alternatively, any vector with no projection onto $\mathcal{A}$ requires said vector be perpendicular to $\mathcal{A}$. 
Just as in Subsection~\ref{appendix:geoalg:rotations}, we therefore identify this decomposition into parallel (projected) and perpendicular (rejected) components relative to a plane spanned by $\mathcal{A}$:
\begin{align}
    \mathbf{v}^{\parallel}&=(\mathbf{v}\cdot\mathcal{A})\mathcal{A}^{-1}\\
    \mathbf{v}^{\perp}&=(\mathbf{v}\wedge\mathcal{A})\mathcal{A}^{-1}.
\end{align}

\subsection{Intersection of Planes}\label{appendix:geoalg:intersection}
This section describes the formalism readily displayed in Figure~\ref{fig:geoalg_intuition}. 
It is the most complicated of the operations we implement using geometric algebra and is included for completeness, despite its complexity. 
As with other operations, it is possible to calculate this intersection using conventional linear algebra techniques. 
Indeed, we confirmed this geometric algebra implementation with such a conventional approach. 

Intuitively, a bivector is to a plane what a vector is to a line. 
Just as one can find the intersection of two lines using their representing vectors, one also can find the intersection of two planes using their representing bivectors. 

Consider two bivectors $\mathcal{A}$ and $\mathcal{B}$, each of which represent two different planes. 
If these two planes intersect (which we will assume they do), then there are at most two linearly independent vectors $\mathbf{a}$ and $\mathbf{b}$ that exist each of the respective planes spanned by $\mathcal{A}$ and $\mathcal{B}$, but that do not reside along these two planes' intersection. 
Furthermore, the intersection of $\mathcal{A}$ and $\mathcal{B}$ is a well-defined, one-dimensional subspace necessarily spanned by some vector, a vector that we will call $\mathbf{c}$. 
This means that $\mathcal{A}=\mathbf{a}\wedge\mathbf{a}$ and $\mathcal{B}=\mathbf{c}\wedge\mathbf{b}$, and taken together, these two bivectors span a three-dimensional subspace (because they share the intersection subspace spanned by $\mathbf{c}$). 
This volume and its orientation can be represented using a trivector $\mathbf{a}\wedge\mathbf{b}\wedge\mathbf{c}$. 
Normalizing this then trivector gives us the intersection's unit volume element:
\begin{equation}
    i = \frac{\mathbf{a}\wedge\mathcal{B}}{|\mathbf{a}\wedge\mathcal{B}|}= \frac{\mathbf{b}\wedge\mathcal{A}}{|\mathbf{b}\wedge\mathcal{A}|}.
\end{equation}

We want a formal expression for $\mathbf{c}$ from the starting bivectors $\mathcal{A}$ and $\mathcal{B}$, though. 
To get this, we first find a bivector whose plane is simultaneously perpendicular to both $\mathcal{A}$ and $\mathbf{B}$, which can be done by taking the grade projection of the geometric product $\mathcal{A}\mathcal{B}$ to its bivector component: $\langle\mathcal{A}\mathbf{B}\rangle_2$. 
The vector perpendicular to this bivector $\langle\mathcal{A}\mathbf{B}\rangle_2$, but which still exists within the volume spanned by the intersection unit volume $i$, is the intersection vector $\mathbf{c}$ we are looking for.
It turns out that taking the geometric product of $i$ with $\langle\mathcal{A}\mathbf{B}\rangle_2$ performs exactly this operation:
\begin{equation}
    \mathbf{c} = i\langle\mathcal{A}\mathcal{B}\rangle_2.
\end{equation}
Once way to think about this is that the unit volume $i$ rotates the bivector $\langle\mathcal{A}\mathbf{B}\rangle_2$ by $\pi/2$ and projects out the vector $\mathbf{c}$ perpendicular to the bivector $\langle\mathcal{A}\mathbf{B}\rangle_2$. 

\section{HMC Sample Chains}
\begin{figure}
    \centering
    \includegraphics{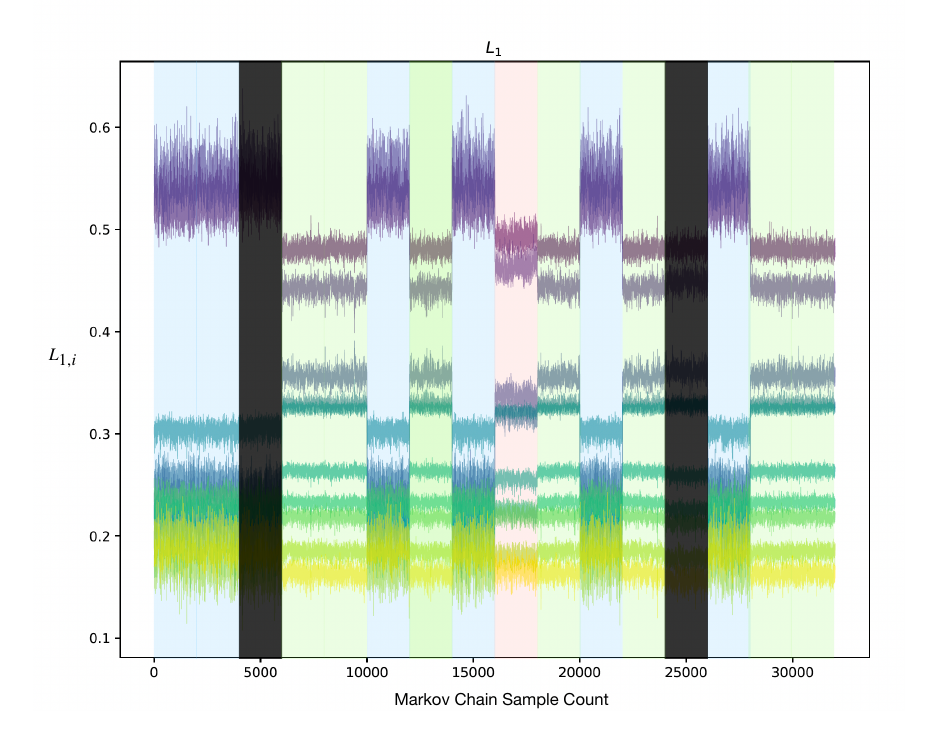}
    \caption{Posterior samplings for each of the $\mathbf{L}_1$ template's ten components. 
    The x-axis labels the 2000 samples for each of the 16 samplers, concatenated in sequence for 32000 total draws.
    The aforementioned groupings are shaded blue, green, and pink.
    The dark gray rectangles mask the samples drawn by two samplers which became lost during sampling and were subsequently excluded from our analysis.}
    \label{fig:twogroups}
\end{figure}

As mentioned and interpreted in Sections~\ref{sec:method} and~\ref{sec:results}, we identify three HMC sampler groupings during post-processing. 
Here we present the HMC posterior sampling chains for the raw $\mathbf{L}_1$ parameters before achromatic offset decorrelation (Appendix~\ref{appendix:degens}) or basis selection (Subsection~\ref{sec:results:colorlaw}). 
As such, the sampler chains seen here are not yet physically interpretable, nor are they used to measure convergence. 
Although also present in post-processed basis representations, the sampler groupings are far more visible before post-processing, providing easier `by eye' identification.
See Figure~\ref{fig:twogroups} for a presentation of the HMC sample chains.

\startlongtable
\begin{deluxetable*}{lrrrrrrr}
\tablecaption{
    Model per-SN parameter median values with 68\textsuperscript{th} percentile upper and lower values. 
    Per-SN parameters $\chi_0 - \langle \chi_0 \rangle$, $s_1$, $c_1$, and $c_2$ have been transformed to magnitudes. The parameter $t_0$
    is the date corresponding to maximum phase node of the light curve model
    for the third band centered at $4084$~\r{A}.
    }
\tablehead{\colhead{SN} & \colhead{$\chi_0 - \langle \chi_0 \rangle$} & \colhead{$s_1$} & \colhead{$c_1$} & \colhead{$c_2$} & \colhead{$t_0$ (MJD)}}
\label{table:table1}
\startdata
SNF20060512-001 & $-0.04\pm0.03$ & $-0.02\pm0.01$ & $-0.33\pm0.09$ & $0.12\pm0.03$ & $53882.5\pm0.12$ \\
SNF20060526-003 & $-0.06\pm0.03$ & $-0.07\pm0.01$ & $-0.31\pm0.09$ & $-0.04\pm0.02$ & $53893.4\pm0.21$ \\
SNF20061020-000 & $0.30\pm0.03$ & $0.21\pm0.01$ & $0.44\pm0.11$ & $0.08\pm0.03$ & $54037.9\pm0.19$ \\
SNF20061021-003 & $-0.08\pm0.03$ & $-0.05\pm0.01$ & $0.38\pm0.08$ & $-0.04\pm0.02$ & $54039.7\pm0.13$ \\
SNF20061022-005 & $-0.11\pm0.03$ & $-0.08\pm0.02$ & $-0.34\pm0.10$ & $0.15\pm0.03$ & $54040.8\pm0.17$ \\
SNF20061111-002 & $0.08\pm0.03$ & $-0.01\pm0.01$ & $-0.30\pm0.12$ & $-0.05\pm0.03$ & $54060.8\pm0.18$ \\
SNF20070424-003 & $0.08\pm0.03$ & $-0.05\pm0.01$ & $-0.26\pm0.09$ & $-0.07\pm0.02$ & $54225.8\pm0.16$ \\
SNF20070506-006 & $-0.19\pm0.03$ & $-0.07\pm0.01$ & $-0.94\pm0.08$ & $0.09\pm0.02$ & $54245.9\pm0.10$ \\
SNF20070630-006 & $-0.09\pm0.02$ & $-0.04\pm0.01$ & $-0.16\pm0.08$ & $-0.06\pm0.02$ & $54294.9\pm0.11$ \\
SNF20070717-003 & $-0.12\pm0.02$ & $0.03\pm0.01$ & $0.90\pm0.08$ & $-0.12\pm0.02$ & $54310.3\pm0.11$ \\
SNF20070802-000 & $-0.19\pm0.03$ & $-0.03\pm0.01$ & $1.00\pm0.09$ & $-0.15\pm0.02$ & $54326.7\pm0.13$ \\
SNF20070818-001 & $-0.24\pm0.03$ & $-0.04\pm0.01$ & $0.80\pm0.10$ & $-0.16\pm0.01$ & $54339.0\pm0.18$ \\
SNF20070831-015 & $-0.28\pm0.03$ & $-0.15\pm0.01$ & $-0.23\pm0.11$ & $-0.03\pm0.03$ & $54352.4\pm0.26$ \\
SNF20070902-018 & $0.11\pm0.03$ & $0.08\pm0.01$ & $0.66\pm0.10$ & $0.04\pm0.03$ & $54353.2\pm0.17$ \\
SNF20080507-000 & $-0.30\pm0.02$ & $-0.06\pm0.01$ & $0.44\pm0.08$ & $0.04\pm0.02$ & $54601.7\pm0.16$ \\
SNF20080510-001 & $-0.01\pm0.03$ & $-0.05\pm0.01$ & $-0.55\pm0.08$ & $-0.05\pm0.02$ & $54607.5\pm0.12$ \\
SNF20080510-005 & $-0.00\pm0.03$ & $-0.11\pm0.01$ & $-0.75\pm0.11$ & $-0.02\pm0.03$ & $54608.1\pm0.18$ \\
SNF20080512-010 & $0.25\pm0.03$ & $0.16\pm0.02$ & $-0.24\pm0.09$ & $0.12\pm0.02$ & $54603.9\pm0.12$ \\
SNF20080514-002 & $0.50\pm0.02$ & $0.20\pm0.01$ & $-0.73\pm0.05$ & $0.14\pm0.01$ & $54613.6\pm0.06$ \\
SNF20080522-000 & $-0.27\pm0.02$ & $-0.08\pm0.01$ & $-0.87\pm0.08$ & $0.20\pm0.02$ & $54624.3\pm0.10$ \\
SNF20080522-011 & $-0.11\pm0.02$ & $-0.05\pm0.01$ & $-1.02\pm0.07$ & $-0.05\pm0.02$ & $54619.1\pm0.12$ \\
SNF20080531-000 & $0.08\pm0.02$ & $0.05\pm0.01$ & $0.03\pm0.07$ & $-0.08\pm0.02$ & $54625.9\pm0.11$ \\
SNF20080614-010 & $0.24\pm0.03$ & $0.22\pm0.01$ & $0.55\pm0.10$ & $0.08\pm0.03$ & $54637.7\pm0.10$ \\
SNF20080620-000 & $0.13\pm0.03$ & $0.09\pm0.01$ & $0.64\pm0.10$ & $-0.03\pm0.02$ & $54642.7\pm0.16$ \\
SNF20080626-002 & $-0.16\pm0.02$ & $-0.08\pm0.01$ & $-0.47\pm0.08$ & $-0.08\pm0.02$ & $54654.0\pm0.15$ \\
SNF20080714-008 & $-0.17\pm0.02$ & $0.00\pm0.01$ & $1.68\pm0.07$ & $-0.20\pm0.02$ & $54670.0\pm0.15$ \\
SNF20080725-004 & $-0.07\pm0.03$ & $-0.10\pm0.01$ & $0.05\pm0.08$ & $-0.10\pm0.02$ & $54680.8\pm0.20$ \\
SNF20080803-000 & $-0.09\pm0.03$ & $-0.05\pm0.01$ & $0.88\pm0.10$ & $-0.04\pm0.03$ & $54693.1\pm0.13$ \\
SNF20080810-001 & $0.17\pm0.02$ & $0.14\pm0.01$ & $0.10\pm0.07$ & $0.11\pm0.02$ & $54700.9\pm0.08$ \\
SNF20080821-000 & $-0.13\pm0.02$ & $-0.06\pm0.01$ & $-0.23\pm0.08$ & $0.01\pm0.02$ & $54710.1\pm0.15$ \\
SNF20080825-010 & $-0.04\pm0.03$ & $0.11\pm0.01$ & $-0.20\pm0.08$ & $0.08\pm0.02$ & $54715.1\pm0.10$ \\
SNF20080909-030 & $-0.15\pm0.05$ & $-0.04\pm0.01$ & $-0.29\pm0.17$ & $0.05\pm0.05$ & $54733.8\pm0.16$ \\
SNF20080913-031 & $-0.19\pm0.07$ & $0.00\pm0.02$ & $0.55\pm0.25$ & $-0.09\pm0.07$ & $54734.6\pm0.40$ \\
SNF20080918-000 & $-0.15\pm0.03$ & $-0.11\pm0.01$ & $1.10\pm0.11$ & $-0.09\pm0.03$ & $54736.8\pm0.17$ \\
CSS130502\_01 & $0.29\pm0.03$ & $-0.01\pm0.01$ & $-1.23\pm0.08$ & $0.07\pm0.02$ & $56425.0\pm0.14$ \\
LSQ12fhe & $-0.45\pm0.03$ & $-0.07\pm0.01$ & $-0.41\pm0.12$ & $0.22\pm0.03$ & $56214.9\pm0.19$ \\
LSQ14cnm & $-0.00\pm0.03$ & $-0.00\pm0.01$ & $-0.88\pm0.08$ & $-0.03\pm0.02$ & $56833.9\pm0.14$ \\
PTF09dlc & $-0.00\pm0.02$ & $-0.03\pm0.01$ & $-0.56\pm0.07$ & $-0.08\pm0.02$ & $55077.4\pm0.13$ \\
PTF09dnl & $-0.23\pm0.02$ & $-0.07\pm0.01$ & $0.26\pm0.06$ & $-0.05\pm0.02$ & $55077.7\pm0.16$ \\
PTF09fox & $0.01\pm0.02$ & $-0.08\pm0.01$ & $-0.83\pm0.07$ & $-0.05\pm0.01$ & $55135.8\pm0.14$ \\
PTF09foz & $-0.06\pm0.02$ & $0.11\pm0.01$ & $0.48\pm0.07$ & $-0.00\pm0.02$ & $55133.8\pm0.11$ \\
PTF10hmv & $0.11\pm0.02$ & $-0.09\pm0.01$ & $1.04\pm0.06$ & $-0.01\pm0.02$ & $55355.3\pm0.11$ \\
PTF10icb & $0.49\pm0.02$ & $0.01\pm0.01$ & $0.32\pm0.07$ & $0.03\pm0.02$ & $55363.7\pm0.09$ \\
PTF10mwb & $0.21\pm0.02$ & $0.10\pm0.01$ & $0.19\pm0.06$ & $-0.02\pm0.02$ & $55392.6\pm0.08$ \\
PTF10nlg & $0.03\pm0.02$ & $-0.03\pm0.01$ & $1.67\pm0.07$ & $-0.20\pm0.01$ & $55394.8\pm0.16$ \\
PTF10qyz & $0.36\pm0.03$ & $0.10\pm0.01$ & $-0.43\pm0.10$ & $-0.08\pm0.02$ & $55429.0\pm0.12$ \\
PTF10wnm & $-0.00\pm0.02$ & $-0.09\pm0.01$ & $-0.66\pm0.08$ & $0.03\pm0.02$ & $55480.5\pm0.17$ \\
PTF10wof & $0.02\pm0.03$ & $-0.06\pm0.01$ & $0.12\pm0.10$ & $-0.12\pm0.02$ & $55476.7\pm0.28$ \\
PTF11bgv & $0.22\pm0.03$ & $0.03\pm0.01$ & $0.30\pm0.08$ & $0.14\pm0.02$ & $55645.6\pm0.12$ \\
PTF11bju & $-0.23\pm0.03$ & $-0.05\pm0.01$ & $-0.14\pm0.09$ & $0.26\pm0.02$ & $55652.7\pm0.14$ \\
PTF11bnx & $-0.00\pm0.03$ & $0.03\pm0.01$ & $1.44\pm0.09$ & $-0.16\pm0.02$ & $55655.6\pm0.16$ \\
PTF11mkx & $-0.50\pm0.04$ & $-0.00\pm0.01$ & $-0.33\pm0.13$ & $0.27\pm0.02$ & $55837.8\pm0.10$ \\
PTF11mty & $-0.04\pm0.02$ & $-0.07\pm0.01$ & $-0.82\pm0.08$ & $0.02\pm0.02$ & $55840.2\pm0.12$ \\
PTF12dxm & $0.34\pm0.03$ & $0.22\pm0.01$ & $0.89\pm0.09$ & $0.02\pm0.02$ & $56054.8\pm0.09$ \\
PTF12fuu & $0.18\pm0.02$ & $-0.05\pm0.01$ & $-1.28\pm0.07$ & $0.01\pm0.02$ & $56115.2\pm0.09$ \\
PTF12grk & $0.10\pm0.03$ & $0.12\pm0.01$ & $0.77\pm0.10$ & $-0.16\pm0.01$ & $56137.0\pm0.10$ \\
PTF12iiq & $0.07\pm0.04$ & $0.12\pm0.01$ & $1.17\pm0.13$ & $-0.17\pm0.02$ & $56183.6\pm0.16$ \\
PTF13anh & $-0.09\pm0.03$ & $0.05\pm0.01$ & $0.51\pm0.12$ & $-0.15\pm0.01$ & $56416.1\pm0.16$ \\
SN2004ef & $0.07\pm0.02$ & $0.14\pm0.01$ & $1.28\pm0.06$ & $-0.08\pm0.02$ & $53265.0\pm0.08$ \\
SN2005hc & $0.00\pm0.03$ & $-0.07\pm0.01$ & $-0.61\pm0.09$ & $-0.03\pm0.02$ & $53670.5\pm0.18$ \\
SN2005hj & $-0.23\pm0.03$ & $-0.13\pm0.01$ & $-0.57\pm0.09$ & $0.11\pm0.02$ & $53677.9\pm0.23$ \\
SN2006cj & $0.02\pm0.02$ & $-0.06\pm0.01$ & $-0.84\pm0.06$ & $0.04\pm0.02$ & $53882.4\pm0.11$ \\
SN2007bd & $0.16\pm0.03$ & $0.03\pm0.02$ & $-0.58\pm0.10$ & $-0.02\pm0.02$ & $54207.6\pm0.16$ \\
SN2010dt & $0.03\pm0.03$ & $0.00\pm0.01$ & $-0.44\pm0.09$ & $-0.10\pm0.01$ & $55363.4\pm0.11$ \\
SN2011bc & $-0.05\pm0.02$ & $0.04\pm0.01$ & $1.18\pm0.08$ & $-0.05\pm0.02$ & $55669.6\pm0.15$ \\
SN2011be & $0.22\pm0.03$ & $-0.03\pm0.01$ & $-1.20\pm0.08$ & $0.08\pm0.02$ & $55656.9\pm0.12$ \\
SN2011ga & $0.05\pm0.02$ & $0.02\pm0.01$ & $-0.43\pm0.08$ & $-0.03\pm0.02$ & $55832.7\pm0.13$ \\
SN2011gf & $0.07\pm0.03$ & $-0.07\pm0.01$ & $-1.54\pm0.07$ & $-0.04\pm0.01$ & $55833.5\pm0.12$ \\
SN2011hr & $-0.25\pm0.03$ & $-0.03\pm0.01$ & $0.12\pm0.08$ & $0.25\pm0.02$ & $55891.0\pm0.10$ \\
SN2012cg & $0.63\pm0.02$ & $0.00\pm0.01$ & $0.67\pm0.07$ & $0.05\pm0.02$ & $56084.5\pm0.10$ \\
SN2012dn & $-0.08\pm0.03$ & $-0.03\pm0.01$ & $0.23\pm0.09$ & $0.20\pm0.02$ & $56135.1\pm0.06$ \\
SN2012fr & $-0.56\pm0.02$ & $-0.09\pm0.01$ & $-0.60\pm0.05$ & $-0.09\pm0.01$ & $56247.1\pm0.05$ \\
SN2013bt & $0.40\pm0.04$ & $0.22\pm0.01$ & $-0.28\pm0.11$ & $0.12\pm0.03$ & $56411.4\pm0.11$
\enddata
\end{deluxetable*}

\bibliography{main}{}

\begin{thebibliography}{}
\expandafter\ifx\csname natexlab\endcsname\relax\def\natexlab#1{#1}\fi
\providecommand{\url}[1]{\href{#1}{#1}}
\providecommand{\dodoi}[1]{doi:~\href{http://doi.org/#1}{\nolinkurl{#1}}}
\providecommand{\doeprint}[1]{\href{http://ascl.net/#1}{\nolinkurl{http://ascl.net/#1}}}
\providecommand{\doarXiv}[1]{\href{https://arxiv.org/abs/#1}{\nolinkurl{https://arxiv.org/abs/#1}}}

\bibitem[{{Aldering} {et~al.}(2002){Aldering}, {Adam}, {Antilogus}, {Astier},
  {Bacon}, {Bongard}, {Bonnaud}, {Copin}, {Hardin}, {Henault}, {Howell},
  {Lemonnier}, {Levy}, {Loken}, {Nugent}, {Pain}, {Pecontal}, {Pecontal},
  {Perlmutter}, {Quimby}, {Schahmaneche}, {Smadja}, \&
  {Wood-Vasey}}]{Aldering2002}
{Aldering}, G., {Adam}, G., {Antilogus}, P., {et~al.} 2002, in Society of
  Photo-Optical Instrumentation Engineers (SPIE) Conference Series, Vol. 4836,
  Survey and Other Telescope Technologies and Discoveries, ed. J.~A. {Tyson} \&
  S.~{Wolff}, 61--72, \dodoi{10.1117/12.458107}

\bibitem[{{Aldering} {et~al.}(2020){Aldering}, {Antilogus}, {Aragon}, {Bailey},
  {Baltay}, {Bongard}, {Boone}, {Buton}, {Chotard}, {Copin}, {Dixon},
  {Fakhouri}, {Feindt}, {Fouchez}, {Gangler}, {Hayden}, {Hillebrandt}, {Kim},
  {Kowalski}, {K{\"u}sters}, {L{\'e}get}, {Lin}, {Lombardo}, {Mondon},
  {Nordin}, {Pain}, {Pecontal}, {Pereira}, {Perlmutter}, {Ponder},
  {Pruzhinskaya}, {Rabinowitz}, {Rigault}, {Rubin}, {Runge}, {Saunders},
  {Says}, {Smadja}, {Suzuki}, {Tao}, {Taubenberger}, {Thomas}, {Vincenzi},
  {Weaver}, \& {Nearby Supernova Factory Collaboration}}]{Aldering2020}
{Aldering}, G., {Antilogus}, P., {Aragon}, C., {et~al.} 2020, Research Notes of
  the American Astronomical Society, 4, 63, \dodoi{10.3847/2515-5172/ab8fa5}

\bibitem[{{Bailey} {et~al.}(2009){Bailey}, {Aldering}, {Antilogus}, {Aragon},
  {Baltay}, {Bongard}, {Buton}, {Childress}, {Chotard}, {Copin}, {Gangler},
  {Loken}, {Nugent}, {Pain}, {Pecontal}, {Pereira}, {Perlmutter}, {Rabinowitz},
  {Rigaudier}, {Runge}, {Scalzo}, {Smadja}, {Swift}, {Tao}, {Thomas}, {Wu}, \&
  {Nearby Supernova Factory}}]{Bailey2009}
{Bailey}, S., {Aldering}, G., {Antilogus}, P., {et~al.} 2009, \aap, 500, L17,
  \dodoi{10.1051/0004-6361/200911973}

\bibitem[{{Betoule} {et~al.}(2014){Betoule}, {Kessler}, {Guy}, {Mosher},
  {Hardin}, {Biswas}, {Astier}, {El-Hage}, {Konig}, {Kuhlmann}, {Marriner},
  {Pain}, {Regnault}, {Balland}, {Bassett}, {Brown}, {Campbell}, {Carlberg},
  {Cellier-Holzem}, {Cinabro}, {Conley}, {D'Andrea}, {DePoy}, {Doi}, {Ellis},
  {Fabbro}, {Filippenko}, {Foley}, {Frieman}, {Fouchez}, {Galbany}, {Goobar},
  {Gupta}, {Hill}, {Hlozek}, {Hogan}, {Hook}, {Howell}, {Jha}, {Le Guillou},
  {Leloudas}, {Lidman}, {Marshall}, {M{\"o}ller}, {Mour{\~a}o}, {Neveu},
  {Nichol}, {Olmstead}, {Palanque-Delabrouille}, {Perlmutter}, {Prieto},
  {Pritchet}, {Richmond}, {Riess}, {Ruhlmann-Kleider}, {Sako}, {Schahmaneche},
  {Schneider}, {Smith}, {Sollerman}, {Sullivan}, {Walton}, \&
  {Wheeler}}]{Betoule14}
{Betoule}, M., {Kessler}, R., {Guy}, J., {et~al.} 2014, \aap, 568, A22,
  \dodoi{10.1051/0004-6361/201423413}

\bibitem[{{Blondin} {et~al.}(2012){Blondin}, {Matheson}, {Kirshner}, {Mandel},
  {Berlind}, {Calkins}, {Challis}, {Garnavich}, {Jha}, {Modjaz}, {Riess}, \&
  {Schmidt}}]{Blondin2012}
{Blondin}, S., {Matheson}, T., {Kirshner}, R.~P., {et~al.} 2012, \aj, 143, 126,
  \dodoi{10.1088/0004-6256/143/5/126}

\bibitem[{{Bohlin}(2014)}]{Bohlin2014}
{Bohlin}, R.~C. 2014, \aj, 147, 127, \dodoi{10.1088/0004-6256/147/6/127}

\bibitem[{{Bongard} {et~al.}(2011){Bongard}, {Soulez}, {Thi{\'e}baut}, \&
  {Pecontal}}]{Bongard2011}
{Bongard}, S., {Soulez}, F., {Thi{\'e}baut}, {\'E}., \& {Pecontal}, {\'E}.
  2011, \mnras, 418, 258, \dodoi{10.1111/j.1365-2966.2011.19480.x}

\bibitem[{{Boone} {et~al.}(2021{\natexlab{a}}){Boone}, {Aldering}, {Antilogus},
  {Aragon}, {Bailey}, {Baltay}, {Bongard}, {Buton}, {Copin}, {Dixon},
  {Fouchez}, {Gangler}, {Gupta}, {Hayden}, {Hillebrandt}, {Kim}, {Kowalski},
  {K{\"u}sters}, {L{\'e}get}, {Mondon}, {Nordin}, {Pain}, {Pecontal},
  {Pereira}, {Perlmutter}, {Ponder}, {Rabinowitz}, {Rigault}, {Rubin}, {Runge},
  {Saunders}, {Smadja}, {Suzuki}, {Tao}, {Taubenberger}, {Thomas}, \&
  {Vincenzi}}]{Boone2021b}
{Boone}, K., {Aldering}, G., {Antilogus}, P., {et~al.} 2021{\natexlab{a}},
  \apj, 912, 71, \dodoi{10.3847/1538-4357/abec3b}

\bibitem[{{Boone} {et~al.}(2021{\natexlab{b}}){Boone}, {Aldering}, {Antilogus},
  {Aragon}, {Bailey}, {Baltay}, {Bongard}, {Buton}, {Copin}, {Dixon},
  {Fouchez}, {Gangler}, {Gupta}, {Hayden}, {Hillebrandt}, {Kim}, {Kowalski},
  {K{\"u}sters}, {L{\'e}get}, {Mondon}, {Nordin}, {Pain}, {Pecontal},
  {Pereira}, {Perlmutter}, {Ponder}, {Rabinowitz}, {Rigault}, {Rubin}, {Runge},
  {Saunders}, {Smadja}, {Suzuki}, {Tao}, {Taubenberger}, {Thomas}, \&
  {Vincenzi}}]{Boone2021}
---. 2021{\natexlab{b}}, \apj, 912, 70, \dodoi{10.3847/1538-4357/abec3c}

\bibitem[{{Branch} {et~al.}(2009){Branch}, {Chau Dang}, \&
  {Baron}}]{Branch2009}
{Branch}, D., {Chau Dang}, L., \& {Baron}, E. 2009, \pasp, 121, 238,
  \dodoi{10.1086/597788}

\bibitem[{{Branch} {et~al.}(1993){Branch}, {Fisher}, \& {Nugent}}]{Branch1993}
{Branch}, D., {Fisher}, A., \& {Nugent}, P. 1993, \aj, 106, 2383,
  \dodoi{10.1086/116810}

\bibitem[{{Brout} \& {Scolnic}(2020)}]{Brout2020}
{Brout}, D., \& {Scolnic}, D. 2020, arXiv e-prints, arXiv:2004.10206.
\newblock \doarXiv{2004.10206}

\bibitem[{{Burns} {et~al.}(2011){Burns}, {Stritzinger}, {Phillips}, {Kattner},
  {Persson}, {Madore}, {Freedman}, {Boldt}, {Campillay}, {Contreras},
  {Folatelli}, {Gonzalez}, {Krzeminski}, {Morrell}, {Salgado}, \&
  {Suntzeff}}]{Burns2011}
{Burns}, C.~R., {Stritzinger}, M., {Phillips}, M.~M., {et~al.} 2011, \aj, 141,
  19, \dodoi{10.1088/0004-6256/141/1/19}

\bibitem[{{Buton} {et~al.}(2013){Buton}, {Copin}, {Aldering}, {Antilogus},
  {Aragon}, {Bailey}, {Baltay}, {Bongard}, {Canto}, {Cellier-Holzem},
  {Childress}, {Chotard}, {Fakhouri}, {Gangler}, {Guy}, {Hsiao}, {Kerschhaggl},
  {Kowalski}, {Loken}, {Nugent}, {Paech}, {Pain}, {P{\'e}contal}, {Pereira},
  {Perlmutter}, {Rabinowitz}, {Rigault}, {Runge}, {Scalzo}, {Smadja}, {Tao},
  {Thomas}, {Weaver}, {Wu}, \& {Nearby SuperNova Factory}}]{Buton2013}
{Buton}, C., {Copin}, Y., {Aldering}, G., {et~al.} 2013, \aap, 549, A8,
  \dodoi{10.1051/0004-6361/201219834}

\bibitem[{{Cardelli} {et~al.}(1989){Cardelli}, {Clayton}, \&
  {Mathis}}]{Cardelli1989}
{Cardelli}, J.~A., {Clayton}, G.~C., \& {Mathis}, J.~S. 1989, \apj, 345, 245,
  \dodoi{10.1086/167900}

\bibitem[{Carpenter {et~al.}(2017)Carpenter, Gelman, Hoffman, Lee, Goodrich,
  Betancourt, Brubaker, Guo, Li, \& Riddell}]{Stan}
Carpenter, B., Gelman, A., Hoffman, M.~D., {et~al.} 2017, Journal of
  statistical software, 76

\bibitem[{{Chotard} {et~al.}(2011){Chotard}, {Gangler}, {Aldering},
  {Antilogus}, {Aragon}, {Bailey}, {Baltay}, {Bongard}, {Buton}, {Canto},
  {Childress}, {Copin}, {Fakhouri}, {Hsiao}, {Kerschhaggl}, {Kowalski},
  {Loken}, {Nugent}, {Paech}, {Pain}, {Pecontal}, {Pereira}, {Perlmutter},
  {Rabinowitz}, {Runge}, {Scalzo}, {Smadja}, {Tao}, {Thomas}, {Weaver}, {Wu},
  \& {Nearby Supernova Factory}}]{Chotard2011}
{Chotard}, N., {Gangler}, E., {Aldering}, G., {et~al.} 2011, \aap, 529, L4,
  \dodoi{10.1051/0004-6361/201116723}

\bibitem[{{Conley} {et~al.}(2008){Conley}, {Sullivan}, {Hsiao}, {Guy},
  {Astier}, {Balam}, {Balland}, {Basa}, {Carlberg}, {Fouchez}, {Hardin},
  {Howell}, {Hook}, {Pain}, {Perrett}, {Pritchet}, \& {Regnault}}]{Conley2008}
{Conley}, A., {Sullivan}, M., {Hsiao}, E.~Y., {et~al.} 2008, \apj, 681, 482,
  \dodoi{10.1086/588518}

\bibitem[{{Dhawan} {et~al.}(2015){Dhawan}, {Leibundgut}, {Spyromilio}, \&
  {Maguire}}]{Dhawan2015}
{Dhawan}, S., {Leibundgut}, B., {Spyromilio}, J., \& {Maguire}, K. 2015,
  \mnras, 448, 1345, \dodoi{10.1093/mnras/stu2716}

\bibitem[{{Fakhouri} {et~al.}(2015){Fakhouri}, {Boone}, {Aldering},
  {Antilogus}, {Aragon}, {Bailey}, {Baltay}, {Barbary}, {Baugh}, {Bongard},
  {Buton}, {Chen}, {Childress}, {Chotard}, {Copin}, {Fagrelius}, {Feindt},
  {Fleury}, {Fouchez}, {Gangler}, {Hayden}, {Kim}, {Kowalski}, {Leget},
  {Lombardo}, {Nordin}, {Pain}, {Pecontal}, {Pereira}, {Perlmutter},
  {Rabinowitz}, {Ren}, {Rigault}, {Rubin}, {Runge}, {Saunders}, {Scalzo},
  {Smadja}, {Sofiatti}, {Strovink}, {Suzuki}, {Tao}, {Thomas}, {Weaver}, \&
  {Nearby Supernova Factory}}]{Fakhouri2015}
{Fakhouri}, H.~K., {Boone}, K., {Aldering}, G., {et~al.} 2015, \apj, 815, 58,
  \dodoi{10.1088/0004-637X/815/1/58}

\bibitem[{{Foley}(2013)}]{Foley2013b}
{Foley}, R.~J. 2013, \mnras, 435, 273, \dodoi{10.1093/mnras/stt1292}

\bibitem[{{Guy} {et~al.}(2007){Guy}, {Astier}, {Baumont}, {Hardin}, {Pain},
  {Regnault}, {Basa}, {Carlberg}, {Conley}, {Fabbro}, {Fouchez}, {Hook},
  {Howell}, {Perrett}, {Pritchet}, {Rich}, {Sullivan}, {Antilogus}, {Aubourg},
  {Bazin}, {Bronder}, {Filiol}, {Palanque-Delabrouille}, {Ripoche}, \&
  {Ruhlmann-Kleider}}]{Guy2007}
{Guy}, J., {Astier}, P., {Baumont}, S., {et~al.} 2007, \aap, 466, 11,
  \dodoi{10.1051/0004-6361:20066930}

\bibitem[{{Hamuy} {et~al.}(1994){Hamuy}, {Suntzeff}, {Heathcote}, {Walker},
  {Gigoux}, \& {Phillips}}]{Hamuy1994}
{Hamuy}, M., {Suntzeff}, N.~B., {Heathcote}, S.~R., {et~al.} 1994, \pasp, 106,
  566, \dodoi{10.1086/133417}

\bibitem[{{Hamuy} {et~al.}(1992){Hamuy}, {Walker}, {Suntzeff}, {Gigoux},
  {Heathcote}, \& {Phillips}}]{Hamuy1992}
{Hamuy}, M., {Walker}, A.~R., {Suntzeff}, N.~B., {et~al.} 1992, \pasp, 104,
  533, \dodoi{10.1086/133028}

\bibitem[{{Hitzer}(2013)}]{Hitzer2013}
{Hitzer}, E. 2013, arXiv e-prints, arXiv:1306.1660.
\newblock \doarXiv{1306.1660}

\bibitem[{{Hoyle} \& {Fowler}(1960)}]{Hoyle1960}
{Hoyle}, F., \& {Fowler}, W.~A. 1960, \apj, 132, 565, \dodoi{10.1086/146963}

\bibitem[{{Iben} \& {Tutukov}(1984)}]{Iben+Tutukov1984}
{Iben}, I., J., \& {Tutukov}, A.~V. 1984, \apjs, 54, 335,
  \dodoi{10.1086/190932}

\bibitem[{{Ivezi{\'c}} {et~al.}(2019){Ivezi{\'c}}, {Kahn}, {Tyson}, {Abel},
  {Acosta}, {Allsman}, {Alonso}, {AlSayyad}, {Anderson}, {Andrew}, {Angel},
  {Angeli}, {Ansari}, {Antilogus}, {Araujo}, {Armstrong}, {Arndt}, {Astier},
  {Aubourg}, {Auza}, {Axelrod}, {Bard}, {Barr}, {Barrau}, {Bartlett}, {Bauer},
  {Bauman}, {Baumont}, {Bechtol}, {Bechtol}, {Becker}, {Becla}, {Beldica},
  {Bellavia}, {Bianco}, {Biswas}, {Blanc}, {Blazek}, {Blandford}, {Bloom},
  {Bogart}, {Bond}, {Booth}, {Borgland}, {Borne}, {Bosch}, {Boutigny},
  {Brackett}, {Bradshaw}, {Brandt}, {Brown}, {Bullock}, {Burchat}, {Burke},
  {Cagnoli}, {Calabrese}, {Callahan}, {Callen}, {Carlin}, {Carlson},
  {Chandrasekharan}, {Charles-Emerson}, {Chesley}, {Cheu}, {Chiang}, {Chiang},
  {Chirino}, {Chow}, {Ciardi}, {Claver}, {Cohen-Tanugi}, {Cockrum}, {Coles},
  {Connolly}, {Cook}, {Cooray}, {Covey}, {Cribbs}, {Cui}, {Cutri}, {Daly},
  {Daniel}, {Daruich}, {Daubard}, {Daues}, {Dawson}, {Delgado}, {Dellapenna},
  {de Peyster}, {de Val-Borro}, {Digel}, {Doherty}, {Dubois},
  {Dubois-Felsmann}, {Durech}, {Economou}, {Eifler}, {Eracleous}, {Emmons},
  {Fausti Neto}, {Ferguson}, {Figueroa}, {Fisher-Levine}, {Focke}, {Foss},
  {Frank}, {Freemon}, {Gangler}, {Gawiser}, {Geary}, {Gee}, {Geha}, {Gessner},
  {Gibson}, {Gilmore}, {Glanzman}, {Glick}, {Goldina}, {Goldstein}, {Goodenow},
  {Graham}, {Gressler}, {Gris}, {Guy}, {Guyonnet}, {Haller}, {Harris},
  {Hascall}, {Haupt}, {Hernandez}, {Herrmann}, {Hileman}, {Hoblitt}, {Hodgson},
  {Hogan}, {Howard}, {Huang}, {Huffer}, {Ingraham}, {Innes}, {Jacoby}, {Jain},
  {Jammes}, {Jee}, {Jenness}, {Jernigan}, {Jevremovi{\'c}}, {Johns}, {Johnson},
  {Johnson}, {Jones}, {Juramy-Gilles}, {Juri{\'c}}, {Kalirai}, {Kallivayalil},
  {Kalmbach}, {Kantor}, {Karst}, {Kasliwal}, {Kelly}, {Kessler}, {Kinnison},
  {Kirkby}, {Knox}, {Kotov}, {Krabbendam}, {Krughoff}, {Kub{\'a}nek},
  {Kuczewski}, {Kulkarni}, {Ku}, {Kurita}, {Lage}, {Lambert}, {Lange},
  {Langton}, {Le Guillou}, {Levine}, {Liang}, {Lim}, {Lintott}, {Long},
  {Lopez}, {Lotz}, {Lupton}, {Lust}, {MacArthur}, {Mahabal}, {Mandelbaum},
  {Markiewicz}, {Marsh}, {Marshall}, {Marshall}, {May}, {McKercher}, {McQueen},
  {Meyers}, {Migliore}, {Miller}, {Mills}, {Miraval}, {Moeyens}, {Moolekamp},
  {Monet}, {Moniez}, {Monkewitz}, {Montgomery}, {Morrison}, {Mueller},
  {Muller}, {Mu{\~n}oz Arancibia}, {Neill}, {Newbry}, {Nief}, {Nomerotski},
  {Nordby}, {O'Connor}, {Oliver}, {Olivier}, {Olsen}, {O'Mullane}, {Ortiz},
  {Osier}, {Owen}, {Pain}, {Palecek}, {Parejko}, {Parsons}, {Pease},
  {Peterson}, {Peterson}, {Petravick}, {Libby Petrick}, {Petry},
  {Pierfederici}, {Pietrowicz}, {Pike}, {Pinto}, {Plante}, {Plate}, {Plutchak},
  {Price}, {Prouza}, {Radeka}, {Rajagopal}, {Rasmussen}, {Regnault}, {Reil},
  {Reiss}, {Reuter}, {Ridgway}, {Riot}, {Ritz}, {Robinson}, {Roby}, {Roodman},
  {Rosing}, {Roucelle}, {Rumore}, {Russo}, {Saha}, {Sassolas}, {Schalk},
  {Schellart}, {Schindler}, {Schmidt}, {Schneider}, {Schneider}, {Schoening},
  {Schumacher}, {Schwamb}, {Sebag}, {Selvy}, {Sembroski}, {Seppala}, {Serio},
  {Serrano}, {Shaw}, {Shipsey}, {Sick}, {Silvestri}, {Slater}, {Smith},
  {Smith}, {Sobhani}, {Soldahl}, {Storrie-Lombardi}, {Stover}, {Strauss},
  {Street}, {Stubbs}, {Sullivan}, {Sweeney}, {Swinbank}, {Szalay}, {Takacs},
  {Tether}, {Thaler}, {Thayer}, {Thomas}, {Thornton}, {Thukral}, {Tice},
  {Trilling}, {Turri}, {Van Berg}, {Vanden Berk}, {Vetter}, {Virieux},
  {Vucina}, {Wahl}, {Walkowicz}, {Walsh}, {Walter}, {Wang}, {Wang}, {Warner},
  {Wiecha}, {Willman}, {Winters}, {Wittman}, {Wolff}, {Wood-Vasey}, {Wu},
  {Xin}, {Yoachim}, \& {Zhan}}]{Ivezic2019}
{Ivezi{\'c}}, {\v{Z}}., {Kahn}, S.~M., {Tyson}, J.~A., {et~al.} 2019, \apj,
  873, 111, \dodoi{10.3847/1538-4357/ab042c}

\bibitem[{{Jha} {et~al.}(2007){Jha}, {Riess}, \& {Kirshner}}]{Jha2007}
{Jha}, S., {Riess}, A.~G., \& {Kirshner}, R.~P. 2007, \apj, 659, 122,
  \dodoi{10.1086/512054}

\bibitem[{{Jha} {et~al.}(2019){Jha}, {Maguire}, \& {Sullivan}}]{Jha2019}
{Jha}, S.~W., {Maguire}, K., \& {Sullivan}, M. 2019, Nature Astronomy, 3, 706,
  \dodoi{10.1038/s41550-019-0858-0}

\bibitem[{{Kasen} \& {Woosley}(2007)}]{Kasen2007}
{Kasen}, D., \& {Woosley}, S.~E. 2007, \apj, 656, 661, \dodoi{10.1086/510375}

\bibitem[{{Kenworthy} {et~al.}(2021){Kenworthy}, {Jones}, {Dai}, {Kessler},
  {Scolnic}, {Brout}, {Siebert}, {Pierel}, {Dettman}, {Dimitriadis}, {Foley},
  {Jha}, {Pan}, {Riess}, {Rodney}, \& {Rojas-Bravo}}]{Kenworthy2021}
{Kenworthy}, W.~D., {Jones}, D.~O., {Dai}, M., {et~al.} 2021, \apj, 923, 265,
  \dodoi{10.3847/1538-4357/ac30d8}

\bibitem[{{Kim} {et~al.}(1996){Kim}, {Goobar}, \& {Perlmutter}}]{Kim1996}
{Kim}, A., {Goobar}, A., \& {Perlmutter}, S. 1996, \pasp, 108, 190,
  \dodoi{10.1086/133709}

\bibitem[{{Lantz} {et~al.}(2004){Lantz}, {Aldering}, {Antilogus}, {Bonnaud},
  {Capoani}, {Castera}, {Copin}, {Dubet}, {Gangler}, {Henault}, {Lemonnier},
  {Pain}, {Pecontal}, {Pecontal}, \& {Smadja}}]{Lantz2004}
{Lantz}, B., {Aldering}, G., {Antilogus}, P., {et~al.} 2004, in Society of
  Photo-Optical Instrumentation Engineers (SPIE) Conference Series, Vol. 5249,
  Optical Design and Engineering, ed. L.~{Mazuray}, P.~J. {Rogers}, \&
  R.~{Wartmann}, 146--155, \dodoi{10.1117/12.512493}

\bibitem[{{L{\'e}get} {et~al.}(2020){L{\'e}get}, {Gangler}, {Mondon},
  {Aldering}, {Antilogus}, {Aragon}, {Bailey}, {Baltay}, {Barbary}, {Bongard},
  {Boone}, {Buton}, {Chotard}, {Copin}, {Dixon}, {Fagrelius}, {Feindt},
  {Fouchez}, {Hayden}, {Hillebrandt}, {Kim}, {Kowalski}, {Kuesters},
  {Lombardo}, {Lin}, {Nordin}, {Pain}, {Pecontal}, {Pereira}, {Perlmutter},
  {Ponder}, {Pruzhinskaya}, {Rabinowitz}, {Rigault}, {Runge}, {Rubin},
  {Saunders}, {Says}, {Smadja}, {Sofiatti}, {Suzuki}, {Taubenberger}, {Tao}, \&
  {Thomas}}]{Leget2020}
{L{\'e}get}, P.~F., {Gangler}, E., {Mondon}, F., {et~al.} 2020, \aap, 636, A46,
  \dodoi{10.1051/0004-6361/201834954}

\bibitem[{{Mandel} {et~al.}(2017){Mandel}, {Scolnic}, {Shariff}, {Foley}, \&
  {Kirshner}}]{Mandel2017}
{Mandel}, K.~S., {Scolnic}, D.~M., {Shariff}, H., {Foley}, R.~J., \&
  {Kirshner}, R.~P. 2017, \apj, 842, 93, \dodoi{10.3847/1538-4357/aa6038}

\bibitem[{{Mandel} {et~al.}(2022){Mandel}, {Thorp}, {Narayan}, {Friedman}, \&
  {Avelino}}]{Mandel2022}
{Mandel}, K.~S., {Thorp}, S., {Narayan}, G., {Friedman}, A.~S., \& {Avelino},
  A. 2022, \mnras, 510, 3939, \dodoi{10.1093/mnras/stab3496}

\bibitem[{{Nugent} {et~al.}(2002){Nugent}, {Kim}, \& {Perlmutter}}]{Nugent2002}
{Nugent}, P., {Kim}, A., \& {Perlmutter}, S. 2002, \pasp, 114, 803,
  \dodoi{10.1086/341707}

\bibitem[{{Nugent} {et~al.}(1995){Nugent}, {Phillips}, {Baron}, {Branch}, \&
  {Hauschildt}}]{Nugent1995}
{Nugent}, P., {Phillips}, M., {Baron}, E., {Branch}, D., \& {Hauschildt}, P.
  1995, \apjl, 455, L147, \dodoi{10.1086/309846}

\bibitem[{Pedregosa {et~al.}(2011)Pedregosa, Varoquaux, Gramfort, Michel,
  Thirion, Grisel, Blondel, Prettenhofer, Weiss, Dubourg, Vanderplas, Passos,
  Cournapeau, Brucher, Perrot, \& Duchesnay}]{scikit-learn}
Pedregosa, F., Varoquaux, G., Gramfort, A., {et~al.} 2011, Journal of Machine
  Learning Research, 12, 2825

\bibitem[{{Pereira} {et~al.}(2013){Pereira}, {Thomas}, {Aldering}, {Antilogus},
  {Baltay}, {Benitez-Herrera}, {Bongard}, {Buton}, {Canto}, {Cellier-Holzem},
  {Chen}, {Childress}, {Chotard}, {Copin}, {Fakhouri}, {Fink}, {Fouchez},
  {Gangler}, {Guy}, {Hillebrandt}, {Hsiao}, {Kerschhaggl}, {Kowalski},
  {Kromer}, {Nordin}, {Nugent}, {Paech}, {Pain}, {P{\'e}contal}, {Perlmutter},
  {Rabinowitz}, {Rigault}, {Runge}, {Saunders}, {Smadja}, {Tao},
  {Taubenberger}, {Tilquin}, \& {Wu}}]{Pereira2013}
{Pereira}, R., {Thomas}, R.~C., {Aldering}, G., {et~al.} 2013, \aap, 554, A27,
  \dodoi{10.1051/0004-6361/201221008}

\bibitem[{{Perlmutter} {et~al.}(1997){Perlmutter}, {Gabi}, {Goldhaber},
  {Goobar}, {Groom}, {Hook}, {Kim}, {Kim}, {Lee}, {Pain}, {Pennypacker},
  {Small}, {Ellis}, {McMahon}, {Boyle}, {Bunclark}, {Carter}, {Irwin},
  {Glazebrook}, {Newberg}, {Filippenko}, {Matheson}, {Dopita}, \&
  {Couch}}]{Perlmutter1997}
{Perlmutter}, S., {Gabi}, S., {Goldhaber}, G., {et~al.} 1997, \apj, 483, 565,
  \dodoi{10.1086/304265}

\bibitem[{{Perlmutter} {et~al.}(1999){Perlmutter}, {Aldering}, {Goldhaber},
  {Knop}, {Nugent}, {Castro}, {Deustua}, {Fabbro}, {Goobar}, {Groom}, {Hook},
  {Kim}, {Kim}, {Lee}, {Nunes}, {Pain}, {Pennypacker}, {Quimby}, {Lidman},
  {Ellis}, {Irwin}, {McMahon}, {Ruiz-Lapuente}, {Walton}, {Schaefer}, {Boyle},
  {Filippenko}, {Matheson}, {Fruchter}, {Panagia}, {Newberg}, {Couch}, \& {The
  Supernova Cosmology Project}}]{Perlmutter99}
{Perlmutter}, S., {Aldering}, G., {Goldhaber}, G., {et~al.} 1999, \apj, 517,
  565, \dodoi{10.1086/307221}

\bibitem[{{Phillips}(1993)}]{Phillips93}
{Phillips}, M.~M. 1993, \apjl, 413, L105, \dodoi{10.1086/186970}

\bibitem[{{Pierel} {et~al.}(2022){Pierel}, {Jones}, {Kenworthy}, {Dai},
  {Kessler}, {Ashall}, {Do}, {Peterson}, {Shappee}, {Siebert}, {Barna},
  {Brink}, {Burke}, {Calamida}, {Camacho-Neves}, {Jaeger}, {Filippenko},
  {Foley}, {Galbany}, {Fox}, {Gomez}, {Hiramatsu}, {Hounsell}, {Howell}, {Jha},
  {Kwok}, {P{\'e}rez-Fournon}, {Poidevin}, {Rest}, {Rubin}, {Scolnic},
  {Shirley}, {Strolger}, {Tinyanont}, \& {Wang}}]{Pierel2022}
{Pierel}, J.~D.~R., {Jones}, D.~O., {Kenworthy}, W.~D., {et~al.} 2022, \apj,
  939, 11, \dodoi{10.3847/1538-4357/ac93f9}

\bibitem[{{Riess} {et~al.}(1998{\natexlab{a}}){Riess}, {Nugent}, {Filippenko},
  {Kirshner}, \& {Perlmutter}}]{Riess1998}
{Riess}, A.~G., {Nugent}, P., {Filippenko}, A.~V., {Kirshner}, R.~P., \&
  {Perlmutter}, S. 1998{\natexlab{a}}, \apj, 504, 935, \dodoi{10.1086/306106}

\bibitem[{{Riess} {et~al.}(1996){Riess}, {Press}, \& {Kirshner}}]{Riess1996}
{Riess}, A.~G., {Press}, W.~H., \& {Kirshner}, R.~P. 1996, \apj, 473, 88,
  \dodoi{10.1086/178129}

\bibitem[{{Riess} {et~al.}(1998{\natexlab{b}}){Riess}, {Filippenko}, {Challis},
  {Clocchiatti}, {Diercks}, {Garnavich}, {Gilliland}, {Hogan}, {Jha},
  {Kirshner}, {Leibundgut}, {Phillips}, {Reiss}, {Schmidt}, {Schommer},
  {Smith}, {Spyromilio}, {Stubbs}, {Suntzeff}, \& {Tonry}}]{Riess98}
{Riess}, A.~G., {Filippenko}, A.~V., {Challis}, P., {et~al.}
  1998{\natexlab{b}}, \aj, 116, 1009, \dodoi{10.1086/300499}

\bibitem[{{Robert Kern and the Clifford Team}(2021)}]{clifford}
{Robert Kern and the Clifford Team}. 2021, clifford: Geometric Algebra for
  Python, 1.4.0.
\newblock \url{https://github.com/pygae/clifford}

\bibitem[{{Rose} {et~al.}(2020){Rose}, {Rubin}, {Strolger}, \&
  {Garnavich}}]{Rose2020}
{Rose}, B.~M., {Rubin}, D., {Strolger}, L., \& {Garnavich}, P.~M. 2020, arXiv
  e-prints, arXiv:2012.01460.
\newblock \doarXiv{2012.01460}

\bibitem[{{Rubin} {et~al.}(2015){Rubin}, {Aldering}, {Barbary}, {Boone},
  {Chappell}, {Currie}, {Deustua}, {Fagrelius}, {Fruchter}, {Hayden}, {Lidman},
  {Nordin}, {Perlmutter}, {Saunders}, {Sofiatti}, \& {Supernova Cosmology
  Project}}]{Rubin2015}
{Rubin}, D., {Aldering}, G., {Barbary}, K., {et~al.} 2015, \apj, 813, 137,
  \dodoi{10.1088/0004-637X/813/2/137}

\bibitem[{{Rubin} {et~al.}(2022){Rubin}, {Aldering}, {Antilogus}, {Aragon},
  {Bailey}, {Baltay}, {Bongard}, {Boone}, {Buton}, {Copin}, {Dixon}, {Fouchez},
  {Gangler}, {Gupta}, {Hayden}, {Hillebrandt}, {Kim}, {Kowalski},
  {K{\"u}sters}, {L{\'e}get}, {Mondon}, {Nordin}, {Pain}, {Pecontal},
  {Pereira}, {Perlmutter}, {Ponder}, {Rabinowitz}, {Rigault}, {Runge},
  {Saunders}, {Smadja}, {Suzuki}, {Tao}, {Taubenberger}, {Thomas}, {Vincenzi},
  \& {Nearby Supernova Factory}}]{Rubin2022}
{Rubin}, D., {Aldering}, G., {Antilogus}, P., {et~al.} 2022, \apjs, 263, 1,
  \dodoi{10.3847/1538-4365/ac7b7f}

\bibitem[{{Saunders} {et~al.}(2018){Saunders}, {Aldering}, {Antilogus},
  {Bailey}, {Baltay}, {Barbary}, {Baugh}, {Boone}, {Bongard}, {Buton}, {Chen},
  {Chotard}, {Copin}, {Dixon}, {Fagrelius}, {Fakhouri}, {Feindt}, {Fouchez},
  {Gangler}, {Hayden}, {Hillebrandt}, {Kim}, {Kowalski}, {K{\"u}sters},
  {Leget}, {Lombardo}, {Nordin}, {Pain}, {Pecontal}, {Pereira}, {Perlmutter},
  {Rabinowitz}, {Rigault}, {Rubin}, {Runge}, {Smadja}, {Sofiatti}, {Suzuki},
  {Tao}, {Taubenberger}, {Thomas}, {Vincenzi}, \& {Nearby Supernova
  Factory}}]{Saunders2018}
{Saunders}, C., {Aldering}, G., {Antilogus}, P., {et~al.} 2018, \apj, 869, 167,
  \dodoi{10.3847/1538-4357/aaec7e}

\bibitem[{{Scalzo} {et~al.}(2014){Scalzo}, {Ruiter}, \& {Sim}}]{Scalzo2014}
{Scalzo}, R.~A., {Ruiter}, A.~J., \& {Sim}, S.~A. 2014, \mnras, 445, 2535,
  \dodoi{10.1093/mnras/stu1808}

\bibitem[{{Scolnic} {et~al.}(2014){Scolnic}, {Riess}, {Foley}, {Rest},
  {Rodney}, {Brout}, \& {Jones}}]{Scolnic2014}
{Scolnic}, D.~M., {Riess}, A.~G., {Foley}, R.~J., {et~al.} 2014, \apj, 780, 37,
  \dodoi{10.1088/0004-637X/780/1/37}

\bibitem[{{Scolnic} {et~al.}(2018){Scolnic}, {Jones}, {Rest}, {Pan},
  {Chornock}, {Foley}, {Huber}, {Kessler}, {Narayan}, {Riess}, {Rodney},
  {Berger}, {Brout}, {Challis}, {Drout}, {Finkbeiner}, {Lunnan}, {Kirshner},
  {Sand ers}, {Schlafly}, {Smartt}, {Stubbs}, {Tonry}, {Wood-Vasey}, {Foley},
  {Hand}, {Johnson}, {Burgett}, {Chambers}, {Draper}, {Hodapp}, {Kaiser},
  {Kudritzki}, {Magnier}, {Metcalfe}, {Bresolin}, {Gall}, {Kotak}, {McCrum}, \&
  {Smith}}]{Scolnic18}
{Scolnic}, D.~M., {Jones}, D.~O., {Rest}, A., {et~al.} 2018, \apj, 859, 101,
  \dodoi{10.3847/1538-4357/aab9bb}

\bibitem[{{Soker}(2019)}]{Soker2019}
{Soker}, N. 2019, \nar, 87, 101535, \dodoi{10.1016/j.newar.2020.101535}

\bibitem[{{Stein} {et~al.}(2022){Stein}, {Seljak}, {Bohm}, {Aldering},
  {Antilogus}, {Aragon}, {Bailey}, {Baltay}, {Bongard}, {Boone}, {Buton},
  {Copin}, {Dixon}, {Fouchez}, {Gangler}, {Gupta}, {Hayden}, {Hillebrandt},
  {Karmen}, {Kim}, {Kowalski}, {Kusters}, {Leget}, {Mondon}, {Nordin}, {Pain},
  {Pecontal}, {Pereira}, {Perlmutter}, {Ponder}, {Rabinowitz}, {Rigault},
  {Rubin}, {Runge}, {Saunders}, {Smadja}, {Suzuki}, {Tao}, {Thomas}, \&
  {Vincenzi}}]{Stein2022}
{Stein}, G., {Seljak}, U., {Bohm}, V., {et~al.} 2022, arXiv e-prints,
  arXiv:2207.07645.
\newblock \doarXiv{2207.07645}

\bibitem[{{Thorp} {et~al.}(2021){Thorp}, {Mandel}, {Jones}, {Ward}, \&
  {Narayan}}]{Thorp2021}
{Thorp}, S., {Mandel}, K.~S., {Jones}, D.~O., {Ward}, S.~M., \& {Narayan}, G.
  2021, \mnras, 508, 4310, \dodoi{10.1093/mnras/stab2849}

\bibitem[{{Tripp}(1998)}]{Tripp1998}
{Tripp}, R. 1998, \aap, 331, 815

\bibitem[{{Vehtari} {et~al.}(2019){Vehtari}, {Gelman}, {Simpson}, {Carpenter},
  \& {B{\"u}rkner}}]{Vehtari2019}
{Vehtari}, A., {Gelman}, A., {Simpson}, D., {Carpenter}, B., \& {B{\"u}rkner},
  P.-C. 2019, arXiv e-prints, arXiv:1903.08008.
\newblock \doarXiv{1903.08008}

\bibitem[{{Wang} {et~al.}(2009){Wang}, {Filippenko}, {Ganeshalingam}, {Li},
  {Silverman}, {Wang}, {Chornock}, {Foley}, {Gates}, {Macomber}, {Serduke},
  {Steele}, \& {Wong}}]{Wang2009}
{Wang}, X., {Filippenko}, A.~V., {Ganeshalingam}, M., {et~al.} 2009, \apjl,
  699, L139, \dodoi{10.1088/0004-637X/699/2/L139}

\bibitem[{{Weingartner} \& {Draine}(2001)}]{Weingartner2001}
{Weingartner}, J.~C., \& {Draine}, B.~T. 2001, \apj, 548, 296,
  \dodoi{10.1086/318651}

\bibitem[{{Whelan} \& {Iben}(1973)}]{Whelan+Iben1973}
{Whelan}, J., \& {Iben}, Icko, J. 1973, \apj, 186, 1007, \dodoi{10.1086/152565}

\end{thebibliography}
\bibliographystyle{aasjournal}

\end{document}